\documentclass{aastex701}

\begin{document}

\title{Newly Identified Planetary Nebula Candidates in M31 using Deep-Learning Search of Hubble Space Telescope Images}

\author[0009-0001-4184-9786]{Yiğit C. Atılgan}
\affiliation{Space Science and Solar Energy Research and Application Center (UZAYMER), University of Çukurova, 01330, Adana, Türkiye}
\email{jeff.mangum@nrao.edu}

\author[0009-0001-0652-7634]{Aleyna Demirci}
\affiliation{Space Science and Solar Energy Research and Application Center (UZAYMER), University of Çukurova, 01330, Adana, Türkiye}
\email{1aleynadmrc@gmail.com}

\author[0000-0001-8721-5036]{Nazım Aksaker}
\affiliation{Space Science and Solar Energy Research and Application Center (UZAYMER), University of Çukurova, 01330, Adana, Türkiye}
\affiliation{Adana Organised Industrial Zones Vocational School of Technical Science, Çukurova University, Adana 01410, Türkiye}
\email{naksaker@cu.edu.tr}

\author[0000-0002-4450-5735]{Ammar Abbas Elmas}
\affiliation{Department of Computer Engineering, Faculty of Engineering, University of Çukurova, Adana, Türkiye}
\email{aelmas@cu.edu.tr}

\author[0000-0001-7323-6864]{Mehmet Sarıgül}
\affiliation{Department of Computer Engineering, Faculty of Engineering, University of Çukurova, Adana, Türkiye}
\email{msarigul@cu.edu.tr}

\author[0000-0001-9533-9805]{Aysun Akyüz}
\affiliation{Space Science and Solar Energy Research and Application Center (UZAYMER), University of Çukurova, 01330, Adana, Türkiye}
\affiliation{Department of Physics, University of Çukurova, Adana 01330, Türkiye}
\email{aakyuz@cu.edu.tr}

\begin{abstract}
The \textit{Hubble Space Telescope} (\textit{HST}) has imaged approximately a quarter of a billion stars in M31’s disc, yet only slightly over a thousand planetary nebulae (PNe) are known, mostly from ground-based surveys. We show that deep learning can find them in broad-band images
alone. In what is the first such search, we scan the Panchromatic Hubble Andromeda Treasury (PHAT) and the Panchromatic Hubble Andromeda Southern Treasury (PHAST) mosaics,
nearly one square degree of the disk, where a PN is a faint blue point among crowded stars.
Because the existing catalogs are incomplete, we train four Faster R-CNN detectors with
partial supervision, so that uncataloged PNe are not taught as background, and combine their
rankings into one candidate list, an ensemble. On spatially separate test fields, this list places 73\% of the
known PNe in its top 200 (with a total of over 9,000 detections), 4--6 times the purity of classical color--magnitude selection.
Newer detection transformers and segmentation networks trained on the same data do not surpass our
detectors. The ranking also follows classifications that the detectors never saw: emission-line objects that the literature calls PNe enter the list at 47\%,
against 0.6\% for those called \ion{H}{2} regions, which makes them candidates for
reclassification. Among the top 400 candidates, 175 appear in no published catalog, 12 of them
flagged as cosmic-ray residuals. One of them, a point
source with [\ion{O}{3}] emission in nine independent exposures, meets the PN criteria of the three
major PN studies of M31, so by those criteria it is a PN, yet it appears in no published catalog. Most of the
others lack strong [\ion{O}{3}], as expected if narrow-band surveys have found the bright
PNe, although some may be PNe with weak [\ion{O}{3}]. These results support the near
completeness of ground-based surveys for bright PNe and extend it from the PHAT area to PHAST.
Finally, we identify evaluation pitfalls for detectors of rare objects and recommend simple remedies.
\end{abstract}



\section{Introduction}
\label{sec:intro}

Planetary nebulae (PNe) mark a short phase, lasting a few $10^4$\,yr, in the late evolution of
low- and intermediate-mass stars of about 0.8--8\,$M_\odot$ \citep{2022PASP..134b2001K}. Because
much of their light comes out in a few emission lines, above all in [\ion{O}{3}]\,$\lambda5007$,
single PNe remain visible even in other galaxies, which makes them useful in several ways. Their
luminosity function (PNLF) has a bright cutoff that serves as a standard candle
\citep{1989ApJ...339...39J,1989ApJ...339...53C,2012Ap&SS.341..151C}. The physics behind this cutoff
 is now also being studied in the Milky Way, where individual PNe are resolved. There, three PNe within 1\,mag of
the cutoff have properties similar to those of the brightest PNe in the Large Magellanic Cloud and
M31 \citep{2026arXiv260817380S}. In M31, the velocities of PNe also trace the
kinematics of the disk and the bulge \citep{2006MNRAS.369...97H,2006MNRAS.369..120M}, while their
number per unit light of the parent population, usually written as $\alpha$, is used to study
stellar populations \citep{2006MNRAS.368..877B}. Distance, kinematic and population studies
therefore all depend on a PN catalog that is both complete and clean.

Such catalogs are usually built from narrow-band [\ion{O}{3}] imaging or from slitless
spectroscopy. In M31, this work goes back to \citet{1978ApJS...38..351F} and
\citet{1989ApJ...339...53C}. Later, the Planetary Nebula Spectrograph (PN.S) survey of
\citet{2006MNRAS.369..120M} found 2615 PNe in M31, and the MegaCam survey of
\citet{2019A&A...624A.132B} listed 4289. All of these surveys were made from the ground, however,
so their completeness and positional accuracy are limited in the crowded inner disk and bulge. Their
samples can also be contaminated by compact \ion{H}{2} regions, symbiotic stars, supernova remnants,
novae and background emission-line galaxies, mostly because objects are classified on only one or
two criteria \citep{2010PASA...27..129F,2016JPhCS.728c2008P}. Many other emission-line sources
still lack a firm class. In the area we use, for example, SIMBAD lists nearly two thousand of them
as objects of type EmO.

\begin{figure*}
    \centering
    \includegraphics[width=\textwidth]{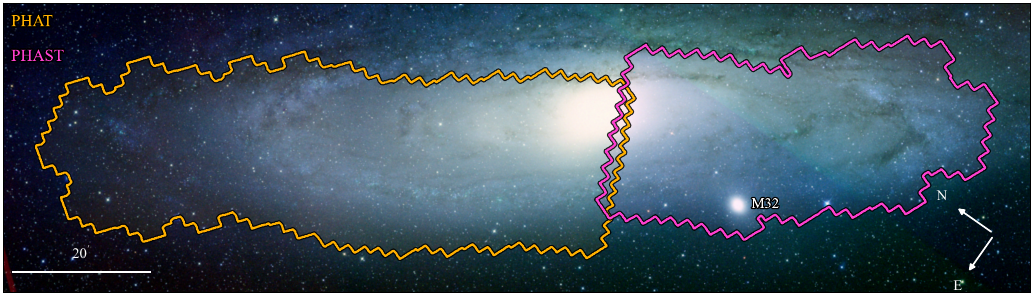}
    \caption{The PHAT and PHAST bricks on M31. The background is a color composite of the second
    Digitized Sky Survey (DSS2), with the near-infrared, red and blue plates in the red, green and
    blue channels, obtained through SkyView \citep{1998IAUS..179..465M}. The outlines show the areas
    of PHAT (gold, 23 bricks) and PHAST (pink, 13 bricks), each the union of the ACS/WFC F475W
    pointings of the survey, taken from the MAST archive. The two surveys overlap in a narrow strip
    south of the bulge. The image is rotated so that the bricks run horizontally, and the arrows mark
    north and east. The bar is $20\arcmin$ long.}
    \label{fig:m31}
\end{figure*}

Images from the \textit{Hubble Space Telescope} (\textit{HST}) are much sharper. The Panchromatic
Hubble Andromeda Treasury \citep[PHAT;][]{2012ApJS..200...18D} used \textit{HST} to image about a
third of the star-forming disk of M31, in six filters from the ultraviolet to the near-infrared. Its
contiguous area of about 0.5\,deg$^2$ is divided into 23 `bricks' of 18 pointings each. Its southern
extension \citep[PHAST;][]{2025ApJ...979...35C} later added about 0.45\,deg$^2$ in 13 bricks
(Figure~\ref{fig:m31}), using
four of these filters, two ultraviolet and two optical. Together the two surveys cover about two
thirds of the star-forming disk and resolve some $2\times10^8$ stars. Even in their combined
mosaics, which have $0\farcs05$ pixels, PNe remain unresolved point sources, but their color sets
them apart.
Most of their blue light comes from emission lines ([\ion{O}{3}]\,$\lambda\lambda4959,5007$ and
H$\beta$), whereas most of their red light is continuum. PNe therefore look blue, although in
faint and low-excitation PNe the continuum also adds noticeably to the blue light
\citep{2014ApJ...792..121V}.
\citet{2014ApJ...792..121V} used this color to find the PHAT counterparts of known PNe and then
selected new candidates from the same color--magnitude region. None of these candidates showed
[\ion{O}{3}] emission in ground-based images, so the authors concluded that ground-based surveys are
complete to F475W\,$\lesssim$\,24 in the PHAT area they studied. Their test, however, left out the
bulge, where 7 of 16 color-selected candidates had earlier proved to be PNe in follow-up spectra.
Color selection has also worked in the outer parts of M31, where \citet{2014AJ....147...16K}
selected candidates from SDSS colors and about 90\% of those with spectra were PNe. A blind search
with color alone is much harder, because PNe are rare among blue sources. In our mosaics, for
example, a color box that holds two thirds of the known PNe contains of order $3\times10^5$
detected sources, while the area is expected to hold only a few thousand PNe.

Deep learning (DL) is now common in galaxy surveys and offers a different approach \citep{2023PASA...40....1H}. Among DL models, object detectors find and outline
many objects in one image, and several of them have been adapted to astronomy, such as ClaRAN
\citep{2019MNRAS.482.1211W}, Astro R-CNN \citep{2019MNRAS.490.3952B} and DeepDISC
\citep{2023MNRAS.526.1122M}. Detectors have also been trained for Galactic PNe.
\citet{2024MNRAS.528.4733S} trained a Mask R-CNN \citep{2017arXiv170306870H} with a Swin
Transformer backbone \citep{2021arXiv210314030L} on IPHAS H$\alpha$ images and then applied it to
VPHAS+. Of the 31 best candidates observed spectroscopically, 22 (71\%) were true, likely or
possible PNe \citep{2024A&A...692A.103L}. Three members of our ensemble use the same backbone.
More recently, \citet{2026AJ....172..238S} trained a You Only Look Once (YOLO) detector on IPHAS
and VPHAS+ images, using as labels boxes drawn from the positions and diameters of PNe in the
Hong Kong/AAO/Strasbourg H$\alpha$ (HASH) catalog \citep{2016JPhCS.728c2008P}. Over the full
surveys it returned 35\,665 detections that do not match known PNe, and extended \ion{H}{2} regions
were the most common false positives.
Other machine learning (ML) work on PNe classifies known objects from
image cutouts or infrared colors
\citep{2020Galax...8...88A,2019MNRAS.488.3238A,2019MNRAS.483.5077A}. In M31, convolutional neural
networks (CNNs) have classified cutouts in searches for star clusters
\citep{2022A&A...658A..51W,2025ApJS..278...16Z}. For extragalactic PNe, automated methods work on
spectral data cubes, either with rules \citep{2026MNRAS.548ag661Y} or with neural networks that
separate PNe from \ion{H}{2} regions and supernova remnants
\citep{2025A&A...694A.212B,2025A&A...697A.148B}. None of these methods, however, detects PNe in
images, and we are not aware of any study that detects extragalactic PNe in broad-band
\textit{HST} images with DL.

Applying a detector to PNe in M31 raises five difficulties. First, the labels are incomplete, so a
PN missing from the catalogs looks like background in the image. Learning from labeled positives and
unlabeled data in this way is known as positive--unlabeled (PU) learning
\citep{2018arXiv181104820B,2025A&A...703A.240K}. Second, there are few positives, only 635 labeled
PNe. Third, the fields are crowded, and fourth, regions covered by few exposures retain cosmic-ray
residuals. Fifth, the evaluation itself has pitfalls: in an M31 star cluster search, for instance,
precision was 99.7\% on an artificially balanced validation set but only 54\% on a test with the
natural class ratio \citep{2025ApJS..278...16Z}. Similarly, the YOLO detector for Galactic PNe
reached 98.7\% precision on test images that each held at least one PN, and its authors expect
lower precision over the full survey, where PNe are rare \citep{2026AJ....172..238S}. For rare
objects, the area under the receiver operating characteristic curve (AUROC) can also rank methods
differently from the success rate within a limited inspection budget
\citep{2019A&A...625A.119M,2026A&A...711A..28E}.

\begin{figure*}
    \centering
    \includegraphics[width=\textwidth]{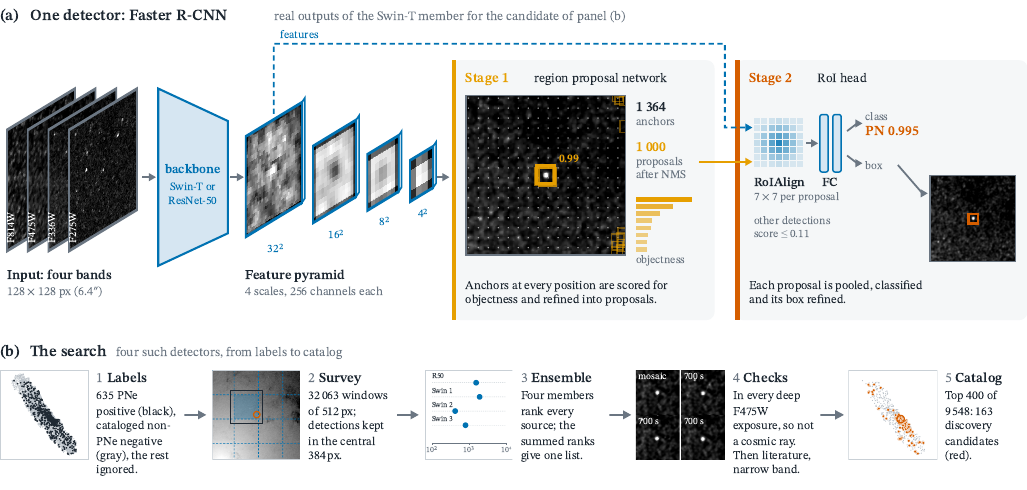}
    \caption{Overview of the method, following one discovery candidate.
    (a) One detector of the ensemble, a Faster R-CNN with a Swin-T backbone, and its real outputs
    for a 128-pixel crop around the candidate (Section~\ref{sec:detector}). The four bands, each with
    the same display stretch, pass through the backbone into a feature pyramid of four scales,
    shown as the mean of each map over its 256 channels. In stage 1, the region proposal network
    scores an anchor at every position of the pyramid and keeps 1000 proposals; the boxes are the
    14 with the highest objectness, the darkest one at 0.99. In stage 2, the RoI head pools the
    features inside each proposal, classifies it and refines its box, giving the candidate a PN
    score of 0.995. (b) The search. (1) Known PNe are positives and cataloged non-PNe negatives;
    the rest of the image is ignored (Section~\ref{sec:partial}). (2) Survey windows on the F475W
    mosaic around the candidate (circle). The window drawn in black keeps only the detections in its
    shaded center, and the dashed lines show the centers of the neighboring windows
    (Section~\ref{sec:survey}). (3) The rank of the candidate in the list of each ensemble member.
    The list ranks the candidates by the sum of these ranks. (4) The candidate in the mosaic and in
    three independent 700\,s F475W exposures (Section~\ref{sec:cosmic}). (5) The discovery
    candidates of the top 400 (red) among the known PNe (Section~\ref{sec:catalog}).}
    \label{fig:overview}
\end{figure*}

To keep unlabeled PNe from being taught as background, we
train Faster R-CNN detectors with partial supervision, and we combine detectors with two different
backbones into one ensemble. We evaluate them without leakage between training and test data, using
a balanced split by spatial groups and a score that counts distinct sky objects within a fixed
budget. Alongside this score, we count unlabeled contaminants by their color. Every setting is run
with several seeds, and the detector is compared with classical color--magnitude selection.
Finally, we deliver a candidate catalog, check it for cosmic rays against independent exposures, and
test every new candidate against independent narrow-band [\ion{O}{3}] and H$\alpha$ data. We also
measure what the detector relies on.
Figure~\ref{fig:overview} shows these steps in order.
Section~\ref{sec:data} describes the data, and each later section then gives one step together
with its result. Section~\ref{sec:method} builds the detector and chooses the ensemble on validation
data, Section~\ref{sec:testing} tests it on held-out data, and Section~\ref{sec:surveysec} surveys
M31 and checks the candidates. Finally, Section~\ref{sec:behavior} asks what the detector relies
on, and Section~\ref{sec:disc} discusses the results.

\section{Data}
\label{sec:data}

\subsection{HST mosaics}
\label{sec:hst}

We use the combined drizzled mosaics \texttt{m31-all-part1} to \texttt{part4} released with the
PHAST high-level science products
(\dataset[doi:10.17909/aa43-mt38]{https://doi.org/10.17909/aa43-mt38})\footnote{\url{https://archive.stsci.edu/hlsp/phast}},
which place the PHAT and PHAST observations in one frame.
From these mosaics we take the ACS/WFC bands F475W and F814W (Table~\ref{tab:data}). F475W
contains the [\ion{O}{3}] and H$\beta$ lines, whereas F814W records mostly continuum, so PNe are
very blue in F475W$-$F814W. The archive F814W mosaic is offset from the F475W pixel grid by 2.0
pixels, and we shifted it onto that grid, leaving an offset below 0.03 pixels. The two
programs also differ in the number of exposures, which matters for cosmic rays. PHAST has two
exposures per ACS filter, and most of its chip gap is covered by a single exposure
\citep{2025ApJ...979...35C}. PHAT has four long exposures in F475W and three in F814W
per visit, and its chip gap is filled only by overlapping neighboring fields
\citep{2012ApJS..200...18D,2014ApJS..215....9W}. Because a cosmic ray cannot be rejected where only
one exposure covers a position, cosmic-ray residuals remain in the mosaics
(Section~\ref{sec:cosmic}).

Three of the four delivered detectors also see the WFC3/UVIS bands F336W and F275W
(Section~\ref{sec:detector}). For these UV bands we mapped the single-visit drizzled products of the
Hubble Advanced Products (717 visits, from MAST) onto the F475W grid by nearest pixel, after
correcting an offset for each visit. The UV bands cover 92\% of the training patches and 83\% of the
survey area. Since we do not fill missing data, models that use UV were trained only on covered
patches, about 8\% fewer than the others, and are not run where UV is missing.

\begin{table}
    \centering
    \caption{Summary of the data.}
    \label{tab:data}
    \begin{tabular}{ll}
        \hline
        Item & Value \\
        \hline
        Mosaics & PHAST HLSP \texttt{m31-all-part1}--\texttt{4} (PHAT$+$PHAST) \\
        Optical bands & ACS/WFC F475W, F814W (all members) \\
        UV bands & WFC3/UVIS F336W, F275W (three members; 83\% of the survey area) \\
        Pixel scale & $0\farcs05$ \\
        Size of one part & $39\,478 \times 42\,208$ pixels \\
        F475W exposures in the area & 81.7\% PHAT, 18.3\% PHAST \\
        Exposures per ACS filter & PHAST 2, PHAT 4 (F475W) or 3 (F814W) long $+$ 1 short \\
        Positive labels & SIMBAD PNe with $m_{5007}$ from \citet{2006MNRAS.369..120M} (635) \\
        Negative labels & SIMBAD non-PN objects, 88 types (7871) \\
        Narrow-band checks & CFHT MegaCam [\ion{O}{3}] and $g$; SITELLE [\ion{O}{3}] and H$\alpha$ cubes \\
        \hline
    \end{tabular}
\end{table}

\subsection{Labels}
\label{sec:labels}

The positive examples come from SIMBAD \citep{2000A&AS..143....9W}. Of the 1065 SIMBAD PNe inside
the mosaics, 791 have an [\ion{O}{3}] magnitude $m_{5007}$ in \citet{2006MNRAS.369..120M}. Their
catalog positions, however, are offset from the \textit{HST} images by a median of 10.8 pixels
($0\farcs54$). We therefore moved each position to a local peak within $1\farcs25$ whose F475W
brightness matches, within $\pm0.5$\,mag, the value expected from $m_{5007}$
\citep{2014ApJ...792..121V}. Where only one peak qualified, 97.2\% of the moved positions were on the
correct source, while objects with several qualifying peaks were not used. For 476 PNe the position
was taken from the PHAT and PHAST photometric catalogs \citep{2014ApJS..215....9W,2025ApJ...979...35C}.
These catalogs list the positions and magnitudes of point sources, measured by fitting the
point-spread function with DOLPHOT \citep{2000PASP..112.1383D,2016ascl.soft08013D}, so we call them
the DOLPHOT catalogs and a source position in them a DOLPHOT position.
In total this gives 635 positive examples, most of which are confirmed by independent data. For
instance, 98\% have PN.S [\ion{O}{3}] detections with velocities. From a comparison with
\citet{2014ApJ...792..121V} and \citet{2012ApJ...758..133S}, we estimate that 1.5--3.2\% of them are
not PNe.

We did not add EmO objects and possible PNe to the positives, although the literature calls many
of them PNe (Section~\ref{sec:untaught}). The positives follow the SIMBAD classes without change, so
that they can be rebuilt from the database, and they rest on stronger evidence.
A positive is listed as a PN by a median of three independent catalogs, and 32\% have a full
spectrum that classifies them as PNe \citep{2012ApJ...758..133S,2014AJ....147...16K}. Of the 165 EmO objects and possible PNe that \citet{2014ApJ...792..121V} confirmed on \textit{HST}
images, 72\% appear as PNe only in the PN.S catalog, and none has a spectrum that shows a PN.
Adding these 165 objects to the training positives made no measurable difference
(Table~\ref{tab:axes}).

The negative examples are cataloged SIMBAD objects that are not PNe. Of these, we kept those at
least 200 pixels from any positive and drew a sample of 7871 objects stratified over 88 object
types, which we call labeled negatives. Objects of type EmO were kept out of this pool as well,
because of how our training rule treats labeled negatives (Section~\ref{sec:partial}). Under this
rule, they are the only objects the detector is penalized for detecting. An EmO label, however,
only says that the object could not be classified, and some EmO objects are PNe
(Section~\ref{sec:untaught}). A PN suppressed in training cannot be recovered later. A contaminant
that slips through, in contrast, stays visible in the candidate list, where later checks can remove
it.

\subsection{Patches and a balanced spatial split}
\label{sec:split}

\begin{figure*}
    \centering
    \includegraphics[width=\textwidth]{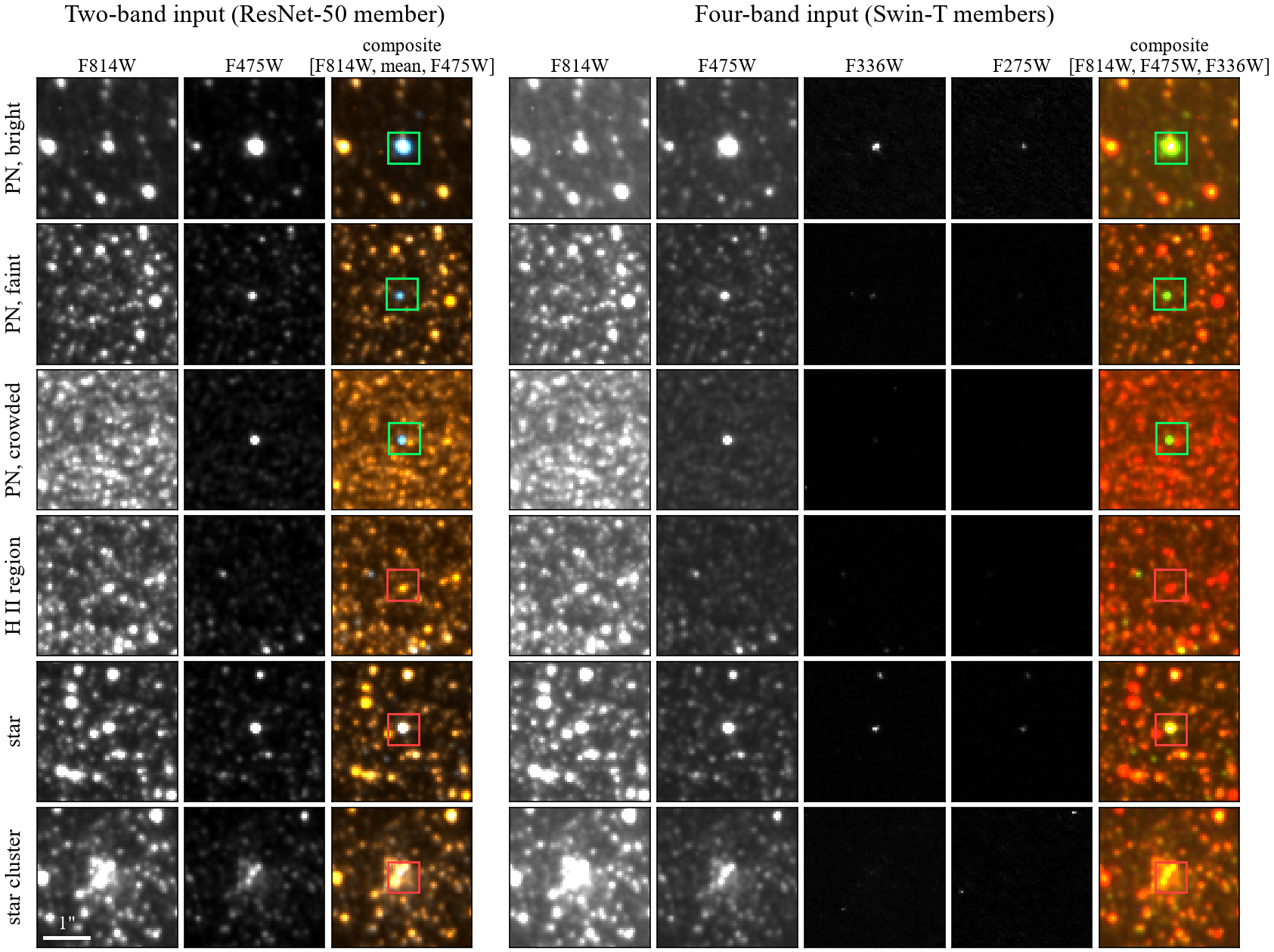}
    \caption{Training examples as the two kinds of members see them. Each row is one object from
    the training split, chosen among patches with UV coverage. The first three rows are PNe
    (bright and isolated, faint, and in a crowded part of the inner disk). The last three are
    labeled negatives (an \ion{H}{2} region, a star and a star cluster). The left block is the
    three-channel input of the ResNet-50 member: F814W, F475W and the color composite of
    [F814W, mean of the two bands, F475W]. The right block is the four-channel input of the Swin
    members: F814W, F475W, F336W, F275W and the composite of the first three. Both use the
    per-patch 1--99 percentile stretch computed on the full 128 pixel crop. The stretch is shared
    by the channels of each input, so the optical channels of the two inputs differ. Only the
    central 64 pixels ($3\farcs2$) are shown. Boxes mark the $14 \times 14$ pixel label boxes
    (green for PNe, red for negatives). The PNe are bright in F475W and faint in F814W.}
    \label{fig:patches}
\end{figure*}

For each source we cut a $512 \times 512$ pixel parent patch (8506 patches) with all four bands,
and training crops are then cut from these parents when they are loaded. Figure~\ref{fig:patches} shows
examples as the model sees them.

A random split by source would leak information in two ways. First, a test window often contains
neighboring PNe, and if sources are split by identity, about 70\% of these neighbors appear in
training in their own patches. For this reason, the unit of our split is a spatial group. Second,
if coverage cuts needed by some experiments (for example UV coverage) are made before the split, the
split depends on the experiment. The split is therefore built once and used by every experiment.
Such spatially blocked splits are recommended for data with spatial structure
\citep{2017Ecogr..40..913R}, and leakage is known to inflate ML results \citep{2022arXiv220707048K}.

The groups are connected components of sources closer than 270 pixels (5066 groups). A
deterministic search assigns them to training, validation and test parts of 68\%, 12\% and 20\%
(Table~\ref{tab:split}).
Because random group splits can differ in PN properties by chance, the search also balances the
parts. It balances them with physical and catalog quantities only, namely $m_{5007}$, fluxes, color,
concentration, sky level, crowding, galactocentric distance, UV coverage and negative type. After
balancing, all 60 standardized mean differences are below 0.10. An automatic check also
confirms that no source, neighbor or crop of the validation and test parts reaches training.

\begin{table}
    \centering
    \caption{The split. Numbers are parent patches and PNe.}
    \label{tab:split}
    \begin{tabular}{lrr}
        \hline
        Part & Patches & PNe \\
        \hline
        Training & 5784 & 432 \\
        Validation & 1018 & 76 \\
        Test & 1704 & 127 \\
        \hline
        Total & 8506 & 635 \\
        \hline
    \end{tabular}
\end{table}



\subsection{Narrow-band data for validation}
\label{sec:nbdata}

To test the candidates with data the detector never saw, we use emission-line data from three
archives (Table~\ref{tab:data}). The main source is CFHT MegaCam \citep{2003SPIE.4841...72B}
imaging in [\ion{O}{3}] and $g$ from the M31 PN survey of \citet{2019A&A...624A.132B}, which
consists of 363 [\ion{O}{3}] exposures ($\sim$98\,ks) and 247 $g$ exposures processed with Elixir
\citep{2004PASP..116..449M}. From every single exposure we cut $20\arcsec$ around each candidate.
These data cover 399 of the top 400 candidates. The second source is data cubes from the imaging
Fourier transform spectrometer SITELLE \citep{2019MNRAS.485.3930D}, reduced with ORBS and ORCS
\citep{2015ASPC..495..327M}. Its SN2 filter contains [\ion{O}{3}] and H$\beta$, while its SN3
filter contains H$\alpha$, [\ion{N}{2}] and [\ion{S}{2}]. The SN3 cube of the center is the one published by
\citet{2018MNRAS.473.4130M}. Of the top 400, 189 lie in an [\ion{O}{3}] cube and 177 in an H$\alpha$
cube.
Finally, the LGGS [\ion{O}{3}] and H$\alpha$ images \citep{2006AJ....131.2478M} give a third
measurement, which we use only as additional information. Beyond these images, we searched all
VizieR catalogs \citep{2000A&AS..143...23O}, among them the LAMOST spectral catalogs
\citep{2012RAA....12.1197C,2012RAA....12..723Z}, and NED within $3\arcsec$ of each candidate. We
also searched the spectral archives of SDSS \citep{2023ApJS..267...44A} and ESO.

\section{The detector}
\label{sec:method}

\subsection{Overview and detector}
\label{sec:detector}

The catalog comes from an ensemble of four Faster R-CNN detectors, one with a ResNet-50 backbone
and three with a Swin Transformer backbone. Section~\ref{sec:selection} explains how they were
chosen, and here we describe what all four share, namely the detection head, the anchors and the
labels.

We use the Faster R-CNN detector \citep{2015arXiv150601497R} as implemented in torchvision
\citep{2019arXiv191201703P}, which works in two stages. First, a region proposal network (RPN)
proposes boxes that may contain an object, relative to reference boxes called anchors that sit at
every image position. A second stage then classifies each proposal and refines its box.

We chose a two-stage detector because we search a large area for rare objects, so that almost every
position in a survey window is background. In a two-stage detector, the proposal stage narrows the
image down to a small set of candidate boxes and thereby removes most of the background. The second
stage then keeps a fixed balance between object and background examples \citep{2017iccv.conf..324L}.
A one-stage detector, by contrast, classifies all positions at once, and the many easy background
examples can dominate its training \citep{2017iccv.conf..324L}. Detectors of the R-CNN family have
also been adapted to astronomy before, in ClaRAN \citep{2019MNRAS.482.1211W}, Astro R-CNN
\citep{2019MNRAS.490.3952B}, DeepDISC \citep{2023MNRAS.526.1122M} and the Galactic PN search of
\citet{2024MNRAS.528.4733S}. We did not measure the benefit of the two stages directly.
The one-stage RetinaNet, which we tested with the same backbone, did no better
(Section~\ref{sec:grid} and Table~\ref{tab:axes}).

In the ResNet-50 member, the image features come from a ResNet-50 network
\citep{2016cvpr.confE...1H} with a feature pyramid network (FPN), which provides feature maps at
several scales \citep{2016arXiv161203144L}. This network starts from weights trained on the COCO
collection of everyday photographs \citep{2014arXiv1405.0312L}, and only its last three blocks are
trained. Its input has three channels, [F814W, (F814W$+$F475W)/2, F475W].

The second backbone is the tiny Swin Transformer \citep[Swin-T;][]{2021arXiv210314030L}, which
computes features with attention inside shifted windows of the image. It starts from torchvision
weights trained on ImageNet, whereas the FPN, the RPN and the box head start at random. Its input
has four channels, [F814W, F475W, F336W, F275W]. Because its first layer was built for three
channels, the fourth channel starts from the mean of the pretrained weights.

Two stretches map the fluxes to the input range. The percentile stretch scales each patch to its
own 1--99 percentile range, shared by its channels. The asinh stretch subtracts the
median of each channel as the sky and then applies an inverse hyperbolic sine with fixed scales,
the same for all patches and both bands, so it keeps the absolute brightness of a source. The
training crops are 64 or 128 pixels wide (Section~\ref{sec:training}). The ResNet-50 member uses the
percentile stretch and 128 pixel crops. The three Swin members use the percentile stretch with 128
and with 64 pixel crops, and the asinh stretch with 128 pixel crops.

Every PN gets a label box of $14 \times 14$ pixels ($0\farcs7$). To compare boxes, we measure their
overlap as intersection over union (IoU), the area they share divided by the area they cover
together. With the default anchors, most PNe would match no anchor, since the best IoU is only 0.19.
We therefore use square anchors of 16 to 256 pixels, which raise the best IoU to 0.73. The final
classifier has two classes, PN and background.

\subsection{Partial supervision}
\label{sec:partial}

In standard training, everything outside a labeled box is background. With an incomplete catalog,
however, this teaches the detector to suppress the PNe missing from it. In our training, a
background anchor or proposal is therefore penalized only if it overlaps the box of a labeled
negative, and it is ignored otherwise. For the same reason, the negatives are cataloged objects rather
than random background, which may hold real PNe (Section~\ref{sec:labels}).
With the overlap threshold set to zero, the rule acts at the level of the crop. In a crop without a
negative box all background is ignored, whereas in a crop with one all background is supervised.
As a result, a median of 61\% of the anchors are ignored. The cost of treating unlabeled objects as
background grows with the fraction of missing labels \citep{2018arXiv180606986W}, and near our
negatives this fraction is small. A PN confirmed by \citet{2014ApJ...792..121V} but missing from our
labels, for example, lies within 64 pixels of a negative in only 0.2\% of cases.

\subsection{Training}
\label{sec:training}

The crop center is shifted at random by up to 0.875 times half the crop size, so that the model does
not look for PNe only at the center. Crops are also rotated and flipped at random. Positives
make up 65\% of the sampled crops, one epoch is 16\,000 crops in batches of 16, and training runs
for 25 epochs. Although the validation loss reached its lowest value at a median epoch of 17, the
model of the last epoch scored as well as the model of the best epoch (Table~\ref{tab:axes}). The
learning rate rises linearly during the first epoch \citep{2017arXiv170602677G} and is then halved
every five epochs. Table~\ref{tab:axes} lists the measurement behind each setting.
All models were trained with PyTorch 2.11 and torchvision 0.26 \citep{2019arXiv191201703P} on a
single NVIDIA RTX 5080 GPU with 16\,GB of memory. On this GPU, a 25-epoch run took 22--37 minutes,
depending on the backbone and the crop size, and a survey of the full mosaics with all four members
took about 1.8 hours.

\subsection{Evaluation}
\label{sec:eval}

Every model is scored on fixed windows of $512 \times 512$ pixels around the validation and test
sources. We use only the 903 and 1512 windows with UV coverage, so that UV experiments are scored on
the same windows as the others. The evaluation is also kept independent of the training settings,
so that the yardstick does not move with the model.

Our main measure, P@200, is the fraction of known PNe among the 200 highest-ranked sky objects. To
compute it, all detections are ranked by score and merged into distinct sky objects within
$0\farcs75$ \citep[position-tolerant matching as in][]{2021MNRAS.500.3821B}. As noted in
Section~\ref{sec:intro}, such a fixed budget suits rare objects better than AUROC. We count distinct
objects because overlapping windows would otherwise count the same PN several times. Since the
windows hold only 71 and 114 distinct known PNe, even a perfect detector would reach a P@200 of
0.355 on validation and 0.570 on test, not 1. The budget is larger than the number of known PNe on
purpose, because unlabeled PNe also take places in the list. With a budget close to the number of
known PNe, they would push known PNe out of it. We therefore also give P@200 as a share of this
ceiling, which equals the fraction of the known PNe in the windows that the top 200 contain.

The measured purity is only a lower limit. Of the objects in the evaluation windows, 89\% have no
label, and many of them may be PNe that no survey has found. We estimate their number from the PNLF
in the form of \citet{1989ApJ...339...53C}, with the cutoff F475W$^* = 20.69$ taken from the F475W
PNLF of \citet{2014ApJ...792..121V}. Their fit was made for F475W $< 24$, and we extend it to 28,
although the mosaics reach this depth only in the outer disk. There, the PHAST photometry is 50\%
complete at F475W 27.7, while near the bulge it is 50\% complete at 26.0 \citep{2025ApJ...979...35C}.
The projection gives about 2800 undiscovered PNe in the area of the mosaics, uncertain by a factor
of two.

For this reason, we report no precision, F1 score or mean average precision for the detector. These
measures count every detection without a label as an error, yet in M31 the PNe without a label
outnumber those with one. A narrow-band survey two magnitudes deeper than earlier ones, for example,
found 1023 PNe in the PHAT area alone \citep{2019A&A...624A.132B}, while our 635 labels cover the PHAT
and PHAST areas together. P@200 shares this bias, so we use it only to rank models on the same fixed
windows. Recall is measured on the known PNe, which are brighter and more central than
those still to be found (Section~\ref{sec:limits}), so it describes the known population only.

P@200 cannot tell an unlabeled star from an unlabeled PN, so two
models with the same P@200 can fill the rest of their lists with different contaminants. We
therefore also count contaminants by color. Each unlabeled object in the top $L$ is matched to the
nearest PHAT DOLPHOT source within $0\farcs3$ \citep{2014ApJS..215....9W}, and it is called hot if
F336W$-$F475W $< -0.5$ and reddish if F475W$-$F814W $> +0.3$. Applied to the known PNe, these cuts
flag 5--6\% as hot and 7--8\% as reddish, which sets the rate of false flags. On validation, the top
200 is longer than the number of findable PNe (71), so beyond rank $\sim$100 every model must fill
its list with non-PNe. We compare contamination mainly within the top 50 to 100.

\subsection{Design of the final experiment}
\label{sec:design}

All numbers in this and the next three sections come from the validation part, except for the
check of the retrained models. The validation part was used to choose the settings and the ensemble,
so these numbers are optimistic, and the test on held-out data follows in Section~\ref{sec:test}.

Before the final experiment, earlier searches varied up to 17 training settings, first one at a
time and then together in a stratified random search \citep{Bergstra2012}. These searches fixed most
settings, such as the number of trained layers, the anchors and the data augmentation
(Table~\ref{tab:axes}). A permutation test \citep{Westfall1993} then left three settings whose
effects could not be ruled out, namely the input bands, the flux stretch and the crop size. The
final experiment is therefore a full grid over these three, extended by a controlled comparison of
backbones and by a four-band arm with the backbone that this comparison kept. Because the effects of
the settings were as large as the scatter between seeds, each setting was run with three seeds. Such
repetition is needed, since in ML comparisons random variation alone can be as large as many
published improvements \citep{2021arXiv210303098B,2021arXiv210908203P}.

\providecommand{\tcell}[2]{\parbox[t]{#1}{\raggedright\strut #2\strut}}
\begin{table*}
    \centering
    \caption{Axes tested before the final experiment and their place in it. Earlier searches
    used different evaluation sets, so the table gives the direction of each
    result and the decision, not absolute scores. The backbone comparison of the final experiment
    is included for completeness. Differences are in P@200 unless stated.}
    \label{tab:axes}
    \begin{tabular}{llll}
        \hline
        \tcell{2.7cm}{Axis} & \tcell{4.2cm}{Tried} & \tcell{6.6cm}{Result} & \tcell{2.9cm}{In the final grid} \\
        \hline
        \tcell{2.7cm}{Backbone and pre-training (earlier searches)} & \tcell{4.2cm}{ResNet-50, ResNet-101, ResNet-34 \citep{2016cvpr.confE...1H} and MobileNetV3 \citep{2019arXiv190502244H}, with COCO, ImageNet \citep{2009cvpr.conf...27D}, random or RadImageNet \citep{Mei2022} weights} & \tcell{6.6cm}{ResNet-101 did not beat ResNet-50. MobileNet collapsed in the usable learning-rate range. ImageNet $+0.012$ ($p = 0.76$). Random weights $-0.065$ ($p = 0.04$) against the pool of pretrained backbones} & \tcell{2.9cm}{controlled comparison in the final experiment} \\
        \tcell{2.7cm}{Backbone and detector (final experiment)} & \tcell{4.2cm}{ResNet-50 with COCO or ImageNet weights, Swin-T, RetinaNet \citep{2017iccv.conf..324L} with gradient clipping \citep{2012arXiv1211.5063P}, ConvNeXt-T \citep{2022cvpr.conf.1167L}} & \tcell{6.6cm}{ConvNeXt-T equivalent to ResNet-50 with COCO, the others not separable. Swin-T kept its score under both stretches and shared the fewest unlabeled detections with ResNet-50 (Section~\ref{sec:grid})} & \tcell{2.9cm}{two backbones, ResNet-50 and Swin-T} \\
        \tcell{2.7cm}{Trained blocks} & \tcell{4.2cm}{0 to 5} & \tcell{6.6cm}{Steady gain from 0 to 3, no difference beyond ($p = 0.75$)} & \tcell{2.9cm}{fixed, 3} \\
        \tcell{2.7cm}{Anchors, label box} & \tcell{4.2cm}{default or square 16--256 pixel anchors, box size} & \tcell{6.6cm}{Best IoU 0.19 with default and 0.73 with square anchors. Box size made no difference ($p = 0.76$)} & \tcell{2.9cm}{fixed} \\
        \tcell{2.7cm}{Crop shift} & \tcell{4.2cm}{amount and formula} & \tcell{6.6cm}{$+0.048$. Without a shift, recall at 48 pixels from the crop center fell from 0.93 to 0.16. The shift scales with crop size (0.875 of half the crop) so that the object stays in the crop} & \tcell{2.9cm}{fixed, 0.875} \\
        \tcell{2.7cm}{Background rule} & \tcell{4.2cm}{overlap threshold 0 or 0.1} & \tcell{6.6cm}{$+0.037$ against ignoring all background. 0 and 0.1 not separable ($-0.011$), external check (Section~\ref{sec:retrain}) weakly favors 0} & \tcell{2.9cm}{fixed, 0} \\
        \tcell{2.7cm}{Training length} & \tcell{4.2cm}{early stopping on the validation loss, fixed 25 epochs} & \tcell{6.6cm}{Early stopping ended seed runs at epochs 6, 6, 7 and 17, before they had converged. In the 72 runs of the final experiment the validation loss was lowest at a median epoch of 17 (quartiles 10 and 22), and the last epoch scored as well as the best ($+0.002$ for ResNet-50, $+0.004$ for Swin-T)} & \tcell{2.9cm}{fixed, 25} \\
        \tcell{2.7cm}{Batch, schedule, share of positives} & \tcell{4.2cm}{several} & \tcell{6.6cm}{no difference} & \tcell{2.9cm}{fixed (choice)} \\
        \tcell{2.7cm}{Learning rate} & \tcell{4.2cm}{$10^{-4}$ to 0.032} & \tcell{6.6cm}{best value depends on the setting} & \tcell{2.9cm}{preliminary step for each grid} \\
        \tcell{2.7cm}{Data augmentation} & \tcell{4.2cm}{group $\times$ probability} & \tcell{6.6cm}{0.7 standard errors, lowest diversity between ensemble members} & \tcell{2.9cm}{closed (eight symmetries)} \\
        \tcell{2.7cm}{Cosmic-ray training} & \tcell{4.2cm}{cosmic rays in the loss (class, ratio, dose)} & \tcell{6.6cm}{cost on the fixed evaluation (Section~\ref{sec:cosmic})} & \tcell{2.9cm}{removed, exposure check at the end} \\
        \tcell{2.7cm}{Stretch} & \tcell{4.2cm}{percentile, asinh, fixed, UV smoothing} & \tcell{6.6cm}{percentile better in crowded fields, asinh ranks the most different unlabeled objects, fixed stretch lost} & \tcell{2.9cm}{open (percentile, asinh)} \\
        \tcell{2.7cm}{Input bands} & \tcell{4.2cm}{2 bands, 3--4 bands with UV, flux difference} & \tcell{6.6cm}{UV gave no clear gain in P@200, so the axis was kept for the final grid. The contamination count later showed that UV halves the hot stars (Section~\ref{sec:grid})} & \tcell{2.9cm}{open (5 levels with ResNet-50, four bands with Swin-T)} \\
        \tcell{2.7cm}{Crop size} & \tcell{4.2cm}{48, 64, 128, 192 pixels} & \tcell{6.6cm}{48 and 192 dropped, 64 and 128 close} & \tcell{2.9cm}{open (64, 128)} \\
        \tcell{2.7cm}{Negative pool} & \tcell{4.2cm}{with or without EmO} & \tcell{6.6cm}{EmO is the only penalized class and contains real PNe} & \tcell{2.9cm}{without EmO (Section~\ref{sec:labels})} \\
        \tcell{2.7cm}{Extra positives} & \tcell{4.2cm}{165 EmO objects and possible PNe confirmed on \textit{HST} by \citet{2014ApJ...792..121V}} & \tcell{6.6cm}{no difference} & \tcell{2.9cm}{not added (Section~\ref{sec:labels})} \\
        \hline
    \end{tabular}
\end{table*}

The main grid used ResNet-50 and crossed five input designs, two stretches (percentile and asinh)
and two crop sizes (64 and 128 pixels), with three seeds each (60 runs). Four of the input designs
have three channels, with F814W first and F475W last, and differ only in the middle channel, which
is the mean of the two bands, the difference F475W$-$F814W, F336W or F275W. The fifth has four
channels, F814W, F475W, F336W and F275W. For each grid, a preliminary step chose the learning rate.

The backbone comparison used the three-channel input with the mean channel, the same four settings
of stretch and crop size and three seeds. In it, four arms were compared with ResNet-50 with COCO
weights, namely ResNet-50 with ImageNet weights, Swin-T, the one-stage detector RetinaNet and
ConvNeXt-T, a CNN built with the design choices of Swin. An arm counted as equivalent to ResNet-50
if its interval, corrected for four comparisons, lay within one seed standard deviation of zero.
Finally, the four-band arm used Swin-T with the two stretches and two crop sizes, three seeds each
(12 runs). Together, the main grid and the four-band arm form the selection pool of 72 runs in 24
settings.

\subsection{Inputs, stretches and backbones}
\label{sec:grid}
\label{sec:gridaxes}
\label{sec:backbones}

The scatter between seeds within the ResNet-50 grid is 0.0237 in P@200. This scatter is not
symmetric, because the settings that replace a channel with a UV band, and the asinh settings,
collapse now and then. Averaged over the other settings and seeds, the input with the mean in
the middle channel is best (0.246), and adding both UV bands as extra channels cannot be told apart
from it ($-0.008$). Replacing the middle channel with a UV band or with the band difference, in
contrast, puts a setting 2.8 to 3.8 standard errors behind. Of the two stretches, the percentile
stretch beats asinh by 0.020 (3.2 standard errors, about four objects in the top 200), while the crop
size makes no measurable difference. The best setting, with the mean channel, the percentile stretch
and 128 pixel crops, reaches 0.258, or 73\% of the ceiling. In earlier work, DeepDISC also found no
single best stretch for ResNet backbones \citep{2023MNRAS.526.1122M}, and per-image normalization,
like our percentile stretch, has been used on \textit{HST} cutouts before \citep{2020A&A...633A.148B}.


In the backbone comparison, the mean P@200 of the settings ranged from 0.218 to 0.258, and no
alternative was shown to be better or worse than ResNet-50 with COCO weights. Relative to it, Swin-T
scored $+0.004$ $[-0.006, +0.014]$, ConvNeXt-T $-0.001$ $[-0.013, +0.010]$, RetinaNet $-0.007$
$[-0.019, +0.005]$ and ResNet-50 with ImageNet weights $-0.010$ $[-0.021, +0.000]$. Only ConvNeXt-T
was equivalent to ResNet-50, whereas for the other three the intervals were too wide to tell. Swin-T,
however, responded differently to the stretch: in the other arms the asinh settings scored 0.015--0.035 below the
percentile settings, while Swin-T scored the same under both.

We also measured how often two models put the same unlabeled objects in their top 200, relative to
what independent choices would give. Since these objects may be contaminants or undiscovered PNe, we
do not call them errors. Seeds of ResNet-50 in the same setting share them 5.4 times more often than
chance, models that differ only in pre-training 3.9 times, and models that differ only in the
detection head 4.2 times. ResNet-50 and Swin-T, in contrast, share them only 2.9 times, the least of
any pair. Which unlabeled objects a model ranks high therefore depends mostly on the backbone.

The four-band Swin-T arm reached the same P@200, a median of 0.260 or 73\% of the ceiling (range
0.245--0.280, 69--79\%). Here the contamination count (Section~\ref{sec:eval}) shows what P@200
misses. With both backbones, UV halves the hot stars among the unlabeled objects in the top 200,
from 49 to 25 for ResNet-50 and from 57 to 26--32 for Swin-T, while the reddish ones do not decrease
(26 to 39). These hot stars are truly rejected rather than moved just below rank 200. Of the 58 hot
stars in the top 200 of the three-channel ResNet-50, for example, 22 are not even in the top 800 of
the four-band Swin-T. The two backbones differ, however, at the top of the list, where the UV bands
cost ResNet-50 some PNe but not Swin-T.

\subsection{Choice of the ensemble}
\label{sec:selection}
\label{sec:enschoice}

The product of this work is a list for follow-up, not a single model. With incomplete labels,
however, a detection without a label cannot be scored as right or wrong (Section~\ref{sec:eval}).
Agreement between detectors, in contrast, gives a signal that needs no labels. If several detectors
that rank different unlabeled objects high all rank a candidate high, the candidate is unlikely to be
the quirk of a single detector. Besides choosing the settings, the grid therefore also had to supply members
whose unlabeled detections differ. For this reason the asinh stretch stayed in the grid, although it
scores lower with ResNet-50 (Section~\ref{sec:grid}).

Candidates were compared on the validation part, pooling the 72 runs of the main grid and the
four-band arm. For this comparison, the spatial groups of the validation part were split at random
into two halves, and one seed per setting was drawn, 1000 times. The candidates were single models
and ensembles of 2 to 5 members, drawn either from all settings or only from settings without UV. In
each repetition, one half chose the members and the other scored them by P@200. More members or UV
input were kept only if the gain in P@200, paired by repetition, had a 95\% interval above zero.

No larger ensemble beat a smaller one, so this rule picked a single model, a ResNet-50 setting, and
adding the UV settings changed P@200 by $-0.002$ $[-0.041, +0.046]$. When the rule did build an
ensemble of four, it most often chose the one we deliver. We still deliver this ensemble, because
of how its members agree, which P@200 does not show.

To see this agreement, we kept, for two members, the objects that both rank in their top 200. Two
seeds of one setting keep 95--97\% of the PNe and 52--66\% of the contaminants. Members that differ
in backbone or stretch keep almost as many PNe, 90--92\%, but only 27--32\% of the contaminants, and
ResNet-50 with Swin-T is the cleanest pair (90\% and 27\%). Models that see the images differently
thus agree on the PNe but not on the contaminants. This finding pointed most clearly to this
ensemble, because summing the ranks of such members puts the agreed candidates first.

The agreement shows at the top of the list. Over all combinations of one seed per member, with
intervals from a bootstrap over spatial groups ($B = 2000$), the delivered ensemble reaches 0.87,
0.81, 0.49 and 0.273 at $L = 20$, 50, 100 and 200. At $L = 50$, single models are behind by 3.8 to
6.4 PNe, with intervals that exclude zero, although at $L = 200$ they cannot be separated from the
ensemble. In the top 50, the ensemble holds 1.1 contaminants, against 5.0 for the single Swin-T model
and 3.0 for the single ResNet-50 model. Its ResNet-50 member also covers the 17\% of the area without
UV data. Throughout the paper, we call its Swin-T member with the percentile stretch and 128 pixel
crops the single model, and we report it for comparison and for analyses that need one model.

\subsection{Retraining on all data}
\label{sec:retrain}

We retrained the four members with the same settings on all labeled data (635 PNe, 47\% more than
training alone). Otherwise the 203 validation and test PNe would never teach the delivered models,
and with so few labels each PN counts. Since no validation part is then left to pick an epoch, the
retrained models keep the weights of their last epoch, which on validation score as well as the
best one (Section~\ref{sec:training}). The survey of the galaxy was made with the retrained
ensemble, whereas all other metrics we report belong to models trained on the training part alone.

To catch a collapse of a retrained model, we use an external check. It asks how well the scores rank
EmO objects that the literature calls PNe or possible PNe (293 objects) above those it calls
\ion{H}{2} regions (425 objects), measured as AUROC. Some of the \ion{H}{2} labels come from
\citet{2011AJ....142..139A}, whose classification is photometric, and the authors note that some
faint compact objects in their catalog may be PNe. On this check, the retrained Swin-T members reach
an AUROC of 0.78--0.83.

We tested what retraining changes on four three-channel ResNet-50 models of the grid, and it
brought no measurable change. Their mean AUROC on the external check is 0.866 retrained and 0.863
trained on the training part alone. We nevertheless use the retrained models, because they use
more data and are not behind on any measure.

That 47\% more data gave no visible gain may have several reasons, which we did not separate. First,
our tests may be too small to see a modest change. The limit may also lie in the images, since recall is
lowest for the faintest PNe (Section~\ref{sec:selfunc}).
Finally, the balanced split makes the added PNe similar to the training PNe, so they do not add the
faint and crowded cases where the detector fails.

\section{Testing the detector}
\label{sec:testing}

\subsection{The test and comparison with classical selection}
\label{sec:test}

On the test part, the delivered ensemble reaches 0.879, 0.844, 0.745 and 0.417 at $L = 20$, 50, 100
and 200. At $L = 200$ this is 73\% of the ceiling of 0.570, against 77\% on validation, so the
ensemble keeps its level on data it never saw. Among the single members, the Swin-T models with 128
pixel crops reach 74\% and 72\% of the ceiling and cannot be told apart from the ensemble. In
contrast, the Swin-T model with 64 pixel crops (68\%) and the ResNet-50 member (66\%) are behind,
with intervals that exclude zero, and so are the other three ResNet-50 settings (60--65\%). Because
the best epoch of each run was chosen on validation, the test values may be slightly optimistic.

To check this ranking, we also ran a five-fold cross-validation of the four member settings, with
folds of spatial groups over the whole labeled set. It gives 69--71\% of the ceiling for every
member. Here the folds differ more than the members, and the difference between Swin-T and
ResNet-50 even changes sign between folds. The lead of the Swin-T models on the test part is thus
not reproduced. It may come from the different measure, weights or split, which we did not
separate.

\begin{figure*}
    \centering
    \includegraphics[width=\textwidth]{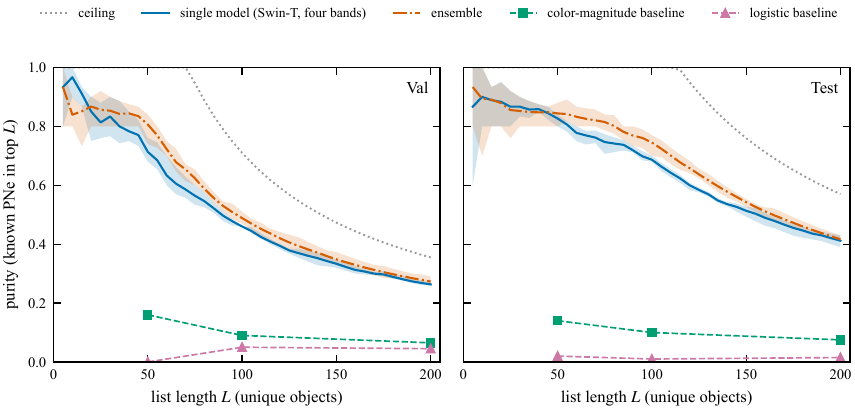}
    \caption{Purity against list length on the fixed evaluation windows, for validation (left)
    and test (right). Purity is the fraction of known PNe among the top $L$ distinct sky
    objects. Single model (Swin-T, four bands, percentile stretch, 128 pixel crops), mean of three
    seeds with the minimum to maximum range as a band. Ensemble, mean over the 81 combinations of one
    seed per member, with the range as a band. The color--magnitude and logistic baselines
    (Section~\ref{sec:test}) are shown at $L = 50$, 100 and 200. The dotted line is the ceiling, the number of distinct PNe in
    the windows divided by $L$. Purity is a lower limit, because most objects in the windows have
    no label (Section~\ref{sec:eval}).}
    \label{fig:purity}
\end{figure*}

We compared the detector with two classical rankings on the same windows. For both, a peak finder
located sources, and aperture photometry measured them in both bands. The color--magnitude baseline
then ranks sources by their Mahalanobis distance to the color--magnitude distribution of the
training PNe, a smooth version of the box of \citet{2014ApJ...792..121V}. The logistic baseline is a logistic regression on hand-made features (fluxes, color, S/N, sharpness, light in the
wings and background). Both were fitted on the training part only and use the same scoring as the
detector.

The peak finder recovers 98--99\% of the known PNe, so the baselines see almost every PN. Their
weakness lies instead in everything else in the same region of color and brightness. Of a pool of
random sources measured without the model, 0.8\% fall in the color and flux range of the PNe
(Section~\ref{sec:bluepool}). Over the mosaics, this amounts to of order $10^5$ sources, against 635
known PNe. A color--magnitude ranking cannot place these sources below the PNe, because by
construction it sees them as PNe. The detector, in contrast, accepts only 2--7\% of them,
against 84--86\% of the known PNe (Section~\ref{sec:bluepool}).

The color--magnitude baseline reaches 0.065 on validation and 0.075 on
test, and the logistic baseline 0.045 and 0.015. The single model and the ensemble, in contrast,
reach 0.26--0.27 and 0.41--0.42, 4--6 times higher (Table~\ref{tab:classical}). Measured
against the ceilings of 0.355 and 0.570, they recover 74--77\% and 72--73\% of the known PNe in the
top 200, whereas the color--magnitude baseline recovers 18\% and 13\%. The logistic baseline,
despite its additional features, is even weaker than the color--magnitude one. This gap does not mean that color is unimportant, only that the detector uses information beyond
the color--magnitude ellipse (Section~\ref{sec:ablation}).

\begin{table}
    \centering
    \caption{Test results. Numbers of known PNe among the top 50, 100 and 200 distinct objects,
    then P@200 and the share of the ceiling that it reaches. The ceiling row is a perfect ranking. The test
    windows hold 114 distinct known PNe, so no method can find more than 114 in the top 200. For
    the single model, the mean over three seeds and, for the top 200, their range. For the
    ensemble, the mean over the 81 combinations of one seed per member and, for the top 200, its
    range.}
    \label{tab:classical}
    \begin{tabular}{lrrrrr}
        \hline
        Method & Top 50 & Top 100 & Top 200 & P@200 & Share of ceiling \\
        \hline
        Ceiling & 50 & 100 & 114 & 0.570 & 100\% \\
        Single model (Swin-T) & 41.4 & 68.7 & 82.4 (78--86) & 0.412 & 72\% \\
        Delivered ensemble & 42.2 & 74.5 & 83.3 (82--86) & 0.417 & 73\% \\
        Color--magnitude baseline & 7 & 10 & 15 & 0.075 & 13\% \\
        Logistic baseline & 1 & 1 & 3 & 0.015 & 3\% \\
        \hline
    \end{tabular}
\end{table}

\section{The survey of M31}
\label{sec:surveysec}

\subsection{Survey and ranking}
\label{sec:survey}

We scanned all four parts of the mosaic with $512 \times 512$ pixel windows. Because detection is
uniform up to 64 pixels from the edge of a window and drops slightly closer to it, we step the
windows by $512 - 2 \times 64 = 384$ pixels and keep a detection only from the window whose center
is nearest to it. In this way every position is seen by exactly one window, at least 64 pixels from
its edge, and repeated looks at the same source cannot inflate false positives. This gives 32\,063
windows, 83\% of them with UV data. Detections above a score of 0.5 were kept, and overlapping
detections of a member were merged by non-maximum suppression.

The raw scores of the members, however, cannot be compared directly. The same score of 0.995, for
example, meant a purity of 91\%, 74\% and 37\% in three members. Each member therefore gets the same
budget per area, set by the working point of 200 detections in 903 validation windows and scaled
with the area of the survey windows. This budget is set separately for the areas with and without
UV, since in the latter only the ResNet-50 member contributes. The candidates are then ranked by the
sum of their rank scores across members, and detections within 15 pixels are merged.

To make the discovery list, we removed detections within 15 pixels of any labeled PN (training,
validation or test). Labeled negatives, in contrast, were kept, so that the list shows how often the
detector still ranks them high. Each candidate was then matched to SIMBAD within 15 pixels. Catalog
positions, however, can be offset by more than this. For the 476 known PNe with a position in the
DOLPHOT catalogs (Section~\ref{sec:labels}), the SIMBAD position lies a median of 11.6 pixels from
the source, and 34\% lie beyond 15 pixels, most of them positions from \citet{1978ApJS...38..351F}.
Candidates without a match therefore get two flags. A candidate within 15 pixels of the catalog
position of a known PN in the DOLPHOT catalogs is marked as that PN. A candidate near a SIMBAD
position corrected for the median offset of its catalog is marked as a possible match, within 15
pixels for any type or within 25 pixels for the sparse types PN, possible PN, EmO, \ion{H}{2} region
and nova. When we repeat this matching with shifted positions, it gives about two chance matches
among the unlabeled candidates of the top 400.

\subsection{Selection function in the survey}
\label{sec:selfunc}

The 203 validation and test PNe were never seen by the training-only models, so we asked how many of
them these models find in the full survey, within the per-area budget. Here a PN counts as found if
a candidate lies within 15 pixels, and intervals are 95\% bootstrap intervals \citep{Efron1979}.

\begin{figure*}
    \centering
    \includegraphics[width=\textwidth]{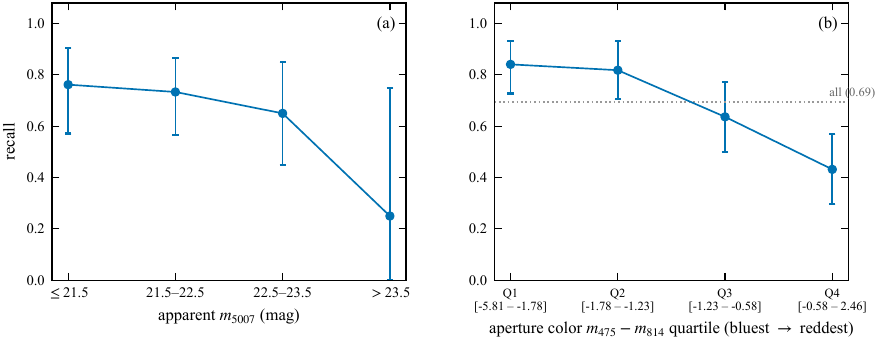}
    \caption{Selection function of the single model, one of the four members. (a) Recall against
    apparent $m_{5007}$ in the fixed evaluation windows, for M31 validation PNe with the
    training-only model. The model did not see these PNe. (b)
    Recall of the training-only model in the full survey within the per-area budget, for the 203
    validation and test PNe, against quarters of our aperture color $m_{475}-m_{814}$. The dotted
    line is the recall of all 203 PNe (0.69). The 27 PNe without a net F814W flux are in no quarter. Error bars are 95\%
    bootstrap intervals. The two panels count a PN as found in different ways, in the evaluation
    window (a) or in the survey list (b).}
    \label{fig:selection}
\end{figure*}

Because the budget is set per member, we measure the selection function on one member, the
training-only single model. It finds 0.695 $[0.631, 0.759]$ of the PNe, and the other three members
find 0.68--0.72. The full list of the training-only ensemble merges the lists of the four members
and is therefore longer, so it finds more, 0.823. Recall of the single model falls toward faint
magnitudes, from 0.88 for $m_{5007} \leq 21.5$ to 0.73, 0.57 and finally 0.31 above 23.5
($n = 13$). In contrast, it depends only weakly on galactocentric radius, crowding and local
background. The strongest dependence is on color, measured in our own aperture photometry as
$m_{475}-m_{814}$, with a recall of 0.84 in the bluest quarter and 0.43 in the reddest. In the top
400 of the ensemble list, too, found and missed PNe differ in $m_{5007}$ (KS
test, $p = 1\times10^{-9}$) and color ($p = 2\times10^{-9}$), but not in radius, background or
crowding ($p > 0.2$). We see two possible reasons why red PNe are missed, but could not tell them
apart. The red missed PNe may have real continuum light, in which case some may not be PNe.
Alternatively, their F814W flux may include light from neighbors, since our F814W aperture
photometry is biased in the inner disk. The detector is therefore strongest where the training
catalog is most complete, at the bright and blue end of the PN population.
Figure~\ref{fig:selection} is the completeness curve that users of the catalog need.

\subsection{Cosmic-ray check with independent exposures}
\label{sec:cosmic}

The survey teams removed cosmic rays before building the mosaics, but this cleaning served their
star catalogs, and residuals remain in the images \citep{2012ApJS..200...18D,2014ApJS..215....9W}.
PHAST notes that many remain where few exposures overlap, such as the chip gap
\citep{2025ApJ...979...35C}, and we find them as well. Because a cosmic ray in the F475W mosaic has
no counterpart in F814W, it looks like a compact blue source, just as a PN does.

A cosmic ray hits the detector during a single readout, so a source that appears in only one of
several independent exposures of the same position is a cosmic ray. For each candidate we therefore
measure the aperture signal-to-noise ratio (S/N) in every single F475W exposure (FLC file) that
covers it. A candidate is called a cosmic ray if three conditions hold. First, the ratio of its
lowest to its highest S/N is at most 0.05. Second, it has at least three deep exposures, of
100\,s or longer, so that the short exposures do not decide. Third, it is seen in only one readout.
The threshold is passed by 99.5\% of empty positions in the same cutouts. Where the exposures do not
decide, we look for the source in F814W, F336W and F275W, which were taken at other times. A source
seen there is real, but its absence proves nothing, because PNe are faint in F814W. Other bands can
therefore only turn an undecided candidate into a real one, never into a cosmic ray.

\begin{figure*}
    \centering
    \includegraphics[width=\textwidth]{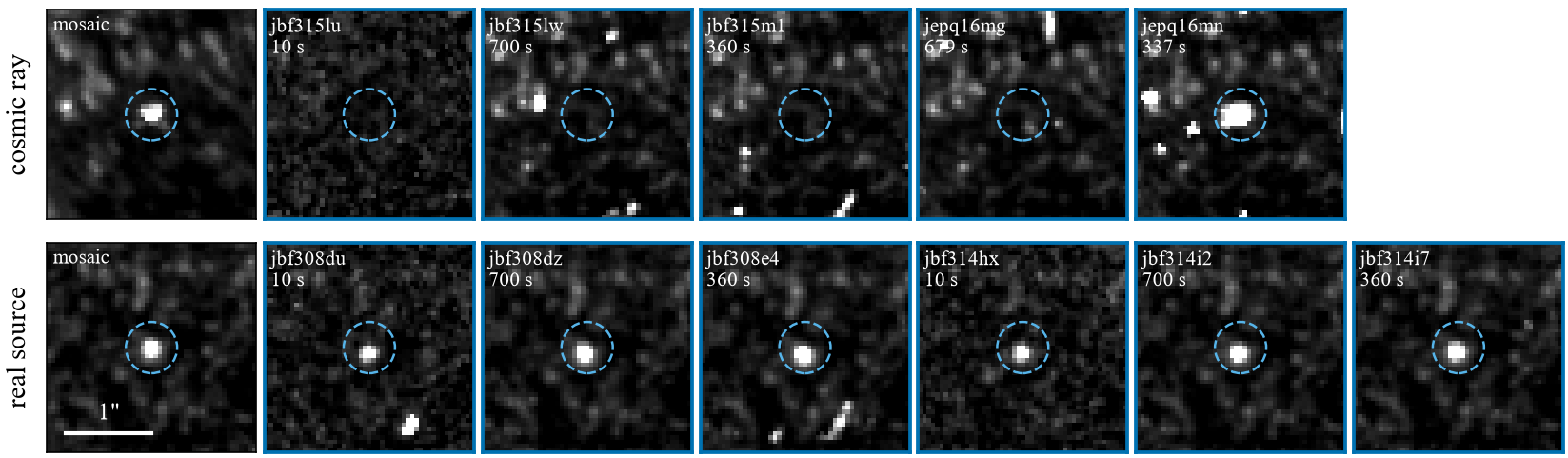}
    \caption{The cosmic-ray check. Each row shows the F475W mosaic (left) and every independent
    F475W exposure that covers the candidate (blue frames), labeled with its dataset name and
    exposure time. The dashed circle marks the candidate. In the top row the source appears
    in only one of five exposures, and the check calls it a cosmic ray (ratio of lowest to highest
    S/N below 0.01). In the bottom row the source is seen in all six exposures (ratio 0.76) and is
    called real. All panels share one stretch in units of the local
    noise.}
    \label{fig:exposures}
\end{figure*}

We tested the check on 100 known PNe from the validation and test parts. None was wrongly flagged
(0/100), and the 95\% Wilson interval \citep{1927JASA...22..209W} is 0--3.7\%. Of the
known PNe, 94.9\% are visible in F336W, but only 2--7\% in F814W, as expected for a source with
little continuum. Figure~\ref{fig:exposures} shows the test for one cosmic ray and one real
source. We do not teach the detector about cosmic rays, because adding them to the training loss
cost performance (Table~\ref{tab:axes}). An existing cleaner is no alternative either, since the
deepCR network \citep{2020ApJ...889...24Z}, trained on single ACS/WFC exposures, performs poorly on
our mosaics (AUROC 0.63).

In the top 400 of the discovery list, the check calls 194 candidates real and 122 more real with
help from another band. It calls 12 cosmic rays, while another 72 could not be decided, mostly
because too few deep exposures cover them. We deleted no row from the catalog, and the result of the
check is carried as a flag instead.

\subsection{The candidate catalog}
\label{sec:catalog}

\begin{figure}
    \centering
    \includegraphics[width=0.5\textwidth]{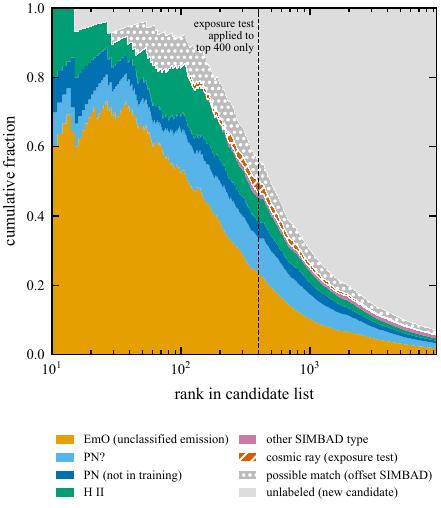}
    \caption{Cumulative composition of the discovery list by rank, with one class per candidate.
    Cosmic rays (hatched) come first and are taken from the exposure check, which was run for the
    top 400 only. Further down, the share of cosmic rays is unknown. Other candidates take
    their SIMBAD type within 15 pixels (EmO, possible PN, PN, \ion{H}{2}, other), a possible match
    (dotted, Section~\ref{sec:survey}) or are unlabeled. PN here means SIMBAD PNe that were not
    training labels, including candidates on the DOLPHOT position of a known PN.}
    \label{fig:composition}
\end{figure}

The discovery list of the retrained ensemble has 9548 candidates, each with a PN score from 0 to 100
(Figure~\ref{fig:composition}), and all labeled PNe were removed from it. EmO objects in SIMBAD lead
the list, with 14 of the top 20 and 52 of the top 100. The top 100 also hold 12
possible PNe, 5 PNe without a training label and 14 \ion{H}{2} regions, as well as 9 possible
matches (Section~\ref{sec:survey}) and 8 candidates without any counterpart. In the top 400 there are
93 EmO objects, 41 possible PNe, 19 PNe, 28 \ion{H}{2} regions, 6 objects of other types, 12 cosmic
rays, 25 possible matches and 176 candidates without a counterpart. These counts, however, must be
read against how many objects of each type the survey area holds without a role in training, namely
1921 EmO objects, 710 possible PNe, 533 PNe and 2675 \ion{H}{2} regions. Relative to these numbers,
the top 400 holds 5.2\%, 5.8\%, 3.4\% and 1.0\% of them (Section~\ref{sec:untaught}).
Down to rank $\sim$600, every candidate was ranked by at least three of the four members. About half of the EmO objects in the list (93 of 177)
are in the top 400, and further down the unlabeled candidates take over.

The list can hold SIMBAD PNe and EmO objects because they were not training positives
(Section~\ref{sec:labels}). A positive needed an [\ion{O}{3}] magnitude from
\citet{2006MNRAS.369..120M} and a unique counterpart on the \textit{HST} image, which not every
SIMBAD PN has. The EmO objects that papers call PNe, on the other hand, rest on weaker evidence than
the positives (Section~\ref{sec:labels}). Where they fall in the list is therefore a test of the
detector on objects it never saw (Section~\ref{sec:untaught}).

For the 214 candidates of the top 400 without a SIMBAD label, or with only an offset match, we
searched all VizieR catalogs and NED within $3\arcsec$, computing the chance of a random match from
the density of each catalog. Each candidate then gets one verdict: known if a catalog lists it
as a PN, a PN candidate or an emission-line object with a spectrum, non-PN if a catalog
classifies it as an \ion{H}{2} region, supernova remnant, nova or star with a spectrum, and
discovery candidate otherwise. Matches to photometric star, variable or X-ray catalogs are noted but do
not change the verdict. In total, this search finds 37 known objects (16 of them with a spectrum)
and 2 non-PNe, which leaves 175 discovery candidates, 12 of which are flagged as cosmic rays.

Most EmO candidates already carry a PN class in the literature, although SIMBAD lists them only as
emission-line objects. Of the 100 EmO objects in the top 400, 93 are called a PN or possible PN by a
paper linked in SIMBAD, one is called both a PN and an \ion{H}{2} region, and none is called only an
\ion{H}{2} region (Section~\ref{sec:untaught}). The evidence behind these classes, however, varies.
Only 10
have a spectrum that separates a PN from other emission-line objects, nine from LAMOST
\citep{2025AJ....169..174C} and one from \citet{2015ApJ...815...69F}, and all ten are PNe or PN
candidates. Another 47 are PNe that \citet{2014ApJ...792..121V} confirmed on \textit{HST} images,
and about 42 have only a single-line [\ion{O}{3}] detection. For 14, one of the sources instead calls
the object a Wolf--Rayet star or an \ion{H}{2} region.

\begin{figure}
    \centering
    \begin{minipage}[b]{0.48\textwidth}
        \centering
        \includegraphics[width=\linewidth]{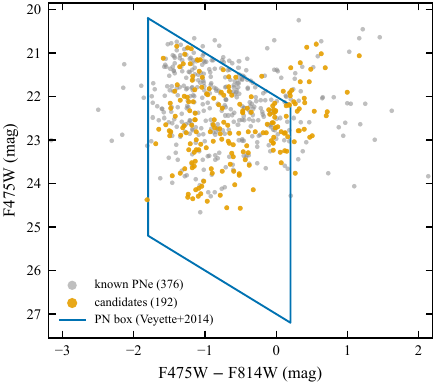}
    \end{minipage}\hfill
    \begin{minipage}[b]{0.48\textwidth}
        \centering
        \includegraphics[width=\linewidth]{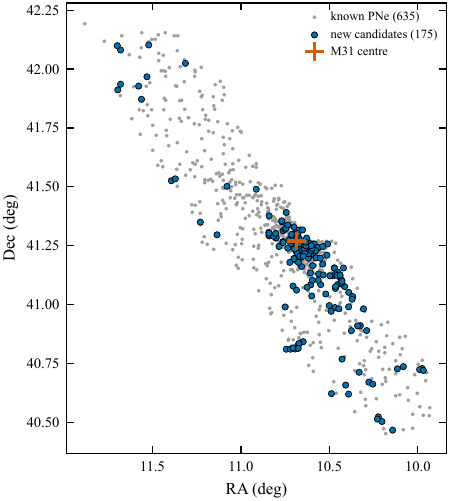}
    \end{minipage}\\[1ex]
    \begin{minipage}[t]{0.48\textwidth}
    \caption{Color--magnitude diagram from PHAT DOLPHOT photometry \citep{2014ApJS..215....9W}.
    Gray points are training PNe with a match (376 of 635). Orange points are the candidates of the
    top 400 that are not flagged as cosmic rays, with a match (192 of 388). The line marks the PN box of \citet{2014ApJ...792..121V},
    $-1.8 <$ F475W$-$F814W $< 0.2$ and $22 <$ F814W $< 27$, drawn in the plane of color and
    F475W.}
    \label{fig:cmd}
    \end{minipage}\hfill
    \begin{minipage}[t]{0.48\textwidth}
    \caption{Discovery candidates on the sky. Gray points are the 635 known PNe used for training.
    Circles are the 175 discovery candidates of the top 400. The cross marks the center of M31.}
    \label{fig:map}
    \end{minipage}
\end{figure}

Most candidates have the colors of PNe. Of the 388 candidates in the top 400 that are not flagged as
cosmic rays, 192 have a PHAT DOLPHOT match within $0\farcs3$ that agrees in F475W, while most of the
others (157) have no DOLPHOT source within $1\arcsec$ because they lie outside the PHAT area. Of the
matched
candidates, 82\% fall in the color--magnitude box of \citet{2014ApJ...792..121V}, as do 81\% of the
known PNe (Figure~\ref{fig:cmd}).
For the PNe in their own sample our box holds 97.6\%. The lower value for all
known PNe comes from objects outside their sample, mostly in the bulge region, many of which are
brighter in F814W than expected. The candidates outside the box lie in the same region, and at the
same deprojected distance from the center they fall outside it about as often as the known PNe, 71\%
against 64\% within $5\arcmin$ and 2\% against 6\% beyond $25\arcmin$. If the candidates had the rate
of the known PNe at their own distances, 32 of the 192 would fall outside the box, and 34 do.
Two effects could make PNe in the bulge redder, dust and blending with unresolved stars. Dust would
also redden F336W$-$F475W, by roughly 1\,mag for the observed shift under a standard extinction law.
Yet the known PNe outside the box are only 0.09\,mag redder in F336W$-$F475W than those inside,
against 1.31\,mag in F475W$-$F814W. Light from unresolved red giants, in contrast, would add mostly
to F814W and the infrared, and in the PHAT infrared band F160W these PNe are 0.83\,mag redder
in F814W$-$F160W. This points to blending rather than dust, although we have not tested it directly.
By the cuts of Section~\ref{sec:eval}, 50 of the discovery
candidates are hot, 29 reddish and 88 in between. In pixel photometry of the whole top 400, the
ensemble has 51 hot and 40 reddish contaminants. The reddish ones come with the Swin-T members and
lie mostly in the bright inner disk. Since their color stays red when measured on the F475W peak,
they are not mismatches.

On the sky (Figure~\ref{fig:map}), the discovery candidates lie closer to the center than the known
PNe (median deprojected radius $15\farcm7$ against $25\farcm9$, KS test $p = 3\times10^{-7}$). Their
$m_{5007}$ estimated from F475W is also fainter (median 22.29 against 21.97, $p = 7\times10^{-8}$),
and their ratio of F475W to F814W flux is lower (median 1.55 against 2.80, $p = 2\times10^{-15}$).
Most of them therefore have more continuum than PNe, which the narrow-band check of
Section~\ref{sec:narrowband} confirms.

\begin{figure*}
    \centering
    \includegraphics[width=\textwidth]{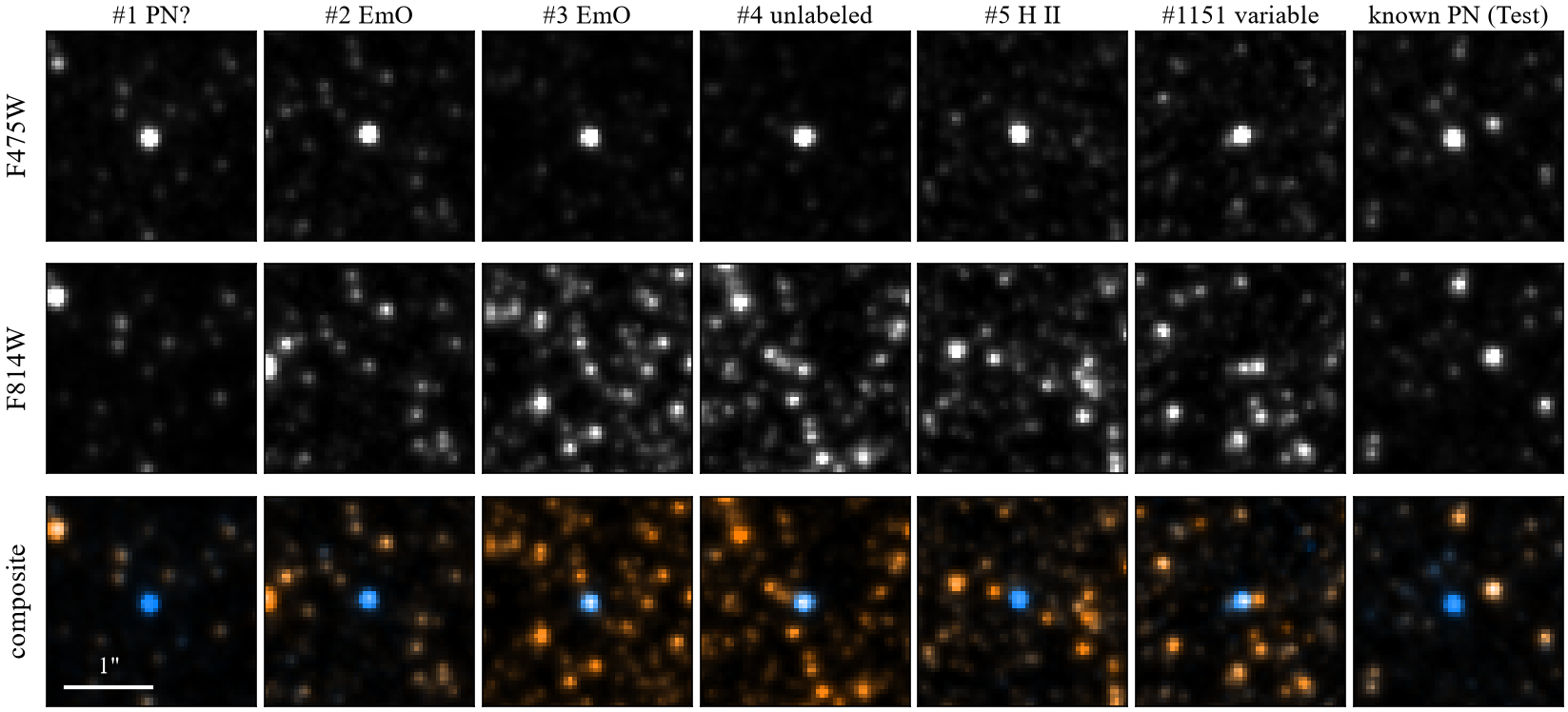}
    \caption{Cutouts from the discovery list. The first four columns are the top four candidates
    called real by the cosmic-ray check, a possible PN, two EmO objects and a source without a SIMBAD
    counterpart. The next two are the highest-ranked \ion{H}{2} region (rank 5) and variable star
    (rank 1151) in the list. The last is the test PN ranked highest by the training-only single
    model, which never saw it. Rows show F475W, F814W and the color composite [F814W, mean, F475W].
    Each panel is 48 pixels ($2\farcs4$) wide and uses a percentile stretch.}
    \label{fig:gallery}
\end{figure*}

Figure~\ref{fig:gallery} shows cutouts of top candidates next to typical contaminants and a known
PN. The targets for spectroscopy then follow from the narrow-band check in the next section, and
Table~\ref{tab:candidates} gives the one candidate that it leaves as a PN.

\begin{table}
    \centering
    \caption{Properties of the new PN. The last column gives the PN criterion of
    \citet[M06]{2006MNRAS.369..120M}, \citet[V14]{2014ApJ...792..121V} or
    \citet[B19]{2019A&A...624A.132B} where one exists, and the source meets every one. The PN score
    (0--100) is from the detector. F475W$-$F814W is from PHAT DOLPHOT, and the size is an upper
    limit on the intrinsic size in F475W. $Z_{\rm PN}$ scores how well the \textit{HST} counterpart
    matches a typical PN, here from the expected F475W and the color only, with the 467 PNe of V14
    as reference. $m_{5007}$ is from the MegaCam [\ion{O}{3}] excess. The [\ion{O}{3}] S/N is that
    of the excess in a $0\farcs8$ aperture, and $m_n-m_b$ is measured in a $1\arcsec$ aperture for
    the 2016 and 2018 epochs. $\log$([\ion{O}{3}]/H$\alpha$) is measured in LGGS and is higher than
    for 94\% of the objects that M06 flagged for a low ratio at the same $m_{5007}$.}
    \label{tab:candidates}
    \small
    \setlength{\tabcolsep}{4pt}
    \begin{tabular}{lll}
        \hline
        Property & Value & PN criterion \\
        \hline
        \multicolumn{3}{l}{\textit{Position and detector}} \\
        RA, Dec (J2000) & 00:44:32.33 $+$41:17:45.4 & \\
        Distance from center & $20\farcm4$ (projected) & \\
        PN score & 88, ranked by all four members & \\
        \hline
        \multicolumn{3}{l}{\textit{HST}} \\
        F475W$-$F814W & $-1.13$ & $-1.8$ to 0.2 (V14) \\
        Shape in F475W & point-like, $\lesssim 0.23$\,pc & point-like (all) \\
        $Z_{\rm PN}$ & 0.73--0.78 & $<1$ (V14), $<1.5$ (B19) \\
        \hline
        \multicolumn{3}{l}{\textit{Narrow band}} \\
        $m_{5007}$ & 22.7 & \\
        {[\ion{O}{3}]} excess & S/N 32, in both epochs & detected (all) \\
        {[\ion{O}{3}]} flux above continuum & 87--96\% & \\
        $m_n-m_b$ & $-2.05$, $-1.84$ & $< -1$ (B19) \\
        $\log$([\ion{O}{3}]/H$\alpha$) & 0.19 & not flagged (M06) \\
        \hline
    \end{tabular}
\end{table}

\subsection{Narrow-band check of the candidates}
\label{sec:narrowband}
\label{sec:nbmethod}

We measured the emission lines of the top 400 in the data of Section~\ref{sec:nbdata}. In MegaCam,
for each candidate we scale all [\ion{O}{3}] and all $g$ exposures to a common level within each band
and combine them with inverse-variance weights. We then subtract the continuum by measuring the excess
[\ion{O}{3}]$-k\,g$ in apertures of $1\farcs5$ and $0\farcs8$, where $k$ scales the $g$ continuum to
the [\ion{O}{3}] band and is taken, together with the noise, from 400 random positions nearby.
To keep a cosmic ray in one exposure from creating a false excess, exposures that deviate by more
than $5\sigma$ from the median are dropped, provided they are a minority. An excess with S/N $> 5$
counts as detected, which happens in less than 1\% of cases at random positions. We convert the
excess to $m_{5007}$ with the PNe of \citet{2006MNRAS.369..120M} detected at S/N $> 10$ ($n = 169$,
scatter of the zero point 0.21\,mag). The local $5\sigma$ limit of a candidate has a median of
$m_{5007} = 24.31$ (10th to 90th percentile 22.78--25.07), so a non-detection says nothing about
fainter PNe. Known PNe with $m_{5007} \leq 24$, on the other hand, are detected in 84--96\% of
cases. Finally,
since the exposures come from two epochs (2016 and 2018), we require a discovery to be detected in
both.

In the SITELLE cubes, each cube was first calibrated in position and velocity on known PNe. At the
position of a candidate, we then take the highest S/N in a window of $7 \times 7$ pixels and over
the velocity channels. [\ion{O}{3}] counts as detected if this S/N exceeds 5 and $\lambda4959$
exceeds 2, and H$\alpha$ if it exceeds 6. Because it is a maximum, this S/N runs high at random sky
pixels (95th percentile 3.3--4.5 for [\ion{O}{3}]). In six H$\alpha$ cubes, the diffuse
H$\alpha$ of the disk gives an S/N above 6 at 5--15\% of random pixels, so there H$\alpha$ alone does
not separate sources. Within 150 pixels of the edge of a cube a non-detection is not reliable, and
we use MegaCam there instead.

Before turning to the discovery candidates, we checked that these measurements work on known
objects. MegaCam finds an [\ion{O}{3}] excess at all 87 EmO objects, all 24 \ion{H}{2} regions and
50 of 53 PNe and possible PNe in the top 400, and at none of the 4 stars of type No*. Where both data
sets exist, SITELLE and MegaCam also agree. Of 73 candidates with SITELLE [\ion{O}{3}], MegaCam
detects 68, and of 116 without it, MegaCam detects 13, 10 of which lie within 120 pixels of a cube
edge. In the cubes added for this work, 47 of the 49 sources with [\ion{O}{3}] also show H$\alpha$ at
a consistent velocity (median difference 24\,km\,s$^{-1}$). All of these are SIMBAD objects or known
from the literature.

These results agree with the literature, since the PNe of M31 were found in [\ion{O}{3}]
(Section~\ref{sec:intro}) and the PN.S survey is complete to $m_{5007} \approx 23.75$ outside the
bright center \citep{2006MNRAS.369..120M}. These surveys have thus labeled the sources that are bright
in [\ion{O}{3}], at least as emission-line objects, and our check confirms it. Every [\ion{O}{3}]
source in the top 400 is already listed, as a PN, a possible PN, an EmO object or an \ion{H}{2}
region, with the one exception described below. Finding many new PNe with strong [\ion{O}{3}] in the
same kind of data would therefore have been unexpected. Our training labels come from these surveys
as well, so the population that the detector learned best is the one they have already cataloged.
Any PN that these surveys missed must instead
be hard to see in [\ion{O}{3}], so each discovery candidate is either such a PN or a contaminant.

The 175 discovery candidates, in contrast, show almost no lines. In SITELLE, none of 94 has
[\ion{O}{3}] and none of 88 has H$\alpha$. Five have an [\ion{O}{3}] S/N above the 95th percentile of
random pixels but below the detection threshold, where 4.7 are expected by chance. For 16
candidates near a cube edge we rely on MegaCam instead. Across all discovery candidates, MegaCam
shows a consistent excess in both epochs for only one, and two more show one in a single epoch and
none in SITELLE.

A missing line, however, only means that the source is fainter in the line than the limit. We
therefore asked whether each candidate would be seen if it were a PN like the known ones. For the
known PNe with a DOLPHOT match and $m_{5007}$ between 22 and 24 ($n = 231$), F475W$-m_{5007}$ has a
5th percentile of $-0.20$, a median of 0.13 and a 95th percentile of 0.41. As such a PN, a
candidate of a given F475W would thus have $m_{5007}$ at most F475W$+0.20$, and if even this value
is brighter than its local MegaCam limit, the PN would be seen. The test assumes that the F475W light
belongs to the candidate. If light from a neighbor contributes, the test becomes stricter against
the PN. We take F475W from DOLPHOT if the match agrees with our own pixel photometry, and from pixel
photometry otherwise.

For 156 candidates a PN like the known ones would be seen, because its [\ion{O}{3}] would lie above
the local limit, by a median of 1.4\,mag (less than 0.3\,mag for 8), and two are borderline. For six,
the MegaCam limit is shallow ($m_{5007}$ limit 21.5--22.6), but SITELLE or LGGS reach the depth
instead. Within a few arcminutes of each, these data detect 77--100\% of the PNe of
\citet{2006MNRAS.369..120M} at the $m_{5007}$ that such a PN would have, and none of the six shows a
line. Eight could not be measured, and three are detected. Most discovery
candidates are therefore unlike the known PNe of M31. This does not rule them out as PNe, however,
because the test inherits a bias from its reference sample. The known PNe were found in
[\ion{O}{3}], so by construction their [\ion{O}{3}] is strong and most of their F475W light is line
emission. A PN with weak [\ion{O}{3}] would therefore look like a non-PN in this test, and the literature
describes such PNe. PNe with cool
central stars, among them very young ones, can have weak or even negligible [\ion{O}{3}]
\citep{2010PASA...27..129F}. Models of the PNLF therefore lower the [\ion{O}{3}] emission of PNe
whose central stars are cooler than about 60\,000\,K, and even so, PNe with weak [\ion{O}{3}] are
more common in the Galaxy than these cool stars explain \citep{1997A&A...321..898M}. Such a PN puts
less of its F475W light into [\ion{O}{3}] than the known PNe do. Its central star heats at
nearly constant luminosity \citep{1997A&A...321..898M}, so while it is still cool it also adds the
most optical continuum to F475W. Our candidates may come from this population, and since a cool
central star is still a hot star, such PNe could hide among the 50 discovery candidates that are hot in UV,
but not among the 29 reddish ones (Section~\ref{sec:catalog}). Light from a neighbor works in the
same direction. We did not measure, however, how many candidates belong to this population. The
others may be hot or reddish stars, or blends, and only a spectrum can tell. Finally, four
candidates that are seen only in F475W have no cosmic-ray verdict and remain open.

\begin{figure*}
    \centering
    \includegraphics[width=\textwidth]{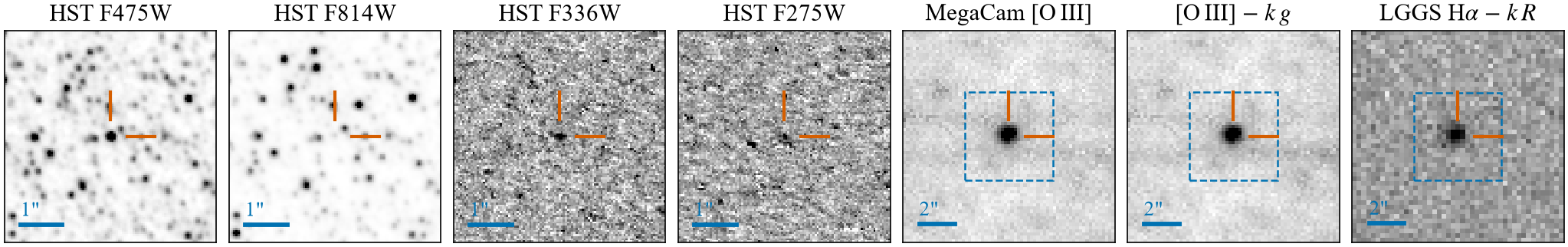}
    \caption{The discovery candidate with an [\ion{O}{3}] excess, rank 86 in the list. From the left,
    \textit{HST} F475W, F814W, F336W and F275W, $5\arcsec$ across, each with its own percentile
    stretch. Then MegaCam [\ion{O}{3}] stacked from nine exposures, the same after subtracting the
    scaled $g$ stack, and LGGS H$\alpha$ after subtracting the scaled $R$ image, $12\arcsec$ across. The
    dashed square marks the area of the \textit{HST} panels and the ticks mark the candidate. North is up
    and east is to the left in all panels. No SITELLE cube covers this position.}
    \label{fig:cand86}
\end{figure*}

One candidate stands out (Figure~\ref{fig:cand86}). It lies at 00:44:32.33 $+$41:17:45.4,
$20\farcm4$ from the center, and all four members rank it. MegaCam shows an [\ion{O}{3}] excess in
each of its nine exposures from two epochs separately (S/N 17--28 in $0\farcs8$). The excess grows
with aperture like a point source, and 87--96\% of its [\ion{O}{3}] flux is excess over the
continuum. LGGS also shows [\ion{O}{3}] and H$\alpha$, with $\log($[\ion{O}{3}]/H$\alpha) = 0.19$, a
value shared in the same field by 32\% of the PNe of \citet{2006MNRAS.369..120M} and 5\% of compact
\ion{H}{2} regions. On \textit{HST}, the source has the color of a PN (F475W$-$F814W $= -1.13$) and
is not resolved in F475W (intrinsic size $\lesssim 0.23$\,pc). Its $m_{5007}$ is about 22.7, yet no
catalog lists an emission-line source at its position. The source lies $13\arcsec$ from a star of
$G = 12.2$, along the east--west direction of the stellar trails in the slitless PN.S images. Along
these trails, PNe of \citet{2006MNRAS.369..120M} are missing within 3--8$\arcsec$ of bright stars (0
found where 8.2 are expected, $p \approx 3\times10^{-4}$), and \citet{2006MNRAS.369..120M} gave the
same reason for eight PNe of \citet{2006MNRAS.369...97H} that they did not find. This likely
explains why the source was missed, but it remains a hypothesis.

The two largest PN surveys of M31 classified PNe from images, with criteria aimed at \ion{H}{2}
regions, their main contaminant, and the candidate passes the criteria of both. The PN.S survey of
\citet{2006MNRAS.369..120M} saw only the [\ion{O}{3}] line. Because even the largest PNe in M31 are
unresolved from the ground, while most \ion{H}{2} regions are extended, they removed sources wider
than the seeing. Since a PN emits most of its light in [\ion{O}{3}], they also removed sources with
a low ratio of [\ion{O}{3}] to H$\alpha+$[\ion{N}{2}], and called the rest probable PNe. Our
candidate is unresolved even on \textit{HST}, and its ratio is shared by a third of their PNe.
\citet{2019A&A...624A.132B} used the same property as a color, since a PN is much brighter in the
narrow [\ion{O}{3}] band than in the broad $g$ band around it. They kept point-like sources with
$m_n - m_b < -1$, where $m_n$ and $m_b$ are the AB magnitudes in the two bands, a cut that
corresponds to an equivalent width of 110\,\AA. Because the MegaCam exposures we use come from their
survey, we could apply their cut to the same data, with the zero points of the image headers checked
on the PNe of \citet{2006MNRAS.369..120M} in these exposures. In a $1\arcsec$ aperture the candidate
has $m_n - m_b = -2.05$ and $-1.84$ in the two epochs, close to the median of $-2.1$ of the known
PNe. In a $2\arcsec$ aperture the values rise to $-1.31$ and $-1.16$, which is still below the cut.
The known PNe change less, so light from neighbors probably enters the larger aperture. With the
relation $m_{5007} = m_n + 2.27$ of \citet{2019A&A...624A.132B}, the candidate has $m_{5007}$ of
22.3--22.9, as found above. Their catalog of 4289 PNe, however, is not public, so we cannot tell
whether they list the source.
\textit{HST} gives a last check, because \citet{2014ApJ...792..121V} found that sources that look extended
from the ground but are point-like on \textit{HST} are good PN candidates, and ours is point-like
with the color of a PN. Both \citet{2014ApJ...792..121V} and \citet{2019A&A...624A.132B} also
scored how well the \textit{HST} counterpart of a PN matches typical PNe in F475W, color, position
and PSF shape. Using the 467 PNe of \citet{2014ApJ...792..121V} as reference, the F475W and color
terms give our candidate a score of 0.73--0.78, within the best class of both papers. The other
terms need data that we do not have. With the ratio of [\ion{O}{3}] to H$\alpha$ measured the same way,
the candidate is also higher than 94\% of the objects that \citet{2006MNRAS.369..120M} flagged as
\ion{H}{2} regions for a low ratio at the same magnitude. By the criteria on which the largest
catalogs of M31 PNe were built, the candidate is therefore a PN, found by the detector and listed in
no published catalog (Table~\ref{tab:candidates}). Most PNe in these catalogs have no spectrum
either, so the same caveat applies to them and to our candidate. With \textit{HST} images,
\citet{2014ApJ...792..121V} showed that 152 of the 711 PNe of \citet{2006MNRAS.369..120M} in the
PHAT area are \ion{H}{2} regions or stars, so a compact \ion{H}{2} region can pass the same tests,
and only a spectrum can rule it out.

\section{What the detector relies on}
\label{sec:behavior}

To see what the detector relies on, we probe it with interventions rather than saliency maps
\citep{2022MNRAS.511.5032B}. We remove parts of the input, a channel-level version of
occlusion \citep{2013arXiv1311.2901Z}, and build comparison samples without asking the model. Some analyses were made on an earlier model, an ensemble of three detectors from our first search, and are marked as such.

\subsection{Removing color}
\label{sec:ablation}

At inference time, without retraining, we filled all three input channels with F475W, which removes
color but keeps brightness and shape. Two variants used F814W or the mean channel instead. We ran
the four three-channel ResNet-50 settings with the mean channel, one of which is the delivered
ResNet-50 member, with three seeds each. Without color, P@200 falls from 0.246 to 0.017 on
validation and from 0.362 to 0.013 on test, and no run exceeds 0.04, even below the
color--magnitude baseline (0.065 and 0.075), and the other two variants collapse further (0.001 and
0.000). Color is therefore necessary but not sufficient. Gray images are foreign to the model,
but a shift of the input alone does not explain the collapse. When only the stretch was
changed and color kept, the asinh models kept more than half of their score. The test does not
tell, however, what a model trained without color would do \citep{2018arXiv180610758H}.

\subsection{Classes the detector was never taught}
\label{sec:untaught}

\begin{figure*}
    \centering
    \includegraphics[width=0.85\textwidth]{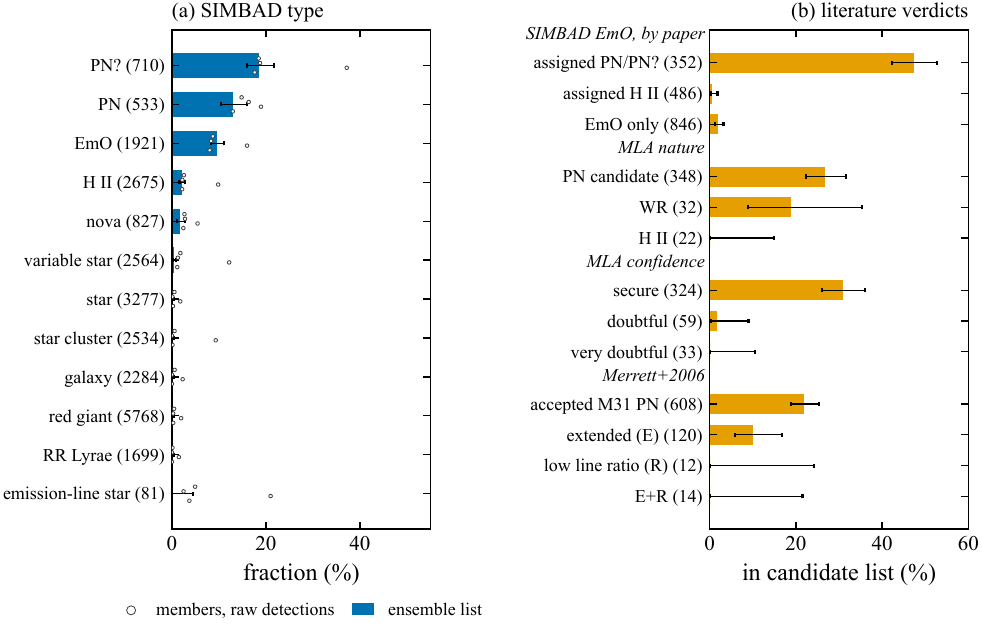}
    \caption{Rates at which objects that played no part in training enter the discovery list.
    (a) By SIMBAD type, matched within 15 pixels. Bars are the rates for the delivered ensemble list.
    Open points are the raw detection rates of the four members (score $\geq 0.5$). The point far to
    the right in most rows is the ResNet-50 member. (b) By the class given in the
    literature. ``Assigned PN/PN?'' are SIMBAD EmO objects that a linked paper explicitly calls a
    PN or possible PN, mostly from [\ion{O}{3}] imaging or spectra. ``PN candidate'' are EmO objects
    whose nature column is empty in the objective-prism catalog of \citet{1993A&AS..102..251M},
    which by that catalog's rule makes them unconfirmed PN candidates. The two groups may partly
    overlap. ``Secure'', ``doubtful'' and ``very doubtful'' are the confidence classes of the same
    catalog. They refer to whether the emission on the prism plates is real, not to the nature of
    the object. The Merrett+2006 classes are from \citet{2006MNRAS.369..120M}. Numbers in brackets
    are the sizes of the classes, excluding objects with a role in training. Black lines are 95\%
    Wilson intervals. Rows without a visible bar are 0/$n$ (for example 0/22 for prism \ion{H}{2}),
    not missing.}
    \label{fig:types}
\end{figure*}

The training had only two classes, PN and negative. EmO objects and possible PNe were never shown,
while \ion{H}{2} regions, stars and galaxies were all negatives. Still, SIMBAD objects that played no
part in training enter the list in a clear order (Figure~\ref{fig:types}a). Possible PNe enter at
18.6\%, PNe without training labels at 12.9\%, EmO objects at 9.6\% and \ion{H}{2} regions at 2.1\%,
whereas stars, galaxies and clusters enter at 0.2\% or less. This order does not arise from chance
coincidences of position, since after all positions were shifted by 50 to 150 pixels, the rates of
the PN-like classes fell to 0.1--0.2\%. Nor does color alone explain it, since many variable stars
score above 0.5 but the ranking pushes them down, and only 0.5\% enter the list.

The list also follows classes given in papers, which the model never saw
(Figure~\ref{fig:types}b). Of the EmO objects that a paper calls a PN or possible PN, 47.4\% enter
(167 of 352), against 0.6\% of those called \ion{H}{2} regions (3 of 486)
\citep{1964ApJ...139.1027B,1978A&AS...31..439P,1992A&AS...92..625W,1993A&AS..102..251M,2006MNRAS.369..120M,2011AJ....142..139A}.
Similarly, in the objective-prism catalog of \citet{1993A&AS..102..251M}, 27\% of the PN candidates
enter and none of the 22 \ion{H}{2} regions, and the list follows the catalog's own confidence
classes. Among the objects of \citet{2006MNRAS.369..120M}, too, accepted PNe enter more often
(21.9\%) than those flagged as extended (10.0\%) or with low line ratios (0 of 26).

Spectra agree with this ordering, since every EmO candidate with a classifying spectrum is a PN or PN
candidate (Section~\ref{sec:catalog}), and all show [\ion{O}{3}] in MegaCam
(Section~\ref{sec:narrowband}), which confirms the emission but cannot separate PNe from \ion{H}{2}
regions. We therefore present EmO candidates as candidates for a new classification.

\subsection{Is it a blue point detector?}
\label{sec:bluepool}

If the model looked only at blueness, its list would fill with random blue points. To test this, we
built a pool without the model. We found local peaks in 300 random windows per part, measured them
with aperture photometry (S/N $\geq 5$), excluded every cataloged object within 30 pixels and drew
40\,000 of the 630\,078 sources. A PN box, the inner 90\% of the 635 PNe in color and F475W flux,
holds 328 of them. A color detector would accept sources in the box about as often as known PNe. At
a score of 0.9, however, the four training-only members accept 2.1--7.3\% of the box and 84--86\% of
the 203 validation and test PNe, so they accept box sources at most a tenth as often. Sources with
stellar colors are accepted at 0.5--2.3\%.

\begin{figure}
    \centering
    \includegraphics[width=0.45\textwidth]{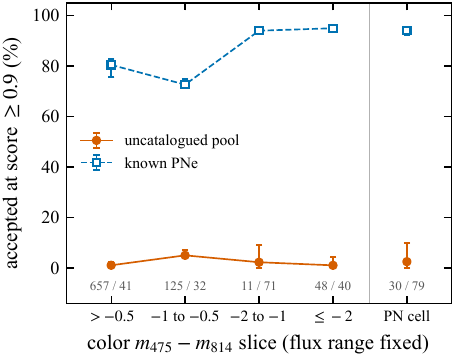}
    \caption{Acceptance of uncataloged blue sources. Acceptance rate at score $\geq 0.9$ in slices
    of color with the flux range fixed, for the pool built without the model and for known PNe in
    the same slice. Points are the means of the four training-only members, bars their range.
    Numbers below give the pool sources and known PNe in each slice. The PN cell is the flux range
    with color $\leq -1$ and F814W S/N $\geq 2$.}
    \label{fig:bluepool}
\end{figure}

The separation holds for the bluest pool sources as well (Figure~\ref{fig:bluepool}). In a PN cell
with the flux range fixed, color $\leq -1.0$ and S/N $\geq 2$ in F814W, three members accept none of
the 30 pool sources, and the member with 64 pixel crops accepts three (10\%, Wilson interval
4--26\%). The same members accept 92--95\% of the 79 known PNe in the cell. The pool, however, may
also contain uncataloged PNe. At the scale of the mosaics, the cell holds about 24\,000 sources, of
which at most about 2400 would be accepted, fewer than the roughly 2800 undiscovered PNe predicted by the PNLF
(Section~\ref{sec:eval}), so even these acceptances may be uncataloged PNe rather than errors.

\subsection{The detector uses its surroundings}
\label{sec:context}

Aperture photometry ignores almost all of a cutout, whereas a detector sees all of it, and context
can inform detection \citep{2003IJCV...53..169T,2020arXiv200609994X}, so we tested this with the
earlier model. Without the model, we described the light in rings 45--110 pixels
around each source in the PN box by its flux, color and smoothness. With spatially blocked
cross-validation, these features predict a score $\geq 0.9$ with an AUROC of 0.67--0.84,
comparable to the distance to the PN locus (0.65--0.75), and the two together give 0.75--0.90.
Accepted sources sit on brighter, redder and smoother light. S/N does not explain this, since
noise is higher there, and cataloged non-PNe from the same area have pool-like surroundings. The
surroundings, however, act as a prior, not as a gate. Known PNe are detected at 88--91\% in the
faintest quarter of surroundings and 97--99\% in the brightest, so the model has not simply learned
to look at the bulge \citep{2020arXiv200407780G}. For a borderline blue point, on the other hand,
the surroundings decide. Because surroundings and galactocentric radius cannot be separated here, we do
not claim that the model recognizes an old population.

\section{Discussion}
\label{sec:disc}

\subsection{Why the detector beats color selection}
\label{sec:why}

On the same windows, the detector reaches 4--6 times the purity of classical selection
(Section~\ref{sec:test}). Color is necessary for this but not sufficient
(Sections~\ref{sec:ablation} and \ref{sec:bluepool}), so the detector must use more. The extra
information may come from the pixel structure in two bands, from shape at the scale of the point
spread function, or from the surroundings (Section~\ref{sec:context}), and we could not separate these
sources in the final models. In the earlier model, excess light in F814W carried the separation,
since PNe are faint there while stars and \ion{H}{2} regions have continuum
(Section~\ref{sec:cosmic}), and the surroundings explained the rest. A role for the surroundings is
plausible, because PNe trace starlight to first order
\citep{1989ApJ...339...53C,2006MNRAS.369..120M,2019A&A...624A.132B}, although the number per unit
light changes between populations \citep{2006MNRAS.368..877B,2026MNRAS.545f2036E}. The model seems to
have learned this trend from pixels alone, so the result is a sanity check on its shortcut rather
than a discovery. Completeness therefore depends on the surroundings and should be weighted by
light, not by area \citep{2026MNRAS.548ag661Y}. This also fits the view that a PN is defined by
several criteria together \citep{2016JPhCS.728c2008P}. The zPN score of \citet{2014ApJ...792..121V},
for example, also combines color, brightness, sharpness and roundness, but only at known positions,
whereas the value of our method lies in a blind search. Several of our results agree with
independent work, such as class verdicts and spectra from the literature (Section~\ref{sec:untaught})
and the narrow-band detections of known objects (Section~\ref{sec:narrowband}).

\subsection{The value of the ensemble}
\label{sec:ensemble}

Ensembles rely on members that make different errors, which for networks trained from scratch comes
from different random starting points \citep{2016arXiv161201474L,2019arXiv191202757F}. Our members,
however, start from pretrained weights, so seeds share most of their unlabeled detections and
diversity has to come from the backbone and the input (Section~\ref{sec:backbones}). The gain of the ensemble lies not in P@200 but at the top of the list and in the contaminants
(Section~\ref{sec:enschoice}). Agreement rejects a contaminant only where the members see it
differently, as for the hot stars that only the UV members can recognize, but not for the reddish
stars that both see alike. Results from lens searches point both ways: in one study, a simple
average of different models did not beat the best one \citep{2025arXiv251023782G}, while in another
a mixed ensemble of different architectures and data did \citep{2026A&A...711A..28E}. Our ensemble,
with two architectures and two inputs, is closer to the second case.

\subsection{Recommendations for evaluating detectors in astronomy}
\label{sec:recommend}

Our evaluation was built against three pitfalls. First, a fixed evaluation set keeps the yardstick
from moving with the model. Second, scoring distinct sky objects keeps overlapping windows from
counting one object several times. Third, splitting by position, in balanced spatial groups, keeps
neighboring objects from leaking from training into the test. A gap between validation and realistic
tests has been reported before \citep{2025ApJS..278...16Z}. Its authors attribute it to class
imbalance, so the link to these pitfalls is our own interpretation.

\subsection{Limits set by labels and data}
\label{sec:labellimit}

Neither 47\% more data (Section~\ref{sec:retrain}) nor extra positives (Table~\ref{tab:axes})
brought a measurable gain, which raises the question of what limits the detector. In eight
ResNet-50 runs of an earlier test on the same validation windows, the unlabeled objects in the top
200 are stable sources, since 27\% of those in the top 200 of one run are in the top 200 of all
seven others. Some of them are real PNe without a label. Of these objects, 5.9\% are PNe that
\citet{2014ApJ...792..121V} confirmed on \textit{HST} but that SIMBAD does not label as PNe, 8.7\%
are EmO objects and 3\% are possible PNe. Most of the rest, however, are hot or reddish stars
(Section~\ref{sec:grid}), which every model has to place in its list once the findable PNe are used
up. This pattern is consistent with a limit set by the information in the images or by the labels.
It remains a hypothesis, though, since we did not measure learning curves.

A newer detector architecture does not lift this limit either. After the ensemble was fixed, we
trained three recent families of models with the same training crops, shifts, rotations and split
as our members: the real-time detection transformers RT-DETR-l and RT-DETR-x
\citep{2024cvpr.conf.1605Z}, the detection transformer RF-DETR \citep{2025arXiv251109554R}, and
U-Net segmentation networks \citep{2015LNCS.9351..234R} with three different encoders that predict
a map of PN centers. Each model was trained once, with a standard optimizer and learning rate and
without tuning. Since the transformers were trained on 128 pixel crops, we applied them to the
evaluation windows in 128 pixel tiles. On the test part, these models reach 42--64\% of the
ceiling, the best being RT-DETR-x, whereas our members reach 66--73\%. None of them is ahead of
the members, so the architecture does not appear to be what limits the detector. One early run also
showed why the training crops are shifted (Section~\ref{sec:training}). When RF-DETR was trained on
windows with the source always at the center, as it is in the evaluation windows, it learned to mark
the center: it found 97--99\% of the PNe there but only 21--33\% of the others, and so reached nearly
the ceiling.

\subsection{Comparison with the completeness result of Veyette et al.}
\label{sec:tension}

\citet{2014ApJ...792..121V} concluded that ground-based surveys are complete to F475W\,$\lesssim$\,24
in the PHAT area outside the bulge (Section~\ref{sec:intro}). Our discovery candidates are as bright
as the known PNe (median F475W 22.6 against 22.0), and for most of them a PN would be detected in
MegaCam. Yet, except for one, none shows a consistent line (Section~\ref{sec:narrowband}). Nor does
any show [\ion{O}{3}] or H$\alpha$ in the SITELLE cubes of the center, which lie outside the area
they studied. Our result therefore agrees with theirs and extends it beyond their bricks, to the
PHAST area and the center. The bright unlabeled sources that the model ranks high are mostly stars
with continuum.

The one discovery with consistent emission, however, lies in bricks they studied. Their PN list is
a match of the PN.S catalog of \citet{2006MNRAS.369..120M} to PHAT, so it inherits what PN.S missed.
If the stellar trails explain this source (Section~\ref{sec:narrowband}), the gap lies in slitless
spectroscopy near bright stars rather than in the depth of the ground-based surveys, and we estimate
that this costs of order 1\% of the PNe. If undiscovered PNe exist in the numbers that the PNLF
predicts (Section~\ref{sec:eval}), most of them must be fainter than the MegaCam limit
($m_{5007} \approx 24$--24.5), where the recall of the detector also falls (Section~\ref{sec:selfunc}).
In the center, the ground-based surveys are limited by poor seeing and high surface brightness
\citep{2006MNRAS.369..120M}, so faint PNe there are missing from both the catalog and the training
labels, and the detector cannot learn them either.

\subsection{Limitations}
\label{sec:limits}

The measured purity is a lower limit (Section~\ref{sec:eval}). Because the training PNe are
concentrated in the inner galaxy and at the bright end, completeness is lower in the outer disk and
at the faint end (Figure~\ref{fig:selection}). In the language of PU learning, the labels thus depend
on the properties of the objects, rather than being selected completely at random
\citep{2018arXiv181104820B,2025A&A...703A.240K}. Since the labels come from [\ion{O}{3}] surveys,
the detector learns the broad-band appearance of PNe that are bright in [\ion{O}{3}], and it is least
sensitive to the faint, red or weak-[\ion{O}{3}] PNe that these surveys miss. The cosmic-ray check needs at least three deep
exposures, so some candidates remain undecided. Our F814W aperture photometry is biased in the inner
disk, although this affects only the analyses. Both backbones sit in the same family of two-stage detectors.
Where UV data are missing (17\% of the area), the ensemble reduces to its ResNet-50 member. The
narrow-band check reaches $m_{5007} \approx 24$--24.5 and is unreliable near the edges of the SITELLE
cubes.
Finally, the advantages of the ensemble were measured on the validation part used to choose it.

\subsection{Future work}
\label{sec:future}

We are also studying M31 with other machine-learning methods, and those results will be presented
in a forthcoming paper. For the present catalog, the next step is spectroscopic follow-up. It should
start with the one discovery with emission lines, where line ratios can separate a PN from a compact
\ion{H}{2} region, and continue with the six candidates that could still be faint PNe and the eight
that MegaCam could not measure (Section~\ref{sec:narrowband}). Beyond these, the list can be
followed deeper, since the members agree down to rank $\sim$600. Completeness could also be
measured by injecting synthetic PNe into real \textit{HST} backgrounds
\citep{2020A&A...633A.148B}, following the light of the galaxy \citep{2026MNRAS.548ag661Y}.
Finally, we plan to apply similar techniques to other galaxies.


\section{Conclusions}
\label{sec:conc}

We searched the PHAT and PHAST mosaics of M31 for planetary nebulae with an ensemble of four Faster
R-CNN detectors, trained with partial supervision so that uncataloged PNe are not taught as
background. On held-out data, the ensemble places 73\% of the findable known PNe in its top 200,
4--6 times the purity of classical color--magnitude selection on the same windows. Color is
necessary for this result but not sufficient, since the detector accepts only a few percent of the
blue sources that a color cut would keep. In the survey of the full mosaics, a single member finds
about 70\% of the PNe it never saw. Its recall falls toward faint magnitudes and red colors, while it
depends only weakly on crowding and galactocentric radius.

The discovery list also orders classes that the detector was never taught. EmO objects that the
literature calls PNe enter it at 47\%, against 0.6\% for those called \ion{H}{2} regions, so we
present the EmO candidates as candidates for a new classification. The known emission-line objects
are concentrated at the top of the discovery list. They make up 78\% of the first 50 candidates and
9\% of ranks 401--1000. Those that the detector misses are mostly fainter in [\ion{O}{3}] or lie in
the crowded center, the same regime in which it misses training PNe. A detector trained on
broad-band images and an incomplete catalog therefore ranks first the PN-like objects that it can
recognize, which shows that the technique works within the range of its labels.
Of the 175 discovery candidates in the top 400, however, only one shows consistent [\ion{O}{3}] emission in narrow-band data. This
source passes the criteria on which the largest PN catalogs of M31 were built, and it was likely
missed behind a stellar trail in the slitless survey, although this remains a hypothesis. Most of
the other candidates are unlike the known PNe of M31 and are probably stars with continuum light,
while some may be PNe with weak [\ion{O}{3}] that only spectra can identify.

The detector learned the PNe that [\ion{O}{3}] surveys find, and in that range the catalogs are
already nearly complete, so few new [\ion{O}{3}]-bright PNe were expected. These results agree with
the completeness of ground-based surveys found by
\citet{2014ApJ...792..121V} and extend it to the PHAST area and the center of the galaxy. If
undiscovered PNe exist in the numbers that the PNLF predicts, most of them must therefore be fainter
than the current narrow-band limits. This work shows that deep learning can recover
extragalactic PN candidates from broad-band images and incomplete labels alone, and that its ranking
follows the classifications in catalogs built by human experts. Our experience also shows that the choice of
evaluation can change the conclusions drawn about detectors of rare objects. For this reason, we
recommend fixed evaluation windows, scoring of distinct sky objects and spatially blocked splits.
\begin{acknowledgments}
This work was funded by the Scientific and Technological Research Council of T\"urkiye
(T\"UB\.{I}TAK) under project 125F455. We used the large language models Gemini 2.5 Pro (Google) and
Claude Opus 4.8, Opus 5 and Opus 5.5 (Anthropic) to help prepare the analysis code and to write this
paper. The authors checked the code, the results and the text, and take full responsibility for them.

This research is based on observations made with the NASA/ESA \textit{Hubble Space Telescope}
obtained from the Space Telescope Science Institute, which is operated by AURA, Inc., under NASA
contract NAS5--26555 (programs 12055--12059, 12070--12076, 12105--12115, 16778 and 16796--16801). It
is also based on observations obtained with MegaPrime/MegaCam and SITELLE \citep{2012SPIE.8446E..0UG}
at the Canada-France-Hawai\textquoteleft i Telescope (CFHT), retrieved from the Canadian Astronomy
Data Centre, and on data from the Astro Data Archive at NSF NOIRLab (program 2000B-0010), LAMOST,
and ESA/Gaia/DPAC \citep{2016A&A...595A...1G,2023A&A...674A...1G}. We used the SIMBAD database and
the VizieR catalogue access tool (CDS, Strasbourg), the NASA/IPAC Extragalactic Database and the
Digitized Sky Surveys. CFHT is located on Maunakea, a mountain of considerable cultural, natural,
and ecological significance and a sacred site to Native Hawaiians.
\end{acknowledgments}

\begin{contribution}

All editors contribute equally to the operation of PASP.


\end{contribution}

%
\facilities{HST(ACS, WFC3), CFHT(MegaPrime, SITELLE), Mayall(Mosaic), LAMOST, Gaia, PO:1.2m, MAST, CDS,
    NED}

\software{PyTorch and torchvision \citep{2019arXiv191201703P},
          NumPy \citep{2020Natur.585..357H},
          SciPy \citep{2020NaMet..17..261V},
          Astropy \citep{2013A&A...558A..33A,2018AJ....156..123A,2022ApJ...935..167A},
          astroquery \citep{2019AJ....157...98G},
          pandas \citep{2010scpy.soft.....M},
          scikit-learn \citep{2011JMLR...12.2825P},
          Matplotlib \citep{2007CSE.....9...90H}
          }



\bibliography{m31}{}

@ARTICLE{2022PASP..134b2001K,
       author = {{Kwitter}, Karen B. and {Henry}, R.~B.~C.},
        title = "{Planetary Nebulae: Sources of Enlightenment}",
      journal = {\pasp},
         year = 2022,
        month = feb,
       volume = {134},
       number = {1032},
          eid = {022001},
        pages = {022001},
          doi = {10.1088/1538-3873/ac32b1},
archivePrefix = {arXiv},
       eprint = {2110.13993},
 primaryClass = {astro-ph.SR},
       adsurl = {https://ui.adsabs.harvard.edu/abs/2022PASP..134b2001K}
}

@ARTICLE{1989ApJ...339...39J,
       author = {{Jacoby}, George H.},
        title = "{Planetary Nebulae as Standard Candles. I. Evolutionary Models}",
      journal = {\apj},
         year = 1989,
        month = apr,
       volume = {339},
        pages = {39},
          doi = {10.1086/167274},
       adsurl = {https://ui.adsabs.harvard.edu/abs/1989ApJ...339...39J}
}

@ARTICLE{1989ApJ...339...53C,
       author = {{Ciardullo}, Robin and {Jacoby}, George H. and {Ford}, Holland C. and {Neill}, James D.},
        title = "{Planetary Nebulae as Standard Candles. II. The Calibration in M31 and Its Companions}",
      journal = {\apj},
         year = 1989,
        month = apr,
       volume = {339},
        pages = {53},
          doi = {10.1086/167275},
       adsurl = {https://ui.adsabs.harvard.edu/abs/1989ApJ...339...53C}
}

@ARTICLE{2012Ap&SS.341..151C,
       author = {{Ciardullo}, Robin},
        title = "{The Planetary Nebula Luminosity Function at the dawn of Gaia}",
      journal = {\apss},
         year = 2012,
        month = sep,
       volume = {341},
       number = {1},
        pages = {151-161},
          doi = {10.1007/s10509-012-1061-2},
archivePrefix = {arXiv},
       eprint = {1203.5551},
 primaryClass = {astro-ph.CO},
       adsurl = {https://ui.adsabs.harvard.edu/abs/2012Ap&SS.341..151C}
}

@ARTICLE{2006MNRAS.368..877B,
       author = {{Buzzoni}, Alberto and {Arnaboldi}, Magda and {Corradi}, Romano L.~M.},
        title = "{Planetary nebulae as tracers of galaxy stellar populations}",
      journal = {\mnras},
         year = 2006,
        month = may,
       volume = {368},
       number = {2},
        pages = {877-894},
          doi = {10.1111/j.1365-2966.2006.10163.x},
archivePrefix = {arXiv},
       eprint = {astro-ph/0602458},
 primaryClass = {astro-ph},
       adsurl = {https://ui.adsabs.harvard.edu/abs/2006MNRAS.368..877B}
}

@ARTICLE{1978ApJS...38..351F,
       author = {{Ford}, H.~C. and {Jacoby}, G.~H.},
        title = "{Planetary nebulae in Local-Group galaxies. VIII. A catalog of planetary nebulae in the Andromeda galaxy.}",
      journal = {\apjs},
         year = 1978,
        month = dec,
       volume = {38},
        pages = {351-356},
          doi = {10.1086/190560},
       adsurl = {https://ui.adsabs.harvard.edu/abs/1978ApJS...38..351F}
}

@ARTICLE{2006MNRAS.369..120M,
       author = {{Merrett}, H.~R. and {Merrifield}, M.~R. and {Douglas}, N.~G. and {Kuijken}, K. and {Romanowsky}, A.~J. and {Napolitano}, N.~R. and {Arnaboldi}, M. and {Capaccioli}, M. and {Freeman}, K.~C. and {Gerhard}, O. and {Coccato}, L. and {Carter}, D. and {Evans}, N.~W. and {Wilkinson}, M.~I. and {Halliday}, C. and {Bridges}, T.~J.},
        title = "{A deep kinematic survey of planetary nebulae in the Andromeda galaxy using the Planetary Nebula Spectrograph}",
      journal = {\mnras},
         year = 2006,
        month = may,
       volume = {369},
       number = {1},
        pages = {120-142},
          doi = {10.1111/j.1365-2966.2006.10268.x},
archivePrefix = {arXiv},
       eprint = {astro-ph/0603125},
 primaryClass = {astro-ph},
       adsurl = {https://ui.adsabs.harvard.edu/abs/2006MNRAS.369..120M}
}

@ARTICLE{2019A&A...624A.132B,
       author = {{Bhattacharya}, Souradeep and {Arnaboldi}, Magda and {Hartke}, Johanna and {Gerhard}, Ortwin and {Comte}, Valentin and {McConnachie}, Alan and {Caldwell}, Nelson},
        title = "{The survey of planetary nebulae in Andromeda (M 31). I. Imaging the disc and halo with MegaCam at the CFHT}",
      journal = {\aap},
         year = 2019,
        month = apr,
       volume = {624},
          eid = {A132},
        pages = {A132},
          doi = {10.1051/0004-6361/201834579},
archivePrefix = {arXiv},
       eprint = {1903.02597},
 primaryClass = {astro-ph.GA},
       adsurl = {https://ui.adsabs.harvard.edu/abs/2019A&A...624A.132B}
}

@ARTICLE{2010PASA...27..129F,
       author = {{Frew}, David J. and {Parker}, Quentin A.},
        title = "{Planetary Nebulae: Observational Properties, Mimics and Diagnostics}",
      journal = {\pasa},
         year = 2010,
        month = may,
       volume = {27},
       number = {2},
        pages = {129-148},
          doi = {10.1071/AS09040},
archivePrefix = {arXiv},
       eprint = {1002.1525},
 primaryClass = {astro-ph.SR},
       adsurl = {https://ui.adsabs.harvard.edu/abs/2010PASA...27..129F}
}

@INPROCEEDINGS{2016JPhCS.728c2008P,
       author = {{Parker}, Quentin A. and {Boji{\v{c}}i{\'c}}, Ivan S. and {Frew}, David J.},
        title = "{HASH: the Hong Kong/AAO/Strasbourg H{\ensuremath{\alpha}} planetary nebula database}",
    booktitle = {Journal of Physics Conference Series},
         year = 2016,
       series = {Journal of Physics Conference Series},
       volume = {728},
        month = jul,
    publisher = {IOP},
          eid = {032008},
        pages = {032008},
          doi = {10.1088/1742-6596/728/3/032008},
archivePrefix = {arXiv},
       eprint = {1603.07042},
 primaryClass = {astro-ph.SR},
       adsurl = {https://ui.adsabs.harvard.edu/abs/2016JPhCS.728c2008P}
}

@ARTICLE{2012ApJS..200...18D,
       author = {{Dalcanton}, Julianne J. and {Williams}, Benjamin F. and {Lang}, Dustin and {Lauer}, Tod R. and {Kalirai}, Jason S. and {Seth}, Anil C. and {Dolphin}, Andrew and {Rosenfield}, Philip and {Weisz}, Daniel R. and {Bell}, Eric F. and {Bianchi}, Luciana C. and {Boyer}, Martha L. and {Caldwell}, Nelson and {Dong}, Hui and {Dorman}, Claire E. and {Gilbert}, Karoline M. and {Girardi}, L{\'e}o and {Gogarten}, Stephanie M. and {Gordon}, Karl D. and {Guhathakurta}, Puragra and {Hodge}, Paul W. and {Holtzman}, Jon A. and {Johnson}, L. Clifton and {Larsen}, S{\o}ren S. and {Lewis}, Alexia and {Melbourne}, Jason L. and {Olsen}, Knut A.~G. and {Rix}, Hans-Walter and {Rosema}, Keith and {Saha}, Abhijit and {Sarajedini}, Ata and {Skillman}, Evan D. and {Stanek}, Krzysztof Z.},
        title = "{The Panchromatic Hubble Andromeda Treasury}",
      journal = {\apjs},
         year = 2012,
        month = jun,
       volume = {200},
       number = {2},
          eid = {18},
        pages = {18},
          doi = {10.1088/0067-0049/200/2/18},
archivePrefix = {arXiv},
       eprint = {1204.0010},
 primaryClass = {astro-ph.CO},
       adsurl = {https://ui.adsabs.harvard.edu/abs/2012ApJS..200...18D}
}

@ARTICLE{2025ApJ...979...35C,
       author = {{Chen}, Zhuo and {Williams}, Benjamin and {Lang}, Dustin and {Dolphin}, Andrew and {Durbin}, Meredith and {Dalcanton}, Julianne J. and {Smercina}, Adam and {Girardi}, L{\'e}o and {Murray}, Claire E. and {Bell}, Eric F. and {Boyer}, Martha L. and {D'Souza}, Richard and {Gilbert}, Karoline and {Gordon}, Karl and {Guhathakurta}, Puragra and {Hammer}, Francois and {Johnson}, L. Clifton and {Lauer}, Tod R. and {Lazzarini}, Margaret and {Murphy}, Jeremiah W. and {Patel}, Ekta and {Quirk}, Amanda and {D{\'\i}az Rodr{\'\i}guez}, Mariangelly and {Roman-Duval}, Julia Christine and {Sanderson}, Robyn E. and {Seth}, Anil and {Wainer}, Tobin M. and {Weisz}, Daniel R.},
        title = "{PHAST. The Panchromatic Hubble Andromeda Southern Treasury. I. Ultraviolet and Optical Photometry of over 90 Million Stars in M31}",
      journal = {\apj},
         year = 2025,
        month = jan,
       volume = {979},
       number = {1},
          eid = {35},
        pages = {35},
          doi = {10.3847/1538-4357/ad7e2b},
       adsurl = {https://ui.adsabs.harvard.edu/abs/2025ApJ...979...35C}
}

@ARTICLE{2014ApJ...792..121V,
       author = {{Veyette}, Mark J. and {Williams}, Benjamin F. and {Dalcanton}, Julianne J. and {Balick}, Bruce and {Caldwell}, Nelson and {Fouesneau}, Morgan and {Girardi}, L{\'e}o and {Gordon}, Karl D. and {Kalirai}, Jason and {Rosenfield}, Philip and {Seth}, Anil C.},
        title = "{Panchromatic Hubble Andromeda Treasury. IX. A Photometric Survey of Planetary Nebulae in M31}",
      journal = {\apj},
         year = 2014,
        month = sep,
       volume = {792},
       number = {2},
          eid = {121},
        pages = {121},
          doi = {10.1088/0004-637X/792/2/121},
archivePrefix = {arXiv},
       eprint = {1407.4391},
 primaryClass = {astro-ph.GA},
       adsurl = {https://ui.adsabs.harvard.edu/abs/2014ApJ...792..121V}
}

@ARTICLE{2014AJ....147...16K,
       author = {{Kniazev}, Alexei Y. and {Grebel}, Eva K. and {Zucker}, Daniel B. and {Rix}, Hans-Walter and {Mart{\'\i}nez-Delgado}, David and {Snedden}, Stephanie A.},
        title = "{A Search for Planetary Nebulae with the Sloan Digital Sky Survey: The Outer Regions of M31}",
      journal = {\aj},
         year = 2014,
        month = jan,
       volume = {147},
       number = {1},
          eid = {16},
        pages = {16},
          doi = {10.1088/0004-6256/147/1/16},
archivePrefix = {arXiv},
       eprint = {1311.1951},
 primaryClass = {astro-ph.GA},
       adsurl = {https://ui.adsabs.harvard.edu/abs/2014AJ....147...16K}
}

@ARTICLE{2026MNRAS.548ag661Y,
       author = {{Yang}, Nancy and {Hartke}, Johanna and {Bureau}, Martin and {Spiniello}, Chiara and {Guit{\'e}}, Louis-Simon and {Flint}, Guy and {Arnaboldi}, Magda and {Ennis}, Ana In{\'e}s and {Martin}, R. Pierre and {Martin}, Thomas and {Robert}, Carmelle and {Rousseau-Nepton}, Laurie and {Valenzuela}, Lucas M. and {Vicens-Mouret}, S{\'e}bastien},
        title = "{An automated method for planetary nebula detection with SIGNALS: first applications to NGC 4214 and NGC 4449}",
      journal = {\mnras},
         year = 2026,
        month = jun,
       volume = {548},
       number = {4},
          eid = {stag661},
        pages = {stag661},
          doi = {10.1093/mnras/stag661},
archivePrefix = {arXiv},
       eprint = {2604.09209},
 primaryClass = {astro-ph.GA},
       adsurl = {https://ui.adsabs.harvard.edu/abs/2026MNRAS.548ag661Y}
}

@ARTICLE{2026A&A...711A..28E,
       author = {{Euclid Collaboration} and {Lines}, N.~E.~P. and {Collett}, T.~E. and {Walmsley}, M. and {Rojas}, K. and {Li}, T. and {Leuzzi}, L. and {Manj{\'o}n-Garc{\'\i}a}, A. and {Vincken}, S.~H. and {Wilde}, J. and {Holloway}, P. and {Verma}, A. and {Metcalf}, R.~B. and {Andika}, I.~T. and {Melo}, A. and {Melchior}, M. and {Dom{\'\i}nguez S{\'a}nchez}, H. and {D{\'\i}az-S{\'a}nchez}, A. and {Acevedo Barroso}, J.~A. and {Cl{\'e}ment}, B. and {Krawczyk}, C. and {Pearce-Casey}, R. and {Serjeant}, S. and {Courbin}, F. and {Despali}, G. and {Gavazzi}, R. and {Schuldt}, S. and {Degaudenzi}, H. and {Ecker}, L.~R. and {Enzi}, W.~J.~R. and {Finner}, K. and {Galan}, A. and {Giocoli}, C. and {Hogg}, N.~B. and {Jahnke}, K. and {Kruk}, S. and {Mahler}, G. and {More}, A. and {Nagam}, B.~C. and {Pearson}, J. and {Sainz de Murieta}, A. and {Scarlata}, C. and {Sluse}, D. and {Sonnenfeld}, A. and {Spiniello}, C. and {Thai}, T.~T. and {Tortora}, C. and {Ulivi}, L. and {Weisenbach}, L. and {Zumalacarregui}, M. and {Aghanim}, N. and {Altieri}, B. and {Amara}, A. and {Andreon}, S. and {Auricchio}, N. and {Aussel}, H. and {Baccigalupi}, C. and {Baldi}, M. and {Balestra}, A. and {Bardelli}, S. and {Battaglia}, P. and {Bender}, R. and {Bernardeau}, F. and {Biviano}, A. and {Bonchi}, A. and {Bonino}, D. and {Branchini}, E. and {Brescia}, M. and {Brinchmann}, J. and {Camera}, S. and {Ca{\~n}as-Herrera}, G. and {Capobianco}, V. and {Carbone}, C. and {Cardone}, V.~F. and {Carretero}, J. and {Casas}, S. and {Castellano}, M. and {Castignani}, G. and {Cavuoti}, S. and {Chambers}, K.~C. and {Cimatti}, A. and {Colodro-Conde}, C. and {Congedo}, G. and {Conselice}, C.~J. and {Conversi}, L. and {Copin}, Y. and {Costille}, A. and {Courtois}, H.~M. and {Cropper}, M. and {Da Silva}, A. and {De Lucia}, G. and {Di Giorgio}, A.~M. and {Dolding}, C. and {Dole}, H. and {Dubath}, F. and {Duncan}, C.~A.~J. and {Dupac}, X. and {Escoffier}, S. and {Fabricius}, M. and {Farina}, M. and {Farinelli}, R. and {Faustini}, F. and {Ferriol}, S. and {Finelli}, F. and {Fotopoulou}, S. and {Frailis}, M. and {Franceschi}, E. and {Fumana}, M. and {Galeotta}, S. and {George}, K. and {Gillard}, W. and {Gillis}, B. and {G{\'o}mez-Alvarez}, P. and {Gracia-Carpio}, J. and {Granett}, B.~R. and {Grazian}, A. and {Grupp}, F. and {Guzzo}, L. and {Gwyn}, S. and {Haugan}, S.~V.~H. and {Holmes}, W. and {Hook}, I.~M. and {Hormuth}, F. and {Hornstrup}, A. and {Hudelot}, P. and {Jhabvala}, M. and {Keih{\"a}nen}, E. and {Kermiche}, S. and {Kiessling}, A. and {Kubik}, B. and {K{\"u}mmel}, M. and {Kunz}, M. and {Kurki-Suonio}, H. and {Le Boulc'h}, Q. and {Le Brun}, A.~M.~C. and {Le Mignant}, D. and {Ligori}, S. and {Lilje}, P.~B. and {Lindholm}, V. and {Lloro}, I. and {Mainetti}, G. and {Maino}, D. and {Maiorano}, E. and {Mansutti}, O. and {Marcin}, S. and {Marggraf}, O. and {Martinelli}, M. and {Martinet}, N. and {Marulli}, F. and {Massey}, R. and {Maurogordato}, S. and {Medinaceli}, E. and {Mei}, S. and {Mellier}, Y. and {Meneghetti}, M. and {Merlin}, E. and {Meylan}, G. and {Mora}, A. and {Moresco}, M. and {Moscardini}, L. and {Nakajima}, R. and {Neissner}, C. and {Nichol}, R.~C. and {Niemi}, S.-M. and {Nightingale}, J.~W. and {Padilla}, C. and {Paltani}, S. and {Pasian}, F. and {Pedersen}, K. and {Percival}, W.~J. and {Pettorino}, V. and {Pires}, S. and {Polenta}, G. and {Poncet}, M. and {Popa}, L.~A. and {Pozzetti}, L. and {Raison}, F. and {Rebolo}, R. and {Renzi}, A. and {Rhodes}, J. and {Riccio}, G. and {Romelli}, E. and {Roncarelli}, M. and {Saglia}, R. and {Sakr}, Z. and {S{\'a}nchez}, A.~G. and {Sapone}, D. and {Sartoris}, B. and {Schewtschenko}, J.~A. and {Schirmer}, M. and {Schneider}, P. and {Schrabback}, T. and {Secroun}, A. and {Seidel}, G. and {Seiffert}, M. and {Serrano}, S. and {Simon}, P. and {Sirignano}, C. and {Sirri}, G. and {Spurio Mancini}, A.},
        title = "{Euclid Quick Data Release (Q1): XXVIII. The Strong Lensing Discovery Engine C ─ Finding lenses with machine learning}",
      journal = {\aap},
         year = 2026,
        month = jun,
       volume = {711},
          eid = {A28},
        pages = {A28},
          doi = {10.1051/0004-6361/202554542},
archivePrefix = {arXiv},
       eprint = {2503.15326},
 primaryClass = {astro-ph.GA},
       adsurl = {https://ui.adsabs.harvard.edu/abs/2026A&A...711A..28E}
}

@ARTICLE{2026MNRAS.545f2036E,
       author = {{Ennis}, Ana In{\'e}s and {Hartke}, Johanna and {Bian}, Fuyan and {Pulsoni}, Claudia and {Spiniello}, Chiara and {Arnaboldi}, Magda and {de Propris}, Roberto},
        title = "{Planetary nebulae as tracers of stellar population properties: a pilot study with MUSE}",
      journal = {\mnras},
         year = 2026,
        month = jan,
       volume = {545},
       number = {2},
          eid = {staf2036},
        pages = {staf2036},
          doi = {10.1093/mnras/staf2036},
archivePrefix = {arXiv},
       eprint = {2511.11479},
 primaryClass = {astro-ph.GA},
       adsurl = {https://ui.adsabs.harvard.edu/abs/2026MNRAS.545f2036E}
}

@ARTICLE{2025A&A...703A.240K,
       author = {{Kliapets}, Mykyta and {Huijse}, Pablo and {Tkachenko}, Andrew and {Kemp}, Alex and {Fritzewski}, Dario J. and {Hey}, Daniel and {Aerts}, Conny},
        title = "{Automated all-sky detection of {\ensuremath{\gamma}} Doradus/{\ensuremath{\delta}} Scuti hybrids in TESS data from positive unlabelled (PU) learning}",
      journal = {\aap},
         year = 2025,
        month = nov,
       volume = {703},
          eid = {A240},
        pages = {A240},
          doi = {10.1051/0004-6361/202556079},
archivePrefix = {arXiv},
       eprint = {2511.20908},
 primaryClass = {astro-ph.SR},
       adsurl = {https://ui.adsabs.harvard.edu/abs/2025A&A...703A.240K}
}

@ARTICLE{2025arXiv251023782G,
       author = {{Gonzalez}, J. and {Collett}, T. and {Rojas}, K. and {Bechtol}, K. and {Acevedo Barroso}, J.~A. and {Melo}, A. and {More}, A. and {Sluse}, D. and {Tortora}, C. and {Holloway}, P. and {Lines}, N.~E.~P. and {Verma}, A.},
        title = "{Does Machine Learning Work? A Comparative Analysis of Strong Gravitational Lens Searches in the Dark Energy Survey}",
      journal = {arXiv e-prints},
         year = 2025,
        month = oct,
          eid = {arXiv:2510.23782},
        pages = {arXiv:2510.23782},
          doi = {10.48550/arXiv.2510.23782},
archivePrefix = {arXiv},
       eprint = {2510.23782},
 primaryClass = {astro-ph.GA},
       adsurl = {https://ui.adsabs.harvard.edu/abs/2025arXiv251023782G}
}

@ARTICLE{2025ApJS..278...16Z,
       author = {{Zhang}, Baisong and {Chen}, Bingqiu and {Yuan}, Haibo and {Chen}, Pinjian and {Wang}, Shoucheng and {Zhang}, Lunwei and {Ren}, Yi and {Guo}, Helong},
        title = "{Identification of Star Clusters in M31 from PAndAS Images Based on Deep Learning}",
      journal = {\apjs},
         year = 2025,
        month = may,
       volume = {278},
       number = {1},
          eid = {16},
        pages = {16},
          doi = {10.3847/1538-4365/adbf89},
archivePrefix = {arXiv},
       eprint = {2503.08130},
 primaryClass = {astro-ph.GA},
       adsurl = {https://ui.adsabs.harvard.edu/abs/2025ApJS..278...16Z}
}

@ARTICLE{2025A&A...697A.148B,
       author = {{Bracci}, Caterina and {Belfiore}, Francesco and {Ginolfi}, Michele and {Feltre}, Anna and {Mannucci}, Filippo and {Marconi}, Alessandro and {Cresci}, Giovanni and {Bertola}, Elena and {Bombini}, Alessandro and {Ceci}, Matteo and {Marconcini}, Cosimo and {Moreschini}, Bianca and {Scialpi}, Martina and {Tozzi}, Giulia and {Ulivi}, Lorenzo and {Venturi}, Giacomo},
        title = "{Classifying spectra of emission-line regions with neural networks: An application to integral field spectroscopic data of M33}",
      journal = {\aap},
         year = 2025,
        month = may,
       volume = {697},
          eid = {A148},
        pages = {A148},
          doi = {10.1051/0004-6361/202554196},
archivePrefix = {arXiv},
       eprint = {2502.20448},
 primaryClass = {astro-ph.GA},
       adsurl = {https://ui.adsabs.harvard.edu/abs/2025A&A...697A.148B}
}

@ARTICLE{2025AJ....169..174C,
       author = {{Chen}, Pinjian and {Chen}, Bingqiu and {Fang}, Xuan and {Yuan}, Haibo and {Zhang}, Baisong and {Zhang}, Xiangwei and {Sun}, Jiarui and {Liu}, Xiaowei},
        title = "{A Comprehensive Catalog of Emission-line Nebulae, Star Clusters, and Supergiants in M31 from the LAMOST Spectroscopic Survey}",
      journal = {\aj},
         year = 2025,
        month = mar,
       volume = {169},
       number = {3},
          eid = {174},
        pages = {174},
          doi = {10.3847/1538-3881/adb038},
archivePrefix = {arXiv},
       eprint = {2501.02206},
 primaryClass = {astro-ph.GA},
       adsurl = {https://ui.adsabs.harvard.edu/abs/2025AJ....169..174C}
}

@ARTICLE{2025A&A...694A.212B,
       author = {{Belfiore}, Francesco and {Ginolfi}, Michele and {Blanc}, Guillermo and {Boquien}, Mederic and {Chevance}, Melanie and {Congiu}, Enrico and {Glover}, Simon C.~O. and {Groves}, Brent and {Klessen}, Ralf S. and {Eduardo M{\'e}ndez-Delgado}, J. and {Williams}, Thomas G.},
        title = "{Machine learning the gap between real and simulated nebulae: A domain-adaptation approach to classify ionised nebulae in nearby galaxies}",
      journal = {\aap},
         year = 2025,
        month = feb,
       volume = {694},
          eid = {A212},
        pages = {A212},
          doi = {10.1051/0004-6361/202451934},
archivePrefix = {arXiv},
       eprint = {2410.16370},
 primaryClass = {astro-ph.GA},
       adsurl = {https://ui.adsabs.harvard.edu/abs/2025A&A...694A.212B}
}

@ARTICLE{2024A&A...692A.103L,
       author = {{Li}, Yushan and {Parker}, Quentin and {Jia}, Peng},
        title = "{First discovery and confirmation of planetary nebula candidates from AI and deep learning techniques applied to VPHAS+ survey data}",
      journal = {\aap},
         year = 2024,
        month = dec,
       volume = {692},
          eid = {A103},
        pages = {A103},
          doi = {10.1051/0004-6361/202450998},
archivePrefix = {arXiv},
       eprint = {2408.09482},
 primaryClass = {astro-ph.GA},
       adsurl = {https://ui.adsabs.harvard.edu/abs/2024A&A...692A.103L}
}

@ARTICLE{2024MNRAS.528.4733S,
       author = {{Sun}, Ruiqi and {Li}, Yushan and {Parker}, Quentin and {Li}, Jiaxin and {Li}, Xu and {Cao}, Liang and {Jia}, Peng},
        title = "{Artificial Intelligence Techniques for Uncovering Resolved Planetary Nebula Candidates from Wide-field VPHAS+ Survey Data}",
      journal = {\mnras},
         year = 2024,
        month = mar,
       volume = {528},
       number = {3},
        pages = {4733-4745},
          doi = {10.1093/mnras/stad3954},
archivePrefix = {arXiv},
       eprint = {2311.02607},
 primaryClass = {astro-ph.IM},
       adsurl = {https://ui.adsabs.harvard.edu/abs/2024MNRAS.528.4733S}
}

@ARTICLE{2023MNRAS.526.1122M,
       author = {{Merz}, Grant and {Liu}, Yichen and {Burke}, Colin J. and {Aleo}, Patrick D. and {Liu}, Xin and {Carrasco Kind}, Matias and {Kindratenko}, Volodymyr and {Liu}, Yufeng},
        title = "{Detection, instance segmentation, and classification for astronomical surveys with deep learning (DEEPDISC): DETECTRON2 implementation and demonstration with Hyper Suprime-Cam data}",
      journal = {\mnras},
         year = 2023,
        month = nov,
       volume = {526},
       number = {1},
        pages = {1122-1137},
          doi = {10.1093/mnras/stad2785},
archivePrefix = {arXiv},
       eprint = {2307.05826},
 primaryClass = {astro-ph.IM},
       adsurl = {https://ui.adsabs.harvard.edu/abs/2023MNRAS.526.1122M}
}

@ARTICLE{2023PASA...40....1H,
       author = {{Huertas-Company}, M. and {Lanusse}, F.},
        title = "{The Dawes Review 10: The impact of deep learning for the analysis of galaxy surveys}",
      journal = {\pasa},
         year = 2023,
        month = jan,
       volume = {40},
          eid = {e001},
        pages = {e001},
          doi = {10.1017/pasa.2022.55},
archivePrefix = {arXiv},
       eprint = {2210.01813},
 primaryClass = {astro-ph.IM},
       adsurl = {https://ui.adsabs.harvard.edu/abs/2023PASA...40....1H}
}

@ARTICLE{2022ApJ...935..167A,
       author = {{Astropy Collaboration} and {Price-Whelan}, Adrian M. and {Lim}, Pey Lian and {Earl}, Nicholas and {Starkman}, Nathaniel and {Bradley}, Larry and {Shupe}, David L. and {Patil}, Aarya A. and {Corrales}, Lia and {Brasseur}, C.~E. and {N{\"o}the}, Maximilian and {Donath}, Axel and {Tollerud}, Erik and {Morris}, Brett M. and {Ginsburg}, Adam and {Vaher}, Eero and {Weaver}, Benjamin A. and {Tocknell}, James and {Jamieson}, William and {van Kerkwijk}, Marten H. and {Robitaille}, Thomas P. and {Merry}, Bruce and {Bachetti}, Matteo and {G{\"u}nther}, H. Moritz and {Aldcroft}, Thomas L. and {Alvarado-Montes}, Jaime A. and {Archibald}, Anne M. and {B{\'o}di}, Attila and {Bapat}, Shreyas and {Barentsen}, Geert and {Baz{\'a}n}, Juanjo and {Biswas}, Manish and {Boquien}, M{\'e}d{\'e}ric and {Burke}, D.~J. and {Cara}, Daria and {Cara}, Mihai and {Conroy}, Kyle E. and {Conseil}, Simon and {Craig}, Matthew W. and {Cross}, Robert M. and {Cruz}, Kelle L. and {D'Eugenio}, Francesco and {Dencheva}, Nadia and {Devillepoix}, Hadrien A.~R. and {Dietrich}, J{\"o}rg P. and {Eigenbrot}, Arthur Davis and {Erben}, Thomas and {Ferreira}, Leonardo and {Foreman-Mackey}, Daniel and {Fox}, Ryan and {Freij}, Nabil and {Garg}, Suyog and {Geda}, Robel and {Glattly}, Lauren and {Gondhalekar}, Yash and {Gordon}, Karl D. and {Grant}, David and {Greenfield}, Perry and {Groener}, Austen M. and {Guest}, Steve and {Gurovich}, Sebastian and {Handberg}, Rasmus and {Hart}, Akeem and {Hatfield-Dodds}, Zac and {Homeier}, Derek and {Hosseinzadeh}, Griffin and {Jenness}, Tim and {Jones}, Craig K. and {Joseph}, Prajwel and {Kalmbach}, J. Bryce and {Karamehmetoglu}, Emir and {Ka{\l}uszy{\'n}ski}, Miko{\l}aj and {Kelley}, Michael S.~P. and {Kern}, Nicholas and {Kerzendorf}, Wolfgang E. and {Koch}, Eric W. and {Kulumani}, Shankar and {Lee}, Antony and {Ly}, Chun and {Ma}, Zhiyuan and {MacBride}, Conor and {Maljaars}, Jakob M. and {Muna}, Demitri and {Murphy}, N.~A. and {Norman}, Henrik and {O'Steen}, Richard and {Oman}, Kyle A. and {Pacifici}, Camilla and {Pascual}, Sergio and {Pascual-Granado}, J. and {Patil}, Rohit R. and {Perren}, Gabriel I. and {Pickering}, Timothy E. and {Rastogi}, Tanuj and {Roulston}, Benjamin R. and {Ryan}, Daniel F. and {Rykoff}, Eli S. and {Sabater}, Jose and {Sakurikar}, Parikshit and {Salgado}, Jes{\'u}s and {Sanghi}, Aniket and {Saunders}, Nicholas and {Savchenko}, Volodymyr and {Schwardt}, Ludwig and {Seifert-Eckert}, Michael and {Shih}, Albert Y. and {Jain}, Anany Shrey and {Shukla}, Gyanendra and {Sick}, Jonathan and {Simpson}, Chris and {Singanamalla}, Sudheesh and {Singer}, Leo P. and {Singhal}, Jaladh and {Sinha}, Manodeep and {Sip{\H{o}}cz}, Brigitta M. and {Spitler}, Lee R. and {Stansby}, David and {Streicher}, Ole and {{\v{S}}umak}, Jani and {Swinbank}, John D. and {Taranu}, Dan S. and {Tewary}, Nikita and {Tremblay}, Grant R. and {de Val-Borro}, Miguel and {Van Kooten}, Samuel J. and {Vasovi{\'c}}, Zlatan and {Verma}, Shresth and {de Miranda Cardoso}, Jos{\'e} Vin{\'\i}cius and {Williams}, Peter K.~G. and {Wilson}, Tom J. and {Winkel}, Benjamin and {Wood-Vasey}, W.~M. and {Xue}, Rui and {Yoachim}, Peter and {Zhang}, Chen and {Zonca}, Andrea and {Astropy Project Contributors}},
        title = "{The Astropy Project: Sustaining and Growing a Community-oriented Open-source Project and the Latest Major Release (v5.0) of the Core Package}",
      journal = {\apj},
         year = 2022,
        month = aug,
       volume = {935},
       number = {2},
          eid = {167},
        pages = {167},
          doi = {10.3847/1538-4357/ac7c74},
archivePrefix = {arXiv},
       eprint = {2206.14220},
 primaryClass = {astro-ph.IM},
       adsurl = {https://ui.adsabs.harvard.edu/abs/2022ApJ...935..167A}
}

@ARTICLE{2022arXiv220707048K,
       author = {{Kapoor}, Sayash and {Narayanan}, Arvind},
        title = "{Leakage and the Reproducibility Crisis in ML-based Science}",
      journal = {arXiv e-prints},
         year = 2022,
        month = jul,
          eid = {arXiv:2207.07048},
        pages = {arXiv:2207.07048},
          doi = {10.48550/arXiv.2207.07048},
archivePrefix = {arXiv},
       eprint = {2207.07048},
 primaryClass = {cs.LG},
       adsurl = {https://ui.adsabs.harvard.edu/abs/2022arXiv220707048K}
}

@ARTICLE{2022MNRAS.511.5032B,
       author = {{Bhambra}, Prabh and {Joachimi}, Benjamin and {Lahav}, Ofer},
        title = "{Explaining deep learning of galaxy morphology with saliency mapping}",
      journal = {\mnras},
         year = 2022,
        month = apr,
       volume = {511},
       number = {4},
        pages = {5032-5041},
          doi = {10.1093/mnras/stac368},
archivePrefix = {arXiv},
       eprint = {2110.08288},
 primaryClass = {astro-ph.IM},
       adsurl = {https://ui.adsabs.harvard.edu/abs/2022MNRAS.511.5032B}
}

@ARTICLE{2022A&A...658A..51W,
       author = {{Wang}, Shoucheng and {Chen}, Bingqiu and {Ma}, Jun and {Long}, Qian and {Yuan}, Haibo and {Liu}, Dezi and {Zhou}, Zhimin and {Liu}, Wei and {Chen}, Jiamin and {He}, Zizhao},
        title = "{Identification of new M 31 star cluster candidates from PAndAS images using convolutional neural networks}",
      journal = {\aap},
         year = 2022,
        month = feb,
       volume = {658},
          eid = {A51},
        pages = {A51},
          doi = {10.1051/0004-6361/202142169},
archivePrefix = {arXiv},
       eprint = {2111.07798},
 primaryClass = {astro-ph.GA},
       adsurl = {https://ui.adsabs.harvard.edu/abs/2022A&A...658A..51W}
}

@ARTICLE{2021arXiv210908203P,
       author = {{Picard}, David},
        title = "{Torch.manual\_seed(3407) is all you need: On the influence of random seeds in deep learning architectures for computer vision}",
      journal = {arXiv e-prints},
         year = 2021,
        month = sep,
          eid = {arXiv:2109.08203},
        pages = {arXiv:2109.08203},
          doi = {10.48550/arXiv.2109.08203},
archivePrefix = {arXiv},
       eprint = {2109.08203},
 primaryClass = {cs.CV},
       adsurl = {https://ui.adsabs.harvard.edu/abs/2021arXiv210908203P}
}

@ARTICLE{2021arXiv210314030L,
       author = {{Liu}, Ze and {Lin}, Yutong and {Cao}, Yue and {Hu}, Han and {Wei}, Yixuan and {Zhang}, Zheng and {Lin}, Stephen and {Guo}, Baining},
        title = "{Swin Transformer: Hierarchical Vision Transformer using Shifted Windows}",
      journal = {arXiv e-prints},
         year = 2021,
        month = mar,
          eid = {arXiv:2103.14030},
        pages = {arXiv:2103.14030},
          doi = {10.48550/arXiv.2103.14030},
archivePrefix = {arXiv},
       eprint = {2103.14030},
 primaryClass = {cs.CV},
       adsurl = {https://ui.adsabs.harvard.edu/abs/2021arXiv210314030L}
}

@ARTICLE{2021arXiv210303098B,
       author = {{Bouthillier}, Xavier and {Delaunay}, Pierre and {Bronzi}, Mirko and {Trofimov}, Assya and {Nichyporuk}, Brennan and {Szeto}, Justin and {Sepah}, Naz and {Raff}, Edward and {Madan}, Kanika and {Voleti}, Vikram and {Ebrahimi Kahou}, Samira and {Michalski}, Vincent and {Serdyuk}, Dmitriy and {Arbel}, Tal and {Pal}, Chris and {Varoquaux}, Ga{\"e}l and {Vincent}, Pascal},
        title = "{Accounting for Variance in Machine Learning Benchmarks}",
      journal = {arXiv e-prints},
         year = 2021,
        month = mar,
          eid = {arXiv:2103.03098},
        pages = {arXiv:2103.03098},
          doi = {10.48550/arXiv.2103.03098},
archivePrefix = {arXiv},
       eprint = {2103.03098},
 primaryClass = {cs.LG},
       adsurl = {https://ui.adsabs.harvard.edu/abs/2021arXiv210303098B}
}

@ARTICLE{2021MNRAS.500.3821B,
       author = {{Bonaldi}, A. and {An}, T. and {Br{\"u}ggen}, M. and {Burkutean}, S. and {Coelho}, B. and {Goodarzi}, H. and {Hartley}, P. and {Sandhu}, P.~K. and {Wu}, C. and {Yu}, L. and {Zhoolideh Haghighi}, M.~H. and {Ant{\'o}n}, S. and {Bagheri}, Z. and {Barbosa}, D. and {Barraca}, J.~P. and {Bartashevich}, D. and {Bergano}, M. and {Bonato}, M. and {Brand}, J. and {de Gasperin}, F. and {Giannetti}, A. and {Dodson}, R. and {Jain}, P. and {Jaiswal}, S. and {Lao}, B. and {Liu}, B. and {Liuzzo}, E. and {Lu}, Y. and {Lukic}, V. and {Maia}, D. and {Marchili}, N. and {Massardi}, M. and {Mohan}, P. and {Morgado}, J.~B. and {Panwar}, M. and {Prabhakar}, P. and {Ribeiro}, V.~A.~R.~M. and {Rygl}, K.~L.~J. and {Sabz Ali}, V. and {Saremi}, E. and {Schisano}, E. and {Sheikhnezami}, S. and {Vafaei Sadr}, A. and {Wong}, A. and {Wong}, O.~I.},
        title = "{Square Kilometre Array Science Data Challenge 1: analysis and results}",
      journal = {\mnras},
         year = 2021,
        month = jan,
       volume = {500},
       number = {3},
        pages = {3821-3837},
          doi = {10.1093/mnras/staa3023},
archivePrefix = {arXiv},
       eprint = {2009.13346},
 primaryClass = {astro-ph.IM},
       adsurl = {https://ui.adsabs.harvard.edu/abs/2021MNRAS.500.3821B}
}

@ARTICLE{2020Galax...8...88A,
       author = {{Awang Iskandar}, Dayang N.~F. and {Zijlstra}, Albert A. and {McDonald}, Iain and {Abdullah}, Rosni and {Fuller}, Gary A. and {Fauzi}, Ahmad H. and {Abdullah}, Johari},
        title = "{Classification of Planetary Nebulae through Deep Transfer Learning}",
      journal = {Galaxies},
         year = 2020,
        month = dec,
       volume = {8},
       number = {4},
          eid = {88},
        pages = {88},
          doi = {10.3390/galaxies8040088},
archivePrefix = {arXiv},
       eprint = {2101.12628},
 primaryClass = {astro-ph.IM},
       adsurl = {https://ui.adsabs.harvard.edu/abs/2020Galax...8...88A}
}

@ARTICLE{2020arXiv200609994X,
       author = {{Xiao}, Kai and {Engstrom}, Logan and {Ilyas}, Andrew and {Madry}, Aleksander},
        title = "{Noise or Signal: The Role of Image Backgrounds in Object Recognition}",
      journal = {arXiv e-prints},
         year = 2020,
        month = jun,
          eid = {arXiv:2006.09994},
        pages = {arXiv:2006.09994},
          doi = {10.48550/arXiv.2006.09994},
archivePrefix = {arXiv},
       eprint = {2006.09994},
 primaryClass = {cs.CV},
       adsurl = {https://ui.adsabs.harvard.edu/abs/2020arXiv200609994X}
}

@ARTICLE{2020arXiv200407780G,
       author = {{Geirhos}, Robert and {Jacobsen}, J{\"o}rn-Henrik and {Michaelis}, Claudio and {Zemel}, Richard and {Brendel}, Wieland and {Bethge}, Matthias and {Wichmann}, Felix A.},
        title = "{Shortcut Learning in Deep Neural Networks}",
      journal = {arXiv e-prints},
         year = 2020,
        month = apr,
          eid = {arXiv:2004.07780},
        pages = {arXiv:2004.07780},
          doi = {10.48550/arXiv.2004.07780},
archivePrefix = {arXiv},
       eprint = {2004.07780},
 primaryClass = {cs.CV},
       adsurl = {https://ui.adsabs.harvard.edu/abs/2020arXiv200407780G}
}

@ARTICLE{2020ApJ...889...24Z,
       author = {{Zhang}, Keming and {Bloom}, Joshua S.},
        title = "{deepCR: Cosmic Ray Rejection with Deep Learning}",
      journal = {\apj},
         year = 2020,
        month = jan,
       volume = {889},
       number = {1},
          eid = {24},
        pages = {24},
          doi = {10.3847/1538-4357/ab3fa6},
archivePrefix = {arXiv},
       eprint = {1907.09500},
 primaryClass = {astro-ph.IM},
       adsurl = {https://ui.adsabs.harvard.edu/abs/2020ApJ...889...24Z}
}

@ARTICLE{2020A&A...633A.148B,
       author = {{Bialopetravi{\v{c}}ius}, J. and {Narbutis}, D.},
        title = "{Deriving star cluster parameters with convolutional neural networks. II. Extinction and cluster-background classification}",
      journal = {\aap},
         year = 2020,
        month = jan,
       volume = {633},
          eid = {A148},
        pages = {A148},
          doi = {10.1051/0004-6361/201936185},
archivePrefix = {arXiv},
       eprint = {1911.10059},
 primaryClass = {astro-ph.GA},
       adsurl = {https://ui.adsabs.harvard.edu/abs/2020A&A...633A.148B}
}

@ARTICLE{2019arXiv191202757F,
       author = {{Fort}, Stanislav and {Hu}, Huiyi and {Lakshminarayanan}, Balaji},
        title = "{Deep Ensembles: A Loss Landscape Perspective}",
      journal = {arXiv e-prints},
         year = 2019,
        month = dec,
          eid = {arXiv:1912.02757},
        pages = {arXiv:1912.02757},
          doi = {10.48550/arXiv.1912.02757},
archivePrefix = {arXiv},
       eprint = {1912.02757},
 primaryClass = {stat.ML},
       adsurl = {https://ui.adsabs.harvard.edu/abs/2019arXiv191202757F}
}

@ARTICLE{2019arXiv191201703P,
       author = {{Paszke}, Adam and {Gross}, Sam and {Massa}, Francisco and {Lerer}, Adam and {Bradbury}, James and {Chanan}, Gregory and {Killeen}, Trevor and {Lin}, Zeming and {Gimelshein}, Natalia and {Antiga}, Luca and {Desmaison}, Alban and {K{\"o}pf}, Andreas and {Yang}, Edward and {DeVito}, Zach and {Raison}, Martin and {Tejani}, Alykhan and {Chilamkurthy}, Sasank and {Steiner}, Benoit and {Fang}, Lu and {Bai}, Junjie and {Chintala}, Soumith},
        title = "{PyTorch: An Imperative Style, High-Performance Deep Learning Library}",
      journal = {arXiv e-prints},
         year = 2019,
        month = dec,
          eid = {arXiv:1912.01703},
        pages = {arXiv:1912.01703},
          doi = {10.48550/arXiv.1912.01703},
archivePrefix = {arXiv},
       eprint = {1912.01703},
 primaryClass = {cs.LG},
       adsurl = {https://ui.adsabs.harvard.edu/abs/2019arXiv191201703P}
}

@ARTICLE{2019MNRAS.490.3952B,
       author = {{Burke}, Colin J. and {Aleo}, Patrick D. and {Chen}, Yu-Ching and {Liu}, Xin and {Peterson}, John R. and {Sembroski}, Glenn H. and {Lin}, Joshua Yao-Yu},
        title = "{Deblending and classifying astronomical sources with Mask R-CNN deep learning}",
      journal = {\mnras},
         year = 2019,
        month = dec,
       volume = {490},
       number = {3},
        pages = {3952-3965},
          doi = {10.1093/mnras/stz2845},
archivePrefix = {arXiv},
       eprint = {1908.02748},
 primaryClass = {astro-ph.IM},
       adsurl = {https://ui.adsabs.harvard.edu/abs/2019MNRAS.490.3952B}
}

@ARTICLE{2019MNRAS.488.3238A,
       author = {{Akras}, Stavros and {Guzman-Ramirez}, Lizette and {Gon{\c{c}}alves}, Denise R.},
        title = "{Compact planetary nebulae: improved IR diagnostic criteria based on classification tree modelling}",
      journal = {\mnras},
         year = 2019,
        month = sep,
       volume = {488},
       number = {3},
        pages = {3238-3250},
          doi = {10.1093/mnras/stz1911},
archivePrefix = {arXiv},
       eprint = {1907.10026},
 primaryClass = {astro-ph.SR},
       adsurl = {https://ui.adsabs.harvard.edu/abs/2019MNRAS.488.3238A}
}

@ARTICLE{2019arXiv190502244H,
       author = {{Howard}, Andrew and {Sandler}, Mark and {Chu}, Grace and {Chen}, Liang-Chieh and {Chen}, Bo and {Tan}, Mingxing and {Wang}, Weijun and {Zhu}, Yukun and {Pang}, Ruoming and {Vasudevan}, Vijay and {Le}, Quoc V. and {Adam}, Hartwig},
        title = "{Searching for MobileNetV3}",
      journal = {arXiv e-prints},
         year = 2019,
        month = may,
          eid = {arXiv:1905.02244},
        pages = {arXiv:1905.02244},
          doi = {10.48550/arXiv.1905.02244},
archivePrefix = {arXiv},
       eprint = {1905.02244},
 primaryClass = {cs.CV},
       adsurl = {https://ui.adsabs.harvard.edu/abs/2019arXiv190502244H}
}

@ARTICLE{2019A&A...625A.119M,
       author = {{Metcalf}, R.~B. and {Meneghetti}, M. and {Avestruz}, C. and {Bellagamba}, F. and {Bom}, C.~R. and {Bertin}, E. and {Cabanac}, R. and {Courbin}, F. and {Davies}, A. and {Decenci{\`e}re}, E. and {Flamary}, R. and {Gavazzi}, R. and {Geiger}, M. and {Hartley}, P. and {Huertas-Company}, M. and {Jackson}, N. and {Jacobs}, C. and {Jullo}, E. and {Kneib}, J.-P. and {Koopmans}, L.~V.~E. and {Lanusse}, F. and {Li}, C.-L. and {Ma}, Q. and {Makler}, M. and {Li}, N. and {Lightman}, M. and {Petrillo}, C.~E. and {Serjeant}, S. and {Sch{\"a}fer}, C. and {Sonnenfeld}, A. and {Tagore}, A. and {Tortora}, C. and {Tuccillo}, D. and {Valent{\'\i}n}, M.~B. and {Velasco-Forero}, S. and {Verdoes Kleijn}, G.~A. and {Vernardos}, G.},
        title = "{The strong gravitational lens finding challenge}",
      journal = {\aap},
         year = 2019,
        month = may,
       volume = {625},
          eid = {A119},
        pages = {A119},
          doi = {10.1051/0004-6361/201832797},
archivePrefix = {arXiv},
       eprint = {1802.03609},
 primaryClass = {astro-ph.GA},
       adsurl = {https://ui.adsabs.harvard.edu/abs/2019A&A...625A.119M}
}

@ARTICLE{2019MNRAS.483.5077A,
       author = {{Akras}, Stavros and {Leal-Ferreira}, Marcelo L. and {Guzman-Ramirez}, Lizette and {Ramos-Larios}, Gerardo},
        title = "{A machine learning approach for identification and classification of symbiotic stars using 2MASS and WISE}",
      journal = {\mnras},
         year = 2019,
        month = mar,
       volume = {483},
       number = {4},
        pages = {5077-5104},
          doi = {10.1093/mnras/sty3359},
archivePrefix = {arXiv},
       eprint = {1901.03016},
 primaryClass = {astro-ph.SR},
       adsurl = {https://ui.adsabs.harvard.edu/abs/2019MNRAS.483.5077A}
}

@ARTICLE{2019MNRAS.482.1211W,
       author = {{Wu}, Chen and {Wong}, Oiwei Ivy and {Rudnick}, Lawrence and {Shabala}, Stanislav S. and {Alger}, Matthew J. and {Banfield}, Julie K. and {Ong}, Cheng Soon and {White}, Sarah V. and {Garon}, Avery F. and {Norris}, Ray P. and {Andernach}, Heinz and {Tate}, Jean and {Lukic}, Vesna and {Tang}, Hongming and {Schawinski}, Kevin and {Diakogiannis}, Foivos I.},
        title = "{Radio Galaxy Zoo: CLARAN - a deep learning classifier for radio morphologies}",
      journal = {\mnras},
         year = 2019,
        month = jan,
       volume = {482},
       number = {1},
        pages = {1211-1230},
          doi = {10.1093/mnras/sty2646},
archivePrefix = {arXiv},
       eprint = {1805.12008},
 primaryClass = {astro-ph.IM},
       adsurl = {https://ui.adsabs.harvard.edu/abs/2019MNRAS.482.1211W}
}

@ARTICLE{2018arXiv181104820B,
       author = {{Bekker}, Jessa and {Davis}, Jesse},
        title = "{Learning from positive and unlabeled data: a survey}",
      journal = {arXiv e-prints},
         year = 2018,
        month = nov,
          eid = {arXiv:1811.04820},
        pages = {arXiv:1811.04820},
          doi = {10.48550/arXiv.1811.04820},
archivePrefix = {arXiv},
       eprint = {1811.04820},
 primaryClass = {cs.LG},
       adsurl = {https://ui.adsabs.harvard.edu/abs/2018arXiv181104820B}
}

@ARTICLE{2018AJ....156..123A,
       author = {{Astropy Collaboration} and {Price-Whelan}, A.~M. and {Sip{\H{o}}cz}, B.~M. and {G{\"u}nther}, H.~M. and {Lim}, P.~L. and {Crawford}, S.~M. and {Conseil}, S. and {Shupe}, D.~L. and {Craig}, M.~W. and {Dencheva}, N. and {Ginsburg}, A. and {VanderPlas}, J.~T. and {Bradley}, L.~D. and {P{\'e}rez-Su{\'a}rez}, D. and {de Val-Borro}, M. and {Aldcroft}, T.~L. and {Cruz}, K.~L. and {Robitaille}, T.~P. and {Tollerud}, E.~J. and {Ardelean}, C. and {Babej}, T. and {Bach}, Y.~P. and {Bachetti}, M. and {Bakanov}, A.~V. and {Bamford}, S.~P. and {Barentsen}, G. and {Barmby}, P. and {Baumbach}, A. and {Berry}, K.~L. and {Biscani}, F. and {Boquien}, M. and {Bostroem}, K.~A. and {Bouma}, L.~G. and {Brammer}, G.~B. and {Bray}, E.~M. and {Breytenbach}, H. and {Buddelmeijer}, H. and {Burke}, D.~J. and {Calderone}, G. and {Cano Rodr{\'\i}guez}, J.~L. and {Cara}, M. and {Cardoso}, J.~V.~M. and {Cheedella}, S. and {Copin}, Y. and {Corrales}, L. and {Crichton}, D. and {D'Avella}, D. and {Deil}, C. and {Depagne}, {\'E}. and {Dietrich}, J.~P. and {Donath}, A. and {Droettboom}, M. and {Earl}, N. and {Erben}, T. and {Fabbro}, S. and {Ferreira}, L.~A. and {Finethy}, T. and {Fox}, R.~T. and {Garrison}, L.~H. and {Gibbons}, S.~L.~J. and {Goldstein}, D.~A. and {Gommers}, R. and {Greco}, J.~P. and {Greenfield}, P. and {Groener}, A.~M. and {Grollier}, F. and {Hagen}, A. and {Hirst}, P. and {Homeier}, D. and {Horton}, A.~J. and {Hosseinzadeh}, G. and {Hu}, L. and {Hunkeler}, J.~S. and {Ivezi{\'c}}, {\v{Z}}. and {Jain}, A. and {Jenness}, T. and {Kanarek}, G. and {Kendrew}, S. and {Kern}, N.~S. and {Kerzendorf}, W.~E. and {Khvalko}, A. and {King}, J. and {Kirkby}, D. and {Kulkarni}, A.~M. and {Kumar}, A. and {Lee}, A. and {Lenz}, D. and {Littlefair}, S.~P. and {Ma}, Z. and {Macleod}, D.~M. and {Mastropietro}, M. and {McCully}, C. and {Montagnac}, S. and {Morris}, B.~M. and {Mueller}, M. and {Mumford}, S.~J. and {Muna}, D. and {Murphy}, N.~A. and {Nelson}, S. and {Nguyen}, G.~H. and {Ninan}, J.~P. and {N{\"o}the}, M. and {Ogaz}, S. and {Oh}, S. and {Parejko}, J.~K. and {Parley}, N. and {Pascual}, S. and {Patil}, R. and {Patil}, A.~A. and {Plunkett}, A.~L. and {Prochaska}, J.~X. and {Rastogi}, T. and {Reddy Janga}, V. and {Sabater}, J. and {Sakurikar}, P. and {Seifert}, M. and {Sherbert}, L.~E. and {Sherwood-Taylor}, H. and {Shih}, A.~Y. and {Sick}, J. and {Silbiger}, M.~T. and {Singanamalla}, S. and {Singer}, L.~P. and {Sladen}, P.~H. and {Sooley}, K.~A. and {Sornarajah}, S. and {Streicher}, O. and {Teuben}, P. and {Thomas}, S.~W. and {Tremblay}, G.~R. and {Turner}, J.~E.~H. and {Terr{\'o}n}, V. and {van Kerkwijk}, M.~H. and {de la Vega}, A. and {Watkins}, L.~L. and {Weaver}, B.~A. and {Whitmore}, J.~B. and {Woillez}, J. and {Zabalza}, V. and {Astropy Contributors}},
        title = "{The Astropy Project: Building an Open-science Project and Status of the v2.0 Core Package}",
      journal = {\aj},
         year = 2018,
        month = sep,
       volume = {156},
       number = {3},
          eid = {123},
        pages = {123},
          doi = {10.3847/1538-3881/aabc4f},
archivePrefix = {arXiv},
       eprint = {1801.02634},
 primaryClass = {astro-ph.IM},
       adsurl = {https://ui.adsabs.harvard.edu/abs/2018AJ....156..123A}
}

@ARTICLE{2018arXiv180610758H,
       author = {{Hooker}, Sara and {Erhan}, Dumitru and {Kindermans}, Pieter-Jan and {Kim}, Been},
        title = "{A Benchmark for Interpretability Methods in Deep Neural Networks}",
      journal = {arXiv e-prints},
         year = 2018,
        month = jun,
          eid = {arXiv:1806.10758},
        pages = {arXiv:1806.10758},
          doi = {10.48550/arXiv.1806.10758},
archivePrefix = {arXiv},
       eprint = {1806.10758},
 primaryClass = {cs.LG},
       adsurl = {https://ui.adsabs.harvard.edu/abs/2018arXiv180610758H}
}

@ARTICLE{2018arXiv180606986W,
       author = {{Wu}, Zhe and {Bodla}, Navaneeth and {Singh}, Bharat and {Najibi}, Mahyar and {Chellappa}, Rama and {Davis}, Larry S.},
        title = "{Soft Sampling for Robust Object Detection}",
      journal = {arXiv e-prints},
         year = 2018,
        month = jun,
          eid = {arXiv:1806.06986},
        pages = {arXiv:1806.06986},
          doi = {10.48550/arXiv.1806.06986},
archivePrefix = {arXiv},
       eprint = {1806.06986},
 primaryClass = {cs.CV},
       adsurl = {https://ui.adsabs.harvard.edu/abs/2018arXiv180606986W}
}

@ARTICLE{2017Ecogr..40..913R,
       author = {{Roberts}, David R. and {Bahn}, Volker and {Ciuti}, Simone and {Boyce}, Mark S. and {Elith}, Jane and {Guillera-Arroita}, Gurutzeta and {Hauenstein}, Severin and {Lahoz-Monfort}, Jos{\'e} J. and {Schr{\"o}der}, Boris and {Thuiller}, Wilfried and {Warton}, David I. and {Wintle}, Brendan A. and {Hartig}, Florian and {Dormann}, Carsten F.},
        title = "{Cross-validation strategies for data with temporal, spatial, hierarchical, or phylogenetic structure}",
      journal = {Ecography},
         year = 2017,
        month = aug,
       volume = {40},
       number = {8},
        pages = {913-929},
          doi = {10.1111/ecog.02881},
       adsurl = {https://ui.adsabs.harvard.edu/abs/2017Ecogr..40..913R}
}

@ARTICLE{2017arXiv170602677G,
       author = {{Goyal}, Priya and {Doll{\'a}r}, Piotr and {Girshick}, Ross and {Noordhuis}, Pieter and {Wesolowski}, Lukasz and {Kyrola}, Aapo and {Tulloch}, Andrew and {Jia}, Yangqing and {He}, Kaiming},
        title = "{Accurate, Large Minibatch SGD: Training ImageNet in 1 Hour}",
      journal = {arXiv e-prints},
         year = 2017,
        month = jun,
          eid = {arXiv:1706.02677},
        pages = {arXiv:1706.02677},
          doi = {10.48550/arXiv.1706.02677},
archivePrefix = {arXiv},
       eprint = {1706.02677},
 primaryClass = {cs.CV},
       adsurl = {https://ui.adsabs.harvard.edu/abs/2017arXiv170602677G}
}

@ARTICLE{2017arXiv170306870H,
       author = {{He}, Kaiming and {Gkioxari}, Georgia and {Doll{\'a}r}, Piotr and {Girshick}, Ross},
        title = "{Mask R-CNN}",
      journal = {arXiv e-prints},
         year = 2017,
        month = mar,
          eid = {arXiv:1703.06870},
        pages = {arXiv:1703.06870},
          doi = {10.48550/arXiv.1703.06870},
archivePrefix = {arXiv},
       eprint = {1703.06870},
 primaryClass = {cs.CV},
       adsurl = {https://ui.adsabs.harvard.edu/abs/2017arXiv170306870H}
}

@ARTICLE{2016arXiv161203144L,
       author = {{Lin}, Tsung-Yi and {Doll{\'a}r}, Piotr and {Girshick}, Ross and {He}, Kaiming and {Hariharan}, Bharath and {Belongie}, Serge},
        title = "{Feature Pyramid Networks for Object Detection}",
      journal = {arXiv e-prints},
         year = 2016,
        month = dec,
          eid = {arXiv:1612.03144},
        pages = {arXiv:1612.03144},
          doi = {10.48550/arXiv.1612.03144},
archivePrefix = {arXiv},
       eprint = {1612.03144},
 primaryClass = {cs.CV},
       adsurl = {https://ui.adsabs.harvard.edu/abs/2016arXiv161203144L}
}

@ARTICLE{2016arXiv161201474L,
       author = {{Lakshminarayanan}, Balaji and {Pritzel}, Alexander and {Blundell}, Charles},
        title = "{Simple and Scalable Predictive Uncertainty Estimation using Deep Ensembles}",
      journal = {arXiv e-prints},
         year = 2016,
        month = dec,
          eid = {arXiv:1612.01474},
        pages = {arXiv:1612.01474},
          doi = {10.48550/arXiv.1612.01474},
archivePrefix = {arXiv},
       eprint = {1612.01474},
 primaryClass = {stat.ML},
       adsurl = {https://ui.adsabs.harvard.edu/abs/2016arXiv161201474L}
}

@software{2016ascl.soft08013D,
       author = {{Dolphin}, Andrew},
        title = "{DOLPHOT: Stellar photometry}",
 howpublished = {Astrophysics Source Code Library, record ascl:1608.013},
         year = 2016,
        month = aug,
          eid = {ascl:1608.013},
archivePrefix = {ascl},
       eprint = {1608.013},
       adsurl = {https://ui.adsabs.harvard.edu/abs/2016ascl.soft08013D}
}

@INPROCEEDINGS{2016cvpr.confE...1H,
       author = {{He}, Kaiming and {Zhang}, Xiangyu and {Ren}, Shaoqing and {Sun}, Jian},
        title = "{Deep Residual Learning for Image Recognition}",
    booktitle = {2016 IEEE Conference on Computer Vision and Pattern Recognition Workshops (CVPRW)},
         year = 2016,
        month = jun,
    publisher = {IEEE},
          eid = {1},
        pages = {1},
          doi = {10.1109/CVPR.2016.90},
archivePrefix = {arXiv},
       eprint = {1512.03385},
 primaryClass = {cs.CV},
       adsurl = {https://ui.adsabs.harvard.edu/abs/2016cvpr.confE...1H}
}

@ARTICLE{2015ApJ...815...69F,
       author = {{Fang}, Xuan and {Garc{\'\i}a-Benito}, Rub{\'e}n and {Guerrero}, Mart{\'\i}n A. and {Liu}, Xiaowei and {Yuan}, Haibo and {Zhang}, Yong and {Zhang}, Bing},
        title = "{Chemical Abundances of Planetary Nebulae in the Substructures of M31}",
      journal = {\apj},
         year = 2015,
        month = dec,
       volume = {815},
       number = {1},
          eid = {69},
        pages = {69},
          doi = {10.1088/0004-637X/815/1/69},
archivePrefix = {arXiv},
       eprint = {1511.02788},
 primaryClass = {astro-ph.GA},
       adsurl = {https://ui.adsabs.harvard.edu/abs/2015ApJ...815...69F}
}

@ARTICLE{2015arXiv150601497R,
       author = {{Ren}, Shaoqing and {He}, Kaiming and {Girshick}, Ross and {Sun}, Jian},
        title = "{Faster R-CNN: Towards Real-Time Object Detection with Region Proposal Networks}",
      journal = {arXiv e-prints},
         year = 2015,
        month = jun,
          eid = {arXiv:1506.01497},
        pages = {arXiv:1506.01497},
          doi = {10.48550/arXiv.1506.01497},
archivePrefix = {arXiv},
       eprint = {1506.01497},
 primaryClass = {cs.CV},
       adsurl = {https://ui.adsabs.harvard.edu/abs/2015arXiv150601497R}
}

@ARTICLE{2014ApJS..215....9W,
       author = {{Williams}, Benjamin F. and {Lang}, Dustin and {Dalcanton}, Julianne J. and {Dolphin}, Andrew E. and {Weisz}, Daniel R. and {Bell}, Eric F. and {Bianchi}, Luciana and {Byler}, Nell and {Gilbert}, Karoline M. and {Girardi}, L{\'e}o and {Gordon}, Karl and {Gregersen}, Dylan and {Johnson}, L.~C. and {Kalirai}, Jason and {Lauer}, Tod R. and {Monachesi}, Antonela and {Rosenfield}, Philip and {Seth}, Anil and {Skillman}, Eva},
        title = "{The Panchromatic Hubble Andromeda Treasury. X. Ultraviolet to Infrared Photometry of 117 Million Equidistant Stars}",
      journal = {\apjs},
         year = 2014,
        month = nov,
       volume = {215},
       number = {1},
          eid = {9},
        pages = {9},
          doi = {10.1088/0067-0049/215/1/9},
archivePrefix = {arXiv},
       eprint = {1409.0899},
 primaryClass = {astro-ph.GA},
       adsurl = {https://ui.adsabs.harvard.edu/abs/2014ApJS..215....9W}
}

@ARTICLE{2014arXiv1405.0312L,
       author = {{Lin}, Tsung-Yi and {Maire}, Michael and {Belongie}, Serge and {Bourdev}, Lubomir and {Girshick}, Ross and {Hays}, James and {Perona}, Pietro and {Ramanan}, Deva and {Zitnick}, C. Lawrence and {Doll{\'a}r}, Piotr},
        title = "{Microsoft COCO: Common Objects in Context}",
      journal = {arXiv e-prints},
         year = 2014,
        month = may,
          eid = {arXiv:1405.0312},
        pages = {arXiv:1405.0312},
          doi = {10.48550/arXiv.1405.0312},
archivePrefix = {arXiv},
       eprint = {1405.0312},
 primaryClass = {cs.CV},
       adsurl = {https://ui.adsabs.harvard.edu/abs/2014arXiv1405.0312L}
}

@ARTICLE{2013arXiv1311.2901Z,
       author = {{Zeiler}, Matthew D and {Fergus}, Rob},
        title = "{Visualizing and Understanding Convolutional Networks}",
      journal = {arXiv e-prints},
         year = 2013,
        month = nov,
          eid = {arXiv:1311.2901},
        pages = {arXiv:1311.2901},
          doi = {10.48550/arXiv.1311.2901},
archivePrefix = {arXiv},
       eprint = {1311.2901},
 primaryClass = {cs.CV},
       adsurl = {https://ui.adsabs.harvard.edu/abs/2013arXiv1311.2901Z}
}

@ARTICLE{2013A&A...558A..33A,
       author = {{Astropy Collaboration} and {Robitaille}, Thomas P. and {Tollerud}, Erik J. and {Greenfield}, Perry and {Droettboom}, Michael and {Bray}, Erik and {Aldcroft}, Tom and {Davis}, Matt and {Ginsburg}, Adam and {Price-Whelan}, Adrian M. and {Kerzendorf}, Wolfgang E. and {Conley}, Alexander and {Crighton}, Neil and {Barbary}, Kyle and {Muna}, Demitri and {Ferguson}, Henry and {Grollier}, Fr{\'e}d{\'e}ric and {Parikh}, Madhura M. and {Nair}, Prasanth H. and {Unther}, Hans M. and {Deil}, Christoph and {Woillez}, Julien and {Conseil}, Simon and {Kramer}, Roban and {Turner}, James E.~H. and {Singer}, Leo and {Fox}, Ryan and {Weaver}, Benjamin A. and {Zabalza}, Victor and {Edwards}, Zachary I. and {Azalee Bostroem}, K. and {Burke}, D.~J. and {Casey}, Andrew R. and {Crawford}, Steven M. and {Dencheva}, Nadia and {Ely}, Justin and {Jenness}, Tim and {Labrie}, Kathleen and {Lim}, Pey Lian and {Pierfederici}, Francesco and {Pontzen}, Andrew and {Ptak}, Andy and {Refsdal}, Brian and {Servillat}, Mathieu and {Streicher}, Ole},
        title = "{Astropy: A community Python package for astronomy}",
      journal = {\aap},
         year = 2013,
        month = oct,
       volume = {558},
          eid = {A33},
        pages = {A33},
          doi = {10.1051/0004-6361/201322068},
archivePrefix = {arXiv},
       eprint = {1307.6212},
 primaryClass = {astro-ph.IM},
       adsurl = {https://ui.adsabs.harvard.edu/abs/2013A&A...558A..33A}
}

@ARTICLE{2012ApJ...758..133S,
       author = {{Sanders}, Nathan E. and {Caldwell}, Nelson and {McDowell}, Jonathan and {Harding}, Paul},
        title = "{The Metallicity Profile of M31 from Spectroscopy of Hundreds of H II Regions and PNe}",
      journal = {\apj},
         year = 2012,
        month = oct,
       volume = {758},
       number = {2},
          eid = {133},
        pages = {133},
          doi = {10.1088/0004-637X/758/2/133},
archivePrefix = {arXiv},
       eprint = {1209.2251},
 primaryClass = {astro-ph.CO},
       adsurl = {https://ui.adsabs.harvard.edu/abs/2012ApJ...758..133S}
}

@ARTICLE{2011AJ....142..139A,
       author = {{Azimlu}, M. and {Marciniak}, R. and {Barmby}, P.},
        title = "{A New Catalog of H II Regions in M31}",
      journal = {\aj},
         year = 2011,
        month = oct,
       volume = {142},
       number = {4},
          eid = {139},
        pages = {139},
          doi = {10.1088/0004-6256/142/4/139},
archivePrefix = {arXiv},
       eprint = {1108.4044},
 primaryClass = {astro-ph.CO},
       adsurl = {https://ui.adsabs.harvard.edu/abs/2011AJ....142..139A}
}

@INPROCEEDINGS{2009cvpr.conf...27D,
       author = {{Deng}, Jia and {Dong}, Wei and {Socher}, Richard and {Li}, Li-Jia and {Kai Li} and {Li Fei-Fei}},
        title = "{ImageNet: A large-scale hierarchical image database}",
    booktitle = {2009 IEEE Conference on Computer Vision and Pattern Recognition},
         year = 2009,
        month = jun,
    publisher = {IEEE},
          eid = {27},
        pages = {27},
          doi = {10.1109/CVPR.2009.5206848},
       adsurl = {https://ui.adsabs.harvard.edu/abs/2009cvpr.conf...27D}
}

@ARTICLE{2006MNRAS.369...97H,
       author = {{Halliday}, C. and {Carter}, D. and {Bridges}, T.~J. and {Jackson}, Z.~C. and {Wilkinson}, M.~I. and {Quinn}, D.~P. and {Evans}, N.~W. and {Douglas}, N.~G. and {Merrett}, H.~R. and {Merrifield}, M.~R. and {Romanowsky}, A.~J. and {Kuijken}, K. and {Irwin}, M.~J.},
        title = "{Planetary nebula velocities in the disc and bulge of M31}",
      journal = {\mnras},
         year = 2006,
        month = jun,
       volume = {369},
       number = {1},
        pages = {97-119},
          doi = {10.1111/j.1365-2966.2006.10364.x},
archivePrefix = {arXiv},
       eprint = {astro-ph/0603706},
 primaryClass = {astro-ph},
       adsurl = {https://ui.adsabs.harvard.edu/abs/2006MNRAS.369...97H}
}

@ARTICLE{2003IJCV...53..169T,
       author = {{Torralba}, Antonio},
        title = "{Contextual Priming for Object Detection}",
      journal = {International Journal of Computer Vision},
         year = 2003,
        month = jul,
       volume = {53},
       number = {2},
        pages = {169-191},
          doi = {10.1023/A:1023052124951},
       adsurl = {https://ui.adsabs.harvard.edu/abs/2003IJCV...53..169T}
}

@ARTICLE{2000PASP..112.1383D,
       author = {{Dolphin}, Andrew E.},
        title = "{WFPC2 Stellar Photometry with HSTPHOT}",
      journal = {\pasp},
         year = 2000,
        month = oct,
       volume = {112},
       number = {776},
        pages = {1383-1396},
          doi = {10.1086/316630},
archivePrefix = {arXiv},
       eprint = {astro-ph/0006217},
 primaryClass = {astro-ph},
       adsurl = {https://ui.adsabs.harvard.edu/abs/2000PASP..112.1383D}
}

@ARTICLE{2000A&AS..143...23O,
       author = {{Ochsenbein}, F. and {Bauer}, P. and {Marcout}, J.},
        title = "{The VizieR database of astronomical catalogues}",
      journal = {\aaps},
         year = 2000,
        month = apr,
       volume = {143},
        pages = {23-32},
          doi = {10.1051/aas:2000169},
archivePrefix = {arXiv},
       eprint = {astro-ph/0002122},
 primaryClass = {astro-ph},
       adsurl = {https://ui.adsabs.harvard.edu/abs/2000A&AS..143...23O}
}

@ARTICLE{2000A&AS..143....9W,
       author = {{Wenger}, M. and {Ochsenbein}, F. and {Egret}, D. and {Dubois}, P. and {Bonnarel}, F. and {Borde}, S. and {Genova}, F. and {Jasniewicz}, G. and {Lalo{\"e}}, S. and {Lesteven}, S. and {Monier}, R.},
        title = "{The SIMBAD astronomical database. The CDS reference database for astronomical objects}",
      journal = {\aaps},
         year = 2000,
        month = apr,
       volume = {143},
        pages = {9-22},
          doi = {10.1051/aas:2000332},
archivePrefix = {arXiv},
       eprint = {astro-ph/0002110},
 primaryClass = {astro-ph},
       adsurl = {https://ui.adsabs.harvard.edu/abs/2000A&AS..143....9W}
}

@ARTICLE{1993A&AS..102..251M,
       author = {{Meyssonnier}, N. and {Lequeux}, J. and {Azzopardi}, M.},
        title = "{An objective-prism survey of emission-line objects in M 31.}",
      journal = {\aaps},
         year = 1993,
        month = dec,
       volume = {102},
        pages = {251-268},
       adsurl = {https://ui.adsabs.harvard.edu/abs/1993A&AS..102..251M}
}

@ARTICLE{1992A&AS...92..625W,
       author = {{Walterbos}, R.~A.~M. and {Braun}, R.},
        title = "{The interstellar medium of M 31: III. Narrow-band imagery in H-alpha and S II .}",
      journal = {\aaps},
         year = 1992,
        month = feb,
       volume = {92},
        pages = {625-682},
       adsurl = {https://ui.adsabs.harvard.edu/abs/1992A&AS...92..625W}
}

@ARTICLE{1978A&AS...31..439P,
       author = {{Pellet}, A. and {Astier}, N. and {Viale}, A. and {Courtes}, G. and {Maucherat}, A. and {Monnet}, G. and {Simien}, F.},
        title = "{A survey of H II regions in M31.}",
      journal = {\aaps},
         year = 1978,
        month = mar,
       volume = {31},
        pages = {439-461},
       adsurl = {https://ui.adsabs.harvard.edu/abs/1978A&AS...31..439P}
}

@ARTICLE{1964ApJ...139.1027B,
       author = {{Baade}, Walter and {Arp}, Halton},
        title = "{Positions of Emission Nebulae in M31.}",
      journal = {\apj},
         year = 1964,
        month = may,
       volume = {139},
        pages = {1027},
          doi = {10.1086/147843},
       adsurl = {https://ui.adsabs.harvard.edu/abs/1964ApJ...139.1027B}
}

@ARTICLE{1927JASA...22..209W,
       author = {{Wilson}, Edwin B.},
        title = "{Probable Inference, the Law of Succession, and Statistical Inference}",
      journal = {Journal of the American Statistical Association},
         year = 1927,
        month = jun,
       volume = {22},
       number = {158},
        pages = {209-212},
          doi = {10.1080/01621459.1927.10502953},
       adsurl = {https://ui.adsabs.harvard.edu/abs/1927JASA...22..209W}
}

@article{Bergstra2012,
  author  = {James Bergstra and Yoshua Bengio},
  title   = {Random Search for Hyper-Parameter Optimization},
  journal = {Journal of Machine Learning Research},
  year    = {2012},
  volume  = {13},
  number  = {10},
  pages   = {281--305},
  url     = {http://jmlr.org/papers/v13/bergstra12a.html}
}

@article{Efron1979,
  title     = {Bootstrap Methods: Another Look at the Jackknife},
  volume    = {7},
  ISSN      = {0090-5364},
  url       = {http://dx.doi.org/10.1214/aos/1176344552},
  DOI       = {10.1214/aos/1176344552},
  number    = {1},
  journal   = {The Annals of Statistics},
  publisher = {Institute of Mathematical Statistics},
  author    = {Efron, B.},
  year      = {1979},
  month     = jan,
  pages     = {1--26}
}

@article{Mei2022,
  title     = {RadImageNet: An Open Radiologic Deep Learning Research Dataset for Effective Transfer Learning},
  volume    = {4},
  ISSN      = {2638-6100},
  url       = {http://dx.doi.org/10.1148/ryai.210315},
  DOI       = {10.1148/ryai.210315},
  number    = {5},
  journal   = {Radiology: Artificial Intelligence},
  publisher = {Radiological Society of North America (RSNA)},
  author    = {Mei, Xueyan and Liu, Zelong and Robson, Philip M. and Marinelli, Brett and Huang, Mingqian and Doshi, Amish and Jacobi, Adam and Cao, Chendi and Link, Katherine E. and Yang, Thomas and Wang, Ying and Greenspan, Hayit and Deyer, Timothy and Fayad, Zahi A. and Yang, Yang},
  year      = {2022},
  month     = sep,
  pages     = {e210315}
}

@book{Westfall1993,
  author    = {Westfall, Peter H. and Young, S. Stanley},
  title     = {Resampling-Based Multiple Testing: Examples and Methods for {p}-Value Adjustment},
  publisher = {Wiley},
  address   = {New York},
  series    = {Wiley Series in Probability and Mathematical Statistics},
  year      = {1993},
  isbn      = {978-0-471-55761-6}
}

@INPROCEEDINGS{2003SPIE.4841...72B,
       author = {{Boulade}, Olivier and {Charlot}, Xavier and {Abbon}, P. and {Aune}, Stephan and {Borgeaud}, Pierre and {Carton}, Pierre-Henri and {Carty}, M. and {Da Costa}, J. and {Deschamps}, H. and {Desforge}, D. and {Eppell{\'e}}, Dominique and {Gallais}, Pascal and {Gosset}, L. and {Granelli}, Remy and {Gros}, Michel and {de Kat}, Jean and {Loiseau}, Denis and {Ritou}, J.-. and {Rouss{\'e}}, Jean Y. and {Starzynski}, Pierre and {Vignal}, Nicolas and {Vigroux}, Laurent G.},
        title = "{MegaCam: the new Canada-France-Hawaii Telescope wide-field imaging camera}",
    booktitle = {Instrument Design and Performance for Optical/Infrared Ground-based Telescopes},
         year = 2003,
       editor = {{Iye}, Masanori and {Moorwood}, Alan F.~M.},
       series = {Society of Photo-Optical Instrumentation Engineers (SPIE) Conference Series},
       volume = {4841},
        month = mar,
    publisher = {SPIE},
        pages = {72-81},
          doi = {10.1117/12.459890},
       adsurl = {https://ui.adsabs.harvard.edu/abs/2003SPIE.4841...72B}
}

@ARTICLE{2004PASP..116..449M,
       author = {{Magnier}, E.~A. and {Cuillandre}, J.-C.},
        title = "{The Elixir System: Data Characterization and Calibration at the Canada-France-Hawaii Telescope}",
      journal = {\pasp},
         year = 2004,
        month = may,
       volume = {116},
       number = {819},
        pages = {449-464},
          doi = {10.1086/420756},
       adsurl = {https://ui.adsabs.harvard.edu/abs/2004PASP..116..449M}
}

@ARTICLE{2019MNRAS.485.3930D,
       author = {{Drissen}, Laurent and {Martin}, Thomas and {Rousseau-Nepton}, Laurie and {Robert}, Carmelle and {Martin}, R. Pierre and {Baril}, Marc and {Prunet}, Simon and {Joncas}, Gilles and {Thibault}, Simon and {Brousseau}, Denis and {Mandar}, Julie and {Grandmont}, Fr{\'e}d{\'e}ric and {Yee}, Howard and {Simard}, Luc},
        title = "{SITELLE: an Imaging Fourier Transform Spectrometer for the Canada-France-Hawaii Telescope}",
      journal = {\mnras},
         year = 2019,
        month = may,
       volume = {485},
       number = {3},
        pages = {3930-3946},
          doi = {10.1093/mnras/stz627},
archivePrefix = {arXiv},
       eprint = {1811.06644},
 primaryClass = {astro-ph.IM},
       adsurl = {https://ui.adsabs.harvard.edu/abs/2019MNRAS.485.3930D}
}

@INPROCEEDINGS{2015ASPC..495..327M,
       author = {{Martin}, T. and {Drissen}, L. and {Joncas}, G.},
        title = "{ORBS, ORCS, OACS, a Software Suite for Data Reduction and Analysis of the Hyperspectral Imagers SITELLE and SpIOMM}",
    booktitle = {Astronomical Data Analysis Software and Systems XXIV (ADASS XXIV)},
         year = 2015,
       editor = {{Taylor}, A.~R. and {Rosolowsky}, E.},
       series = {Astronomical Society of the Pacific Conference Series},
       volume = {495},
        month = sep,
        pages = {327},
       adsurl = {https://ui.adsabs.harvard.edu/abs/2015ASPC..495..327M}
}

@ARTICLE{2018MNRAS.473.4130M,
       author = {{Martin}, Thomas B. and {Drissen}, Laurent and {Melchior}, Anne-Laure},
        title = "{A SITELLE view of M31's central region - I. Calibrations and radial velocity catalogue of nearly 800 emission-line point-like sources}",
      journal = {\mnras},
         year = 2018,
        month = jan,
       volume = {473},
       number = {3},
        pages = {4130-4149},
          doi = {10.1093/mnras/stx2513},
archivePrefix = {arXiv},
       eprint = {1707.01366},
 primaryClass = {astro-ph.GA},
       adsurl = {https://ui.adsabs.harvard.edu/abs/2018MNRAS.473.4130M}
}

@ARTICLE{2006AJ....131.2478M,
       author = {{Massey}, Philip and {Olsen}, K.~A.~G. and {Hodge}, Paul W. and {Strong}, Shay B. and {Jacoby}, George H. and {Schlingman}, Wayne and {Smith}, R.~C.},
        title = "{A Survey of Local Group Galaxies Currently Forming Stars. I. UBVRI Photometry of Stars in M31 and M33}",
      journal = {\aj},
         year = 2006,
        month = may,
       volume = {131},
       number = {5},
        pages = {2478-2496},
          doi = {10.1086/503256},
archivePrefix = {arXiv},
       eprint = {astro-ph/0602128},
 primaryClass = {astro-ph},
       adsurl = {https://ui.adsabs.harvard.edu/abs/2006AJ....131.2478M}
}

@INPROCEEDINGS{2017iccv.conf..324L,
       author = {{Lin}, Tsung-Yi and {Goyal}, Priya and {Girshick}, Ross and {He}, Kaiming and {Dollar}, Piotr},
        title = "{Focal Loss for Dense Object Detection}",
    booktitle = {2017 IEEE International Conference on Computer Vision (ICCV)},
         year = 2017,
        month = oct,
    publisher = {IEEE},
          eid = {324},
        pages = {324},
          doi = {10.1109/ICCV.2017.324},
       adsurl = {https://ui.adsabs.harvard.edu/abs/2017iccv.conf..324L}
}

@INPROCEEDINGS{2022cvpr.conf.1167L,
       author = {{Liu}, Zhuang and {Mao}, Hanzi and {Wu}, Chao-Yuan and {Feichtenhofer}, Christoph and {Darrell}, Trevor and {Xie}, Saining},
        title = "{A ConvNet for the 2020s}",
    booktitle = {2022 IEEE/CVF Conference on Computer Vision and Pattern Recognition Workshops (CVPRW)},
         year = 2022,
        month = jun,
    publisher = {IEEE},
          eid = {1167},
        pages = {1167},
          doi = {10.1109/CVPR52688.2022.01167},
       adsurl = {https://ui.adsabs.harvard.edu/abs/2022cvpr.conf.1167L}
}

@ARTICLE{2012arXiv1211.5063P,
       author = {{Pascanu}, Razvan and {Mikolov}, Tomas and {Bengio}, Yoshua},
        title = "{On the difficulty of training Recurrent Neural Networks}",
      journal = {arXiv e-prints},
         year = 2012,
        month = nov,
          eid = {arXiv:1211.5063},
        pages = {arXiv:1211.5063},
          doi = {10.48550/arXiv.1211.5063},
archivePrefix = {arXiv},
       eprint = {1211.5063},
 primaryClass = {cs.LG},
       adsurl = {https://ui.adsabs.harvard.edu/abs/2012arXiv1211.5063P}
}

@ARTICLE{1997A&A...321..898M,
       author = {{Mendez}, R.~H. and {Soffner}, T.},
        title = "{Improved simulations of the planetary nebula luminosity function.}",
      journal = {\aap},
         year = 1997,
        month = may,
       volume = {321},
        pages = {898-906},
          doi = {10.48550/arXiv.astro-ph/9611128},
archivePrefix = {arXiv},
       eprint = {astro-ph/9611128},
 primaryClass = {astro-ph},
       adsurl = {https://ui.adsabs.harvard.edu/abs/1997A&A...321..898M}
}

@ARTICLE{2012RAA....12.1197C,
       author = {{Cui}, Xiang-Qun and {Zhao}, Yong-Heng and {Chu}, Yao-Quan and {Li}, Guo-Ping and {Li}, Qi and {Zhang}, Li-Ping and {Su}, Hong-Jun and {Yao}, Zheng-Qiu and {Wang}, Ya-Nan and {Xing}, Xiao-Zheng and {Li}, Xin-Nan and {Zhu}, Yong-Tian and {Wang}, Gang and {Gu}, Bo-Zhong and {Luo}, A.-Li and {Xu}, Xin-Qi and {Zhang}, Zhen-Chao and {Liu}, Gen-Rong and {Zhang}, Hao-Tong and {Yang}, De-Hua and {Cao}, Shu-Yun and {Chen}, Hai-Yuan and {Chen}, Jian-Jun and {Chen}, Kun-Xin and {Chen}, Ying and {Chu}, Jia-Ru and {Feng}, Lei and {Gong}, Xue-Fei and {Hou}, Yong-Hui and {Hu}, Hong-Zhuan and {Hu}, Ning-Sheng and {Hu}, Zhong-Wen and {Jia}, Lei and {Jiang}, Fang-Hua and {Jiang}, Xiang and {Jiang}, Zi-Bo and {Jin}, Ge and {Li}, Ai-Hua and {Li}, Yan and {Li}, Ye-Ping and {Liu}, Guan-Qun and {Liu}, Zhi-Gang and {Lu}, Wen-Zhi and {Mao}, Yin-Dun and {Men}, Li and {Qi}, Yong-Jun and {Qi}, Zhao-Xiang and {Shi}, Huo-Ming and {Tang}, Zheng-Hong and {Tao}, Qing-Sheng and {Wang}, Da-Qi and {Wang}, Dan and {Wang}, Guo-Min and {Wang}, Hai and {Wang}, Jia-Ning and {Wang}, Jian and {Wang}, Jian-Ling and {Wang}, Jian-Ping and {Wang}, Lei and {Wang}, Shu-Qing and {Wang}, You and {Wang}, Yue-Fei and {Xu}, Ling-Zhe and {Xu}, Yan and {Yang}, Shi-Hai and {Yu}, Yong and {Yuan}, Hui and {Yuan}, Xiang-Yan and {Zhai}, Chao and {Zhang}, Jing and {Zhang}, Yan-Xia and {Zhang}, Yong and {Zhao}, Ming and {Zhou}, Fang and {Zhou}, Guo-Hua and {Zhu}, Jie and {Zou}, Si-Cheng},
        title = "{The Large Sky Area Multi-Object Fiber Spectroscopic Telescope (LAMOST)}",
      journal = {Research in Astronomy and Astrophysics},
         year = 2012,
        month = sep,
       volume = {12},
       number = {9},
        pages = {1197-1242},
          doi = {10.1088/1674-4527/12/9/003},
       adsurl = {https://ui.adsabs.harvard.edu/abs/2012RAA....12.1197C}
}

@ARTICLE{2012RAA....12..723Z,
       author = {{Zhao}, Gang and {Zhao}, Yong-Heng and {Chu}, Yao-Quan and {Jing}, Yi-Peng and {Deng}, Li-Cai},
        title = "{LAMOST spectral survey {\textemdash} An overview}",
      journal = {Research in Astronomy and Astrophysics},
         year = 2012,
        month = jul,
       volume = {12},
       number = {7},
        pages = {723-734},
          doi = {10.1088/1674-4527/12/7/002},
       adsurl = {https://ui.adsabs.harvard.edu/abs/2012RAA....12..723Z}
}

@ARTICLE{2023ApJS..267...44A,
       author = {{Almeida}, Andr{\'e}s and {Anderson}, Scott F. and {Argudo-Fern{\'a}ndez}, Maria and {Badenes}, Carles and {Barger}, Kat and {Barrera-Ballesteros}, Jorge K. and {Bender}, Chad F. and {Benitez}, Erika and {Besser}, Felipe and {Bird}, Jonathan C. and {Bizyaev}, Dmitry and {Blanton}, Michael R. and {Bochanski}, John and {Bovy}, Jo and {Brandt}, William Nielsen and {Brownstein}, Joel R. and {Buchner}, Johannes and {Bulbul}, Esra and {Burchett}, Joseph N. and {Cano D{\'\i}az}, Mariana and {Carlberg}, Joleen K. and {Casey}, Andrew R. and {Chandra}, Vedant and {Cherinka}, Brian and {Chiappini}, Cristina and {Coker}, Abigail A. and {Comparat}, Johan and {Conroy}, Charlie and {Contardo}, Gabriella and {Cortes}, Arlin and {Covey}, Kevin and {Crane}, Jeffrey D. and {Cunha}, Katia and {Dabbieri}, Collin and {Davidson}, James W. and {Davis}, Megan C. and {de Andrade Queiroz}, Anna Barbara and {De Lee}, Nathan and {M{\'e}ndez Delgado}, Jos{\'e} Eduardo and {Demasi}, Sebastian and {Di Mille}, Francesco and {Donor}, John and {Dow}, Peter and {Dwelly}, Tom and {Eracleous}, Mike and {Eriksen}, Jamey and {Fan}, Xiaohui and {Farr}, Emily and {Frederick}, Sara and {Fries}, Logan and {Frinchaboy}, Peter and {G{\"a}nsicke}, Boris T. and {Ge}, Junqiang and {Gonz{\'a}lez {\'A}vila}, Consuelo and {Grabowski}, Katie and {Grier}, Catherine and {Guiglion}, Guillaume and {Gupta}, Pramod and {Hall}, Patrick and {Hawkins}, Keith and {Hayes}, Christian R. and {Hermes}, J.~J. and {Hern{\'a}ndez-Garc{\'\i}a}, Lorena and {Hogg}, David W. and {Holtzman}, Jon A. and {Ibarra-Medel}, Hector Javier and {Ji}, Alexander and {Jofre}, Paula and {Johnson}, Jennifer A. and {Jones}, Amy M. and {Kinemuchi}, Karen and {Kluge}, Matthias and {Koekemoer}, Anton and {Kollmeier}, Juna A. and {Kounkel}, Marina and {Krishnarao}, Dhanesh and {Krumpe}, Mirko and {Lacerna}, Ivan and {Lago}, Paulo Jakson Assuncao and {Laporte}, Chervin and {Liu}, Chao and {Liu}, Ang and {Liu}, Xin and {Lopes}, Alexandre Roman and {Macktoobian}, Matin and {Majewski}, Steven R. and {Malanushenko}, Viktor and {Maoz}, Dan and {Masseron}, Thomas and {Masters}, Karen L. and {Matijevic}, Gal and {McBride}, Aidan and {Medan}, Ilija and {Merloni}, Andrea and {Morrison}, Sean and {Myers}, Natalie and {M{\'e}sz{\'a}ros}, Szabolcs and {Negrete}, C. Alenka and {Nidever}, David L. and {Nitschelm}, Christian and {Oravetz}, Daniel and {Oravetz}, Audrey and {Pan}, Kaike and {Peng}, Yingjie and {Pinsonneault}, Marc H. and {Pogge}, Rick and {Qiu}, Dan and {Ramirez}, Solange V. and {Rix}, Hans-Walter and {Fern{\'a}ndez Rosso}, Daniela and {Runnoe}, Jessie and {Salvato}, Mara and {Sanchez}, Sebastian F. and {Santana}, Felipe A. and {Saydjari}, Andrew and {Sayres}, Conor and {Schlaufman}, Kevin C. and {Schneider}, Donald P. and {Schwope}, Axel and {Serna}, Javier and {Shen}, Yue and {Sobeck}, Jennifer and {Song}, Ying-Yi and {Souto}, Diogo and {Spoo}, Taylor and {Stassun}, Keivan G. and {Steinmetz}, Matthias and {Straumit}, Ilya and {Stringfellow}, Guy and {S{\'a}nchez-Gallego}, Jos{\'e} and {Taghizadeh-Popp}, Manuchehr and {Tayar}, Jamie and {Thakar}, Ani and {Tissera}, Patricia B. and {Tkachenko}, Andrew and {Hernandez Toledo}, Hector and {Trakhtenbrot}, Benny and {Fern{\'a}ndez-Trincado}, Jos{\'e} G. and {Troup}, Nicholas and {Trump}, Jonathan R. and {Tuttle}, Sarah and {Ulloa}, Natalie and {Vazquez-Mata}, Jose Antonio and {Vera Alfaro}, Pablo and {Villanova}, Sandro and {Wachter}, Stefanie and {Weijmans}, Anne-Marie and {Wheeler}, Adam and {Wilson}, John and {Wojno}, Leigh and {Wolf}, Julien and {Xue}, Xiang-Xiang and {Ybarra}, Jason E. and {Zari}, Eleonora and {Zasowski}, Gail},
        title = "{The Eighteenth Data Release of the Sloan Digital Sky Surveys: Targeting and First Spectra from SDSS-V}",
      journal = {\apjs},
         year = 2023,
        month = aug,
       volume = {267},
       number = {2},
          eid = {44},
        pages = {44},
          doi = {10.3847/1538-4365/acda98},
archivePrefix = {arXiv},
       eprint = {2301.07688},
 primaryClass = {astro-ph.GA},
       adsurl = {https://ui.adsabs.harvard.edu/abs/2023ApJS..267...44A}
}

@ARTICLE{2026AJ....172..238S,
       author = {{S{\'e}randour}, Titouan and {M{\'e}alier}, Anne-Laure and {Colicchio}, Bruno and {Daniel}, Dominique and {Gilberton}, Philippe and {Oros}, Philippe and {Boch}, Thomas and {Abdeslam}, Djaffar Ould and {AIP-PN}},
        title = "{Identifying Planetary Nebulae Candidates from IPHAS and VPHAS+ Surveys via YOLO Deep Learning Models}",
      journal = {\aj},
         year = 2026,
        month = oct,
       volume = {172},
       number = {4},
          eid = {238},
        pages = {238},
          doi = {10.3847/1538-3881/aea417},
       adsurl = {https://ui.adsabs.harvard.edu/abs/2026AJ....172..238S}
}

@ARTICLE{2026arXiv260817380S,
       author = {{Soemitro}, Azlizan A. and {Roth}, Martin M. and {Walsh}, Jeremy R. and {Monreal-Ibero}, Ana and {Pruijt}, Brigitte G.~A. and {Chornay}, Nicholas},
        title = "{Candidates for the most [$O_{III}$] $\lambda5007$-luminous planetary nebula in the Milky Way. I. Integrated light properties of NGC 6572, NGC 6884, and M 1-71}",
      journal = {arXiv e-prints},
         year = 2026,
        month = aug,
          eid = {arXiv:2608.17380},
        pages = {arXiv:2608.17380},
          doi = {10.48550/arXiv.2608.17380},
archivePrefix = {arXiv},
       eprint = {2608.17380},
 primaryClass = {astro-ph.GA},
       adsurl = {https://ui.adsabs.harvard.edu/abs/2026arXiv260817380S}
}

@INPROCEEDINGS{1998IAUS..179..465M,
       author = {{McGlynn}, T. and {Scollick}, K. and {White}, N.},
        title = "{SKYVIEW:The Multi-Wavelength Sky on the Internet}",
    booktitle = {New Horizons from Multi-Wavelength Sky Surveys},
         year = 1998,
       editor = {{McLean}, Brian J. and {Golombek}, Daniel A. and {Hayes}, Jeffrey J.~E. and {Payne}, Harry E.},
       series = {IAU Symposium},
       volume = {179},
        month = jan,
        pages = {465},
       adsurl = {https://ui.adsabs.harvard.edu/abs/1998IAUS..179..465M}
}

@INPROCEEDINGS{2012SPIE.8446E..0UG,
       author = {{Grandmont}, F. and {Drissen}, L. and {Mandar}, Julie and {Thibault}, S. and {Baril}, Marc},
        title = "{Final design of SITELLE: a wide-field imaging Fourier transform spectrometer for the Canada-France-Hawaii Telescope}",
    booktitle = {Ground-based and Airborne Instrumentation for Astronomy IV},
         year = 2012,
       editor = {{McLean}, Ian S. and {Ramsay}, Suzanne K. and {Takami}, Hideki},
       series = {Society of Photo-Optical Instrumentation Engineers (SPIE) Conference Series},
       volume = {8446},
        month = sep,
          eid = {84460U},
        pages = {84460U},
          doi = {10.1117/12.926782},
       adsurl = {https://ui.adsabs.harvard.edu/abs/2012SPIE.8446E..0UG}
}

@ARTICLE{2016A&A...595A...1G,
       author = {{Gaia Collaboration} and {Prusti}, T. and {de Bruijne}, J.~H.~J. and {Brown}, A.~G.~A. and {Vallenari}, A. and {Babusiaux}, C. and {Bailer-Jones}, C.~A.~L. and {Bastian}, U. and {Biermann}, M. and {Evans}, D.~W. and {Eyer}, L. and {Jansen}, F. and {Jordi}, C. and {Klioner}, S.~A. and {Lammers}, U. and {Lindegren}, L. and {Luri}, X. and {Mignard}, F. and {Milligan}, D.~J. and {Panem}, C. and {Poinsignon}, V. and {Pourbaix}, D. and {Randich}, S. and {Sarri}, G. and {Sartoretti}, P. and {Siddiqui}, H.~I. and {Soubiran}, C. and {Valette}, V. and {van Leeuwen}, F. and {Walton}, N.~A. and {Aerts}, C. and {Arenou}, F. and {Cropper}, M. and {Drimmel}, R. and {H{\o}g}, E. and {Katz}, D. and {Lattanzi}, M.~G. and {O'Mullane}, W. and {Grebel}, E.~K. and {Holland}, A.~D. and {Huc}, C. and {Passot}, X. and {Bramante}, L. and {Cacciari}, C. and {Casta{\~n}eda}, J. and {Chaoul}, L. and {Cheek}, N. and {De Angeli}, F. and {Fabricius}, C. and {Guerra}, R. and {Hern{\'a}ndez}, J. and {Jean-Antoine-Piccolo}, A. and {Masana}, E. and {Messineo}, R. and {Mowlavi}, N. and {Nienartowicz}, K. and {Ord{\'o}{\~n}ez-Blanco}, D. and {Panuzzo}, P. and {Portell}, J. and {Richards}, P.~J. and {Riello}, M. and {Seabroke}, G.~M. and {Tanga}, P. and {Th{\'e}venin}, F. and {Torra}, J. and {Els}, S.~G. and {Gracia-Abril}, G. and {Comoretto}, G. and {Garcia-Reinaldos}, M. and {Lock}, T. and {Mercier}, E. and {Altmann}, M. and {Andrae}, R. and {Astraatmadja}, T.~L. and {Bellas-Velidis}, I. and {Benson}, K. and {Berthier}, J. and {Blomme}, R. and {Busso}, G. and {Carry}, B. and {Cellino}, A. and {Clementini}, G. and {Cowell}, S. and {Creevey}, O. and {Cuypers}, J. and {Davidson}, M. and {De Ridder}, J. and {de Torres}, A. and {Delchambre}, L. and {Dell'Oro}, A. and {Ducourant}, C. and {Fr{\'e}mat}, Y. and {Garc{\'\i}a-Torres}, M. and {Gosset}, E. and {Halbwachs}, J.-L. and {Hambly}, N.~C. and {Harrison}, D.~L. and {Hauser}, M. and {Hestroffer}, D. and {Hodgkin}, S.~T. and {Huckle}, H.~E. and {Hutton}, A. and {Jasniewicz}, G. and {Jordan}, S. and {Kontizas}, M. and {Korn}, A.~J. and {Lanzafame}, A.~C. and {Manteiga}, M. and {Moitinho}, A. and {Muinonen}, K. and {Osinde}, J. and {Pancino}, E. and {Pauwels}, T. and {Petit}, J.-M. and {Recio-Blanco}, A. and {Robin}, A.~C. and {Sarro}, L.~M. and {Siopis}, C. and {Smith}, M. and {Smith}, K.~W. and {Sozzetti}, A. and {Thuillot}, W. and {van Reeven}, W. and {Viala}, Y. and {Abbas}, U. and {Abreu Aramburu}, A. and {Accart}, S. and {Aguado}, J.~J. and {Allan}, P.~M. and {Allasia}, W. and {Altavilla}, G. and {{\'A}lvarez}, M.~A. and {Alves}, J. and {Anderson}, R.~I. and {Andrei}, A.~H. and {Anglada Varela}, E. and {Antiche}, E. and {Antoja}, T. and {Ant{\'o}n}, S. and {Arcay}, B. and {Atzei}, A. and {Ayache}, L. and {Bach}, N. and {Baker}, S.~G. and {Balaguer-N{\'u}{\~n}ez}, L. and {Barache}, C. and {Barata}, C. and {Barbier}, A. and {Barblan}, F. and {Baroni}, M. and {Barrado y Navascu{\'e}s}, D. and {Barros}, M. and {Barstow}, M.~A. and {Becciani}, U. and {Bellazzini}, M. and {Bellei}, G. and {Bello Garc{\'\i}a}, A. and {Belokurov}, V. and {Bendjoya}, P. and {Berihuete}, A. and {Bianchi}, L. and {Bienaym{\'e}}, O. and {Billebaud}, F. and {Blagorodnova}, N. and {Blanco-Cuaresma}, S. and {Boch}, T. and {Bombrun}, A. and {Borrachero}, R. and {Bouquillon}, S. and {Bourda}, G. and {Bouy}, H. and {Bragaglia}, A. and {Breddels}, M.~A. and {Brouillet}, N. and {Br{\"u}semeister}, T. and {Bucciarelli}, B. and {Budnik}, F. and {Burgess}, P. and {Burgon}, R. and {Burlacu}, A. and {Busonero}, D. and {Buzzi}, R. and {Caffau}, E. and {Cambras}, J. and {Campbell}, H. and {Cancelliere}, R. and {Cantat-Gaudin}, T. and {Carlucci}, T. and {Carrasco}, J.~M. and {Castellani}, M. and {Charlot}, P. and {Charnas}, J. and {Charvet}, P. and {Chassat}, F. and {Chiavassa}, A. and {Clotet}, M. and {Cocozza}, G. and {Collins}, R.~S. and {Collins}, P. and {Costigan}, G.},
        title = "{The Gaia mission}",
      journal = {\aap},
         year = 2016,
        month = nov,
       volume = {595},
          eid = {A1},
        pages = {A1},
          doi = {10.1051/0004-6361/201629272},
archivePrefix = {arXiv},
       eprint = {1609.04153},
 primaryClass = {astro-ph.IM},
       adsurl = {https://ui.adsabs.harvard.edu/abs/2016A&A...595A...1G}
}

@ARTICLE{2023A&A...674A...1G,
       author = {{Gaia Collaboration} and {Vallenari}, A. and {Brown}, A.~G.~A. and {Prusti}, T. and {de Bruijne}, J.~H.~J. and {Arenou}, F. and {Babusiaux}, C. and {Biermann}, M. and {Creevey}, O.~L. and {Ducourant}, C. and {Evans}, D.~W. and {Eyer}, L. and {Guerra}, R. and {Hutton}, A. and {Jordi}, C. and {Klioner}, S.~A. and {Lammers}, U.~L. and {Lindegren}, L. and {Luri}, X. and {Mignard}, F. and {Panem}, C. and {Pourbaix}, D. and {Randich}, S. and {Sartoretti}, P. and {Soubiran}, C. and {Tanga}, P. and {Walton}, N.~A. and {Bailer-Jones}, C.~A.~L. and {Bastian}, U. and {Drimmel}, R. and {Jansen}, F. and {Katz}, D. and {Lattanzi}, M.~G. and {van Leeuwen}, F. and {Bakker}, J. and {Cacciari}, C. and {Casta{\~n}eda}, J. and {De Angeli}, F. and {Fabricius}, C. and {Fouesneau}, M. and {Fr{\'e}mat}, Y. and {Galluccio}, L. and {Guerrier}, A. and {Heiter}, U. and {Masana}, E. and {Messineo}, R. and {Mowlavi}, N. and {Nicolas}, C. and {Nienartowicz}, K. and {Pailler}, F. and {Panuzzo}, P. and {Riclet}, F. and {Roux}, W. and {Seabroke}, G.~M. and {Sordo}, R. and {Th{\'e}venin}, F. and {Gracia-Abril}, G. and {Portell}, J. and {Teyssier}, D. and {Altmann}, M. and {Andrae}, R. and {Audard}, M. and {Bellas-Velidis}, I. and {Benson}, K. and {Berthier}, J. and {Blomme}, R. and {Burgess}, P.~W. and {Busonero}, D. and {Busso}, G. and {C{\'a}novas}, H. and {Carry}, B. and {Cellino}, A. and {Cheek}, N. and {Clementini}, G. and {Damerdji}, Y. and {Davidson}, M. and {de Teodoro}, P. and {Nu{\~n}ez Campos}, M. and {Delchambre}, L. and {Dell'Oro}, A. and {Esquej}, P. and {Fern{\'a}ndez-Hern{\'a}ndez}, J. and {Fraile}, E. and {Garabato}, D. and {Garc{\'\i}a-Lario}, P. and {Gosset}, E. and {Haigron}, R. and {Halbwachs}, J.-L. and {Hambly}, N.~C. and {Harrison}, D.~L. and {Hern{\'a}ndez}, J. and {Hestroffer}, D. and {Hodgkin}, S.~T. and {Holl}, B. and {Jan{\ss}en}, K. and {Jevardat de Fombelle}, G. and {Jordan}, S. and {Krone-Martins}, A. and {Lanzafame}, A.~C. and {L{\"o}ffler}, W. and {Marchal}, O. and {Marrese}, P.~M. and {Moitinho}, A. and {Muinonen}, K. and {Osborne}, P. and {Pancino}, E. and {Pauwels}, T. and {Recio-Blanco}, A. and {Reyl{\'e}}, C. and {Riello}, M. and {Rimoldini}, L. and {Roegiers}, T. and {Rybizki}, J. and {Sarro}, L.~M. and {Siopis}, C. and {Smith}, M. and {Sozzetti}, A. and {Utrilla}, E. and {van Leeuwen}, M. and {Abbas}, U. and {{\'A}brah{\'a}m}, P. and {Abreu Aramburu}, A. and {Aerts}, C. and {Aguado}, J.~J. and {Ajaj}, M. and {Aldea-Montero}, F. and {Altavilla}, G. and {{\'A}lvarez}, M.~A. and {Alves}, J. and {Anders}, F. and {Anderson}, R.~I. and {Anglada Varela}, E. and {Antoja}, T. and {Baines}, D. and {Baker}, S.~G. and {Balaguer-N{\'u}{\~n}ez}, L. and {Balbinot}, E. and {Balog}, Z. and {Barache}, C. and {Barbato}, D. and {Barros}, M. and {Barstow}, M.~A. and {Bartolom{\'e}}, S. and {Bassilana}, J.-L. and {Bauchet}, N. and {Becciani}, U. and {Bellazzini}, M. and {Berihuete}, A. and {Bernet}, M. and {Bertone}, S. and {Bianchi}, L. and {Binnenfeld}, A. and {Blanco-Cuaresma}, S. and {Blazere}, A. and {Boch}, T. and {Bombrun}, A. and {Bossini}, D. and {Bouquillon}, S. and {Bragaglia}, A. and {Bramante}, L. and {Breedt}, E. and {Bressan}, A. and {Brouillet}, N. and {Brugaletta}, E. and {Bucciarelli}, B. and {Burlacu}, A. and {Butkevich}, A.~G. and {Buzzi}, R. and {Caffau}, E. and {Cancelliere}, R. and {Cantat-Gaudin}, T. and {Carballo}, R. and {Carlucci}, T. and {Carnerero}, M.~I. and {Carrasco}, J.~M. and {Casamiquela}, L. and {Castellani}, M. and {Castro-Ginard}, A. and {Chaoul}, L. and {Charlot}, P. and {Chemin}, L. and {Chiaramida}, V. and {Chiavassa}, A. and {Chornay}, N. and {Comoretto}, G. and {Contursi}, G. and {Cooper}, W.~J. and {Cornez}, T. and {Cowell}, S. and {Crifo}, F. and {Cropper}, M. and {Crosta}, M. and {Crowley}, C. and {Dafonte}, C. and {Dapergolas}, A. and {David}, M. and {David}, P. and {de Laverny}, P. and {De Luise}, F. and {De March}, R.},
        title = "{Gaia Data Release 3. Summary of the content and survey properties}",
      journal = {\aap},
         year = 2023,
        month = jun,
       volume = {674},
          eid = {A1},
        pages = {A1},
          doi = {10.1051/0004-6361/202243940},
archivePrefix = {arXiv},
       eprint = {2208.00211},
 primaryClass = {astro-ph.GA},
       adsurl = {https://ui.adsabs.harvard.edu/abs/2023A&A...674A...1G}
}

@ARTICLE{2019AJ....157...98G,
       author = {{Ginsburg}, Adam and {Sip{\H{o}}cz}, Brigitta M. and {Brasseur}, C.~E. and {Cowperthwaite}, Philip S. and {Craig}, Matthew W. and {Deil}, Christoph and {Guillochon}, James and {Guzman}, Giannina and {Liedtke}, Simon and {Lian Lim}, Pey and {Lockhart}, Kelly E. and {Mommert}, Michael and {Morris}, Brett M. and {Norman}, Henrik and {Parikh}, Madhura and {Persson}, Magnus V. and {Robitaille}, Thomas P. and {Segovia}, Juan-Carlos and {Singer}, Leo P. and {Tollerud}, Erik J. and {de Val-Borro}, Miguel and {Valtchanov}, Ivan and {Woillez}, Julien and {Astroquery Collaboration} and {a subset of astropy Collaboration}},
        title = "{astroquery: An Astronomical Web-querying Package in Python}",
      journal = {\aj},
         year = 2019,
        month = mar,
       volume = {157},
       number = {3},
          eid = {98},
        pages = {98},
          doi = {10.3847/1538-3881/aafc33},
archivePrefix = {arXiv},
       eprint = {1901.04520},
 primaryClass = {astro-ph.IM},
       adsurl = {https://ui.adsabs.harvard.edu/abs/2019AJ....157...98G}
}

@software{2010scpy.soft.....M,
       author = {{McKinney}, Wes},
        title = "{Data Structures for Statistical Computing in Python}",
 howpublished = {Proceedings of the 9th Python in Science Conference},
         year = 2010,
        month = jan,
          doi = {10.25080/Majora-92bf1922-00a},
       adsurl = {https://ui.adsabs.harvard.edu/abs/2010scpy.soft.....M}
}

@ARTICLE{2011JMLR...12.2825P,
       author = {{Pedregosa}, Fabian and {Varoquaux}, Ga{\"e}l and {Gramfort}, Alexandre and {Michel}, Vincent and {Thirion}, Bertrand and {Grisel}, Olivier and {Blondel}, Mathieu and {M{\"u}ller}, Andreas and {Nothman}, Joel and {Louppe}, Gilles and {Prettenhofer}, Peter and {Weiss}, Ron and {Dubourg}, Vincent and {Vanderplas}, Jake and {Passos}, Alexandre and {Cournapeau}, David and {Brucher}, Matthieu and {Perrot}, Matthieu and {Duchesnay}, {\'E}douard},
        title = "{Scikit-learn: Machine Learning in Python}",
      journal = {Journal of Machine Learning Research},
         year = 2011,
        month = oct,
       volume = {12},
        pages = {2825-2830},
          doi = {10.48550/arXiv.1201.0490},
archivePrefix = {arXiv},
       eprint = {1201.0490},
 primaryClass = {cs.LG},
       adsurl = {https://ui.adsabs.harvard.edu/abs/2011JMLR...12.2825P}
}

@ARTICLE{2007CSE.....9...90H,
       author = {{Hunter}, John D.},
        title = "{Matplotlib: A 2D Graphics Environment}",
      journal = {Computing in Science and Engineering},
         year = 2007,
        month = jan,
       volume = {9},
       number = {3},
        pages = {90-95},
          doi = {10.1109/MCSE.2007.55},
       adsurl = {https://ui.adsabs.harvard.edu/abs/2007CSE.....9...90H}
}

@ARTICLE{2020Natur.585..357H,
       author = {{Harris}, Charles R. and {Millman}, K. Jarrod and {van der Walt}, St{\'e}fan J. and {Gommers}, Ralf and {Virtanen}, Pauli and {Cournapeau}, David and {Wieser}, Eric and {Taylor}, Julian and {Berg}, Sebastian and {Smith}, Nathaniel J. and {Kern}, Robert and {Picus}, Matti and {Hoyer}, Stephan and {van Kerkwijk}, Marten H. and {Brett}, Matthew and {Haldane}, Allan and {del R{\'\i}o}, Jaime Fern{\'a}ndez and {Wiebe}, Mark and {Peterson}, Pearu and {G{\'e}rard-Marchant}, Pierre and {Sheppard}, Kevin and {Reddy}, Tyler and {Weckesser}, Warren and {Abbasi}, Hameer and {Gohlke}, Christoph and {Oliphant}, Travis E.},
        title = "{Array programming with NumPy}",
      journal = {\nat},
         year = 2020,
        month = sep,
       volume = {585},
       number = {7825},
        pages = {357-362},
          doi = {10.1038/s41586-020-2649-2},
archivePrefix = {arXiv},
       eprint = {2006.10256},
 primaryClass = {cs.MS},
       adsurl = {https://ui.adsabs.harvard.edu/abs/2020Natur.585..357H}
}

@ARTICLE{2020NaMet..17..261V,
       author = {{Virtanen}, Pauli and {Gommers}, Ralf and {Oliphant}, Travis E. and {Haberland}, Matt and {Reddy}, Tyler and {Cournapeau}, David and {Burovski}, Evgeni and {Peterson}, Pearu and {Weckesser}, Warren and {Bright}, Jonathan and {van der Walt}, St{\'e}fan J. and {Brett}, Matthew and {Wilson}, Joshua and {Millman}, K. Jarrod and {Mayorov}, Nikolay and {Nelson}, Andrew R.~J. and {Jones}, Eric and {Kern}, Robert and {Larson}, Eric and {Carey}, C.~J. and {Polat}, {\.I}lhan and {Feng}, Yu and {Moore}, Eric W. and {VanderPlas}, Jake and {Laxalde}, Denis and {Perktold}, Josef and {Cimrman}, Robert and {Henriksen}, Ian and {Quintero}, E.~A. and {Harris}, Charles R. and {Archibald}, Anne M. and {Ribeiro}, Ant{\^o}nio H. and {Pedregosa}, Fabian and {van Mulbregt}, Paul and {SciPy 1.  0 Contributors}},
        title = "{SciPy 1.0: fundamental algorithms for scientific computing in Python}",
      journal = {Nature Methods},
         year = 2020,
        month = feb,
       volume = {17},
        pages = {261-272},
          doi = {10.1038/s41592-019-0686-2},
archivePrefix = {arXiv},
       eprint = {1907.10121},
 primaryClass = {cs.MS},
       adsurl = {https://ui.adsabs.harvard.edu/abs/2020NaMet..17..261V}
}

@INPROCEEDINGS{2024cvpr.conf.1605Z,
       author = {{Zhao}, Yian and {Lv}, Wenyu and {Xu}, Shangliang and {Wei}, Jinman and {Wang}, Guanzhong and {Dang}, Qingqing and {Liu}, Yi and {Chen}, Jie},
        title = "{DETRs Beat YOLOs on Real-time Object Detection}",
    booktitle = {2024 IEEE/CVF Conference on Computer Vision and Pattern Recognition (CVPR)},
         year = 2024,
        month = jun,
    publisher = {IEEE},
        pages = {16965-16974},
          doi = {10.1109/CVPR52733.2024.01605},
       adsurl = {https://ui.adsabs.harvard.edu/abs/2024cvpr.conf.1605Z}
}

@INCOLLECTION{2015LNCS.9351..234R,
       author = {{Ronneberger}, Olaf and {Fischer}, Philipp and {Brox}, Thomas},
        title = "{U-Net: Convolutional Networks for Biomedical Image Segmentation}",
    booktitle = {Lecture Notes in Computer Science},
         year = 2015,
       volume = {9351},
        pages = {234-241},
          doi = {10.1007/978-3-319-24574-4_28},
       adsurl = {https://ui.adsabs.harvard.edu/abs/2015LNCS.9351..234R}
}

@ARTICLE{2025arXiv251109554R,
       author = {{Robinson}, Isaac and {Robicheaux}, Peter and {Popov}, Matvei and {Ramanan}, Deva and {Peri}, Neehar},
        title = "{RF-DETR: Neural Architecture Search for Real-Time Detection Transformers}",
      journal = {arXiv e-prints},
         year = 2025,
        month = nov,
          eid = {arXiv:2511.09554},
        pages = {arXiv:2511.09554},
          doi = {10.48550/arXiv.2511.09554},
archivePrefix = {arXiv},
       eprint = {2511.09554},
 primaryClass = {cs.CV},
       adsurl = {https://ui.adsabs.harvard.edu/abs/2025arXiv251109554R}
}
\bibliographystyle{aasjournalv7}



\end{document}